\documentclass[12pt]{article}

\usepackage{amsfonts,amsmath}
\usepackage{xcolor}

\usepackage[T1]{fontenc}

\usepackage{amssymb}
\usepackage{ytableau}
\usepackage{booktabs}
\usepackage{mathtools} 
\usepackage{quiver} 
\usepackage{tikz-cd} 
\usepackage{graphicx}
\usepackage{pgfplots}
\pgfplotsset{compat=1.5}
\usepgfplotslibrary{fillbetween} 
\usepackage{tikz}
\usetikzlibrary{arrows.meta,bending,patterns,angles,calc}
\usetikzlibrary{patterns.meta}
\usepackage{siunitx}
\usetikzlibrary{decorations.shapes}
\tikzset{decorate sep/.style 2 args=
{decorate,decoration={shape backgrounds,shape=circle,shape size=#1,shape sep=#2}}}

\pgfplotsset{ignore zero/.style={%
  #1ticklabel={\ifdim\tick pt=0pt \else\pgfmathprintnumber{\tick}\fi}
}} 

\newcommand{\dr}{{{\rm d}}}

\renewcommand{\theequation}{\thesection.\arabic{equation}}

\makeatletter \@addtoreset{equation}{section} \makeatother

{\vspace{3mm} }

\def\al{\alpha}

\def\*{\star}
\def\e{\mathbf{e}}
\def\E2{\mathbf{E}}
\def\w{\mathbf{w}}

\def\u{\mathbf{u}}

\def\rf{\boldsymbol{r}}
\def\phif{\boldsymbol{\phi}}
\def\varphif{\boldsymbol{\varphi}}
\def\vf{\boldsymbol{v}}
\def\s{\overline{s}}

\newcommand{\be}{\begin{equation}}
\newcommand{\ee}{\end{equation}}
\newcommand{\bee}{\begin{eqnarray}}
\newcommand{\beee}{\begin{array}}
\newcommand{\eee}{\end{eqnarray}}
\newcommand{\eeee}{\end{array}}
\newcommand{\ga}{\alpha}
\newcommand{\gb}{\beta}
\newcommand{\gga}{\gamma}

\newcommand{\ua}{\underline{A}}
\newcommand{\ub}{\underline{B}}

\newcommand{\gd}{\delta}
\newcommand{\gl}{\lambda}
\newcommand{\gk}{\varkappa}
\newcommand{\gep}{\epsilon}

\newcommand{\go}{\omega}

\newcommand{\dal}{\dot \alpha}
\newcommand{\dgb}{\dot \beta}
\newcommand{\dgga}{\dot \gamma}

\newcommand{\half}{\frac{1}{2}}

\newcommand{\p}{\partial}

\newcommand{\rom}[1]{\uppercase\expandafter{\romannumeral #1\relax}}

\newcommand{\so}{\mathfrak{so}}
\newcommand{\iso}{\mathfrak{iso}}
\renewcommand{\sl}{\mathfrak{sl}}
\renewcommand{\sp}{\mathfrak{sp}}

\newcommand{\h}{\mathfrak{h}}
\renewcommand{\u}{\mathfrak{u}}

\begin{document}
    
\begin{flushright}
FIAN/TD/11-2026\\
\end{flushright}

\vspace{0.5cm}
\begin{center}
{\large\bf Classifying double copies and multicopies in AdS}

\vspace{1 cm}

\textbf{V.E.~Didenko$^{1}$ and N.K.~Dosmanbetov$^{2}$}\\

\vspace{1 cm}

\textbf{}\textbf{}\\
 \vspace{0.5cm}
 \textit{$^{1}$I.E. Tamm Department of Theoretical Physics,
Lebedev Physical Institute,}\\
 \textit{ Leninsky prospect 53, 119991, Moscow, Russia }\\
 \vspace{0.5cm}
 {\it
			$^2$Moscow Institute of Physics and Technology,\\
			Institutsky lane 9, 141700, Dolgoprudny, Moscow region, Russia}

\par\end{center}

\begin{center}
\vspace{0.6cm}
e-mails: didenko@lpi.ru, dosmanbetov.nk@phystech.edu \\
\par\end{center}

\vspace{0.4cm}

\begin{abstract}
\noindent In this paper, we draw a parallel between solutions of pure three-dimensional gravity with a negative cosmological constant and classical double copies in four dimensions. In the former case, topological solutions, such as the BTZ black hole, deficit angles, and naked singularities, emerge from identifying points in AdS using elements from its isometry algebra  $\so(2,2)$. The type of solution corresponds one-to-one with the orbits of $\so(2,2)$.
We demonstrate how various double copies of four-dimensional AdS gravity similarly arise from the  $\so(2,3)$  isometry elements, which also correspond one-to-one with their orbits through a Penrose-type transform. We classify all such elements and generate the corresponding double copies, which include AdS black holes, black branes, and many others. The double-copy isometries originate from the centralizer of a given AdS isometry, allowing us to define canonical coordinates associated with its Abelian part. Additionally, the two Casimir invariants of  $\so(2,3)$ feature in the metrics via the characteristic polynomial of the AdS symmetry parameter. Our classification naturally extends to higher spins, providing nonequivalent multicopies at the linearized level.

\end{abstract}
\newpage

\tableofcontents
\newpage

\section{Introduction}

The relation between gravity and gauge theories has long been anticipated. One of the first compelling examples of this connection was presented through the KLT relations \cite{Kawai:1985xq}, derived from string theory. These relations demonstrate that asymptotic states of gravity can be represented as tensor products of gauge states. More recently, insights have emerged from the BCJ relations \cite{Bern:2008qj, Bern:2010yg, Bern:2010ue}, which are based on the intriguing color-kinematics duality found in Yang-Mills theory. This duality allows for the reproduction of gravity amplitudes by substituting color factors with kinematic ones in gauge theory, highlighting the so-called double copy structure of gravity. Increasing evidence \cite{Bern:2019prr} suggests that this double copy phenomenon continues to hold true even at the loop level.

The double-copy duality suggests that two theories are equivalent at the level of classical solutions. This means that classical solutions in gravity should be expressible, possibly in a non-local way, using building blocks derived from solutions in gauge theory. A prominent example of this mapping is the Kerr black hole in four-dimensional anti-de Sitter (AdS) space, which can be represented in the well-known Kerr-Schild form \cite{Kerr:1965vyg, Kerr:1965wfc}.

\begin{equation}\label{Kerr-Schild}
    g_{\mu\nu}=\bar g_{\mu\nu}+{2}M\phi_{\mu\nu}\,,\qquad \phi_{\mu\nu}=\phi k_{\mu}k_{\nu}\,.
\end{equation}
Here, $\bar g_{\mu\nu}$ is the background metric (Minkowski or (A)dS),  $M$ is the mass of a black hole, $k_{\mu}$ is the Kerr-Schild vector, which is null and geodesic, while $\phi$ is a certain scalar function. As shown in \cite{Didenko:2008va} (see also \cite{Bahjat-Abbas:2017htu}), the Kerr black hole satisfies some additional conditions, which are properties of the solution itself, rather than direct consequences of the Einstein equations. That is, the field $\phi_{\mu}=\phi k_{\mu}$ is a solution of the Maxwell equations, while $\phi$ satisfies the conformal Klein-Gordon equation. These properties make the Kerr graviton $\phi_{\mu\nu}$ a square of the gauge field $\phi_{\mu}$ up to a factor attributed to a massless scalar. It was also shown that such a doubling takes place at the curvature level. Again, up to a massless factor, schematically put, 
\begin{equation}\label{Wel DC}
    \textnormal{Weyl}\sim(\textnormal{Maxwell})^2\,.
\end{equation}
The latter Eq. \eqref{Wel DC} was, in fact, known long before in \cite{Walker:1970un, Hughston:1972qf}.

That the aforementioned doubling of black holes in terms of a gauge field came from color kinematics duality in the asymptotically flat case was proposed in \cite{Monteiro:2014cda, Luna:2018dpt} and was dubbed the classical double copy: the Kerr-Schild double copy and the Weyl one. In particular, the role of Kerr's function $\phi$ is explained in this context. It was suggested to have an origin from a biadjoint scalar theory, called the zeroth copy, that arises upon replacing kinematical factors with color ones in amplitudes. Its equation of motion is non-linear in general, but it reduces to the free massless scalar in the Abelian case. The fact that the Kerr black hole is described by the Abelian double copy has to do with the Kerr-Schild ``gauge'' \eqref{Kerr-Schild} that linearizes Einstein equations.  These results have caused substantial growth in interest in further research on classical double copy in the literature; see \cite{White:2024pve} for a review and references. 

While the double copy works in any dimension, the case of three space-time dimensions is distinguished in that pure gravity is topological and, therefore, there are no non-trivial scattering amplitudes. Nevertheless, the BTZ black hole \cite{Banados:1992wn} formally admits the double copy decomposition in full analogy with its higher-dimensional counterparts; see \cite{Carrillo-Gonzalez:2017iyj}. From the topological point of view, the BTZ black hole is obtained from empty AdS by identifying its points using the proper isometry of $\so(2,2)$, \cite{Banados:1992gq}. More generally, there is a class of solutions that various $\so(2,2)$ adjoint orbits generate. They were classified in detail in \cite{Banados:1992gq}. More generally, the double copy in curved space-times is reviewed in \cite{CarrilloGonzalez:2026phk}. 

The natural question now is: do the background isometries generate classical double copies in $d>3$? The key difference is that, of course, the higher-dimensional case is not topological; {\it i.e.,} the double copies there are not locally diffeomorphic to an empty background. This implies they cannot be obtained via the global background factorization.\footnote{Solutions of the BTZ type have been considered in $d>3$, see, {\it e.g.,} \cite{Holst:1997tm}, where their global properties were investigated{, and \cite{Banados:1998dc}, where the four-dimensional BTZ-type spacetime obtained by identifications in AdS$_4$ was properly interpreted geometrically as a black hole. {The BTZ$_3$ black hole was described within the unfolded approach in \cite{Didenko:2006zd}.} Both the BTZ$_3$ and BTZ$_4$ metrics were extended by means of the unfolded formalism in \cite{Aros:2019pgj}; see also \cite{Iazeolla:2022dal}}.} However, given the similarities between AdS black holes in $d=3$ and those in $d>3$, it is still possible that certain higher-dimensional solutions are generated by the AdS isometry through one mechanism or another.

That this is indeed the case has been shown to be so in $d=4$, \cite{Didenko:2009tc} for a general AdS isometry $K\in \so(2,3)$. Specifically, assuming there is a map from the AdS global symmetry condition for a parameter $K\in \so(2,3)$ into the Petrov type D Weyl curvature, the integrating flow that provides such a deformation was manifestly constructed. The map obtained not only provides manifest expressions for the Weyl double copy in terms of $K$, but also gives algebraic expressions for the Kerr-Schild vector $k^{\mu}$ and the zeroth copy $\phi$ in a coordinate-independent way.

In \cite{Didenko:2009td}, it was recognized that the integrating flow constructed for the Weyl double copy essentially serves as the AdS analog of the Penrose transform \cite{Eastwood:1981jy} (see also \cite{Adamo:2017qyl}), in which $K$ plays a role as the rank two dual twistor;\footnote{Due to the spinorial isomorphism $\so(2,3)=\sp(4, \mathbb{R})$, the AdS symmetry parameter $K$ can be represented as a four by four symmetric matrix $K_{AB}=K_{BA}$.} {for the twistorial origin of the classical double copy in flat space, see \cite{White:2020sfn} and subsequent developments \cite{Chacon:2021wbr, Chacon:2021hfe, Farnsworth:2021wvs, Luna:2022dxo}.} Some immediate consequences of the Penrose transform are the following: (i) parameters $K$ corresponding to different $\so(2,3)$ orbits provide different double copies. (ii) The algebra of the leftover global symmetries of the generated solutions is given by all elements $\xi\in \so(2,3)$, such that $[\xi, K]=0$, {\it i.e.}, it comes as a centralizer of $K$. 

The aim of our paper is to classify all adjoint orbits $\so(2,3)$ and construct the corresponding double copies. The first part of this problem has been solved in various contexts and is available in the literature; see {\it e.g.,} \cite{Holst:1997tm, Madden:2004yc, Figueroa-OFarrill:2004lpm, Lovrekovic:2025dwd}. We revisit the problem of adjoint orbits following \cite{Holst:1997tm} and use these results for constructing solutions.    

A similar but not quite exhaustive analysis has been carried out in $d=5$ in \cite{Didenko:2011ir}, where a few examples of double copies were constructed in terms of the background Killing vector, which could be space-like, time-like, or null, and also recently in \cite{Easson:2023dbk}.            

Another aspect of classical double copy that hints it might be a phenomenon of a larger scale than the original gauge/gravity duality implies is the puzzling value of Kerr's zeroth copy mass in $d$ dimensional AdS spacetime with cosmological constant $\Lambda$,
\begin{equation}
    m^2=2\Lambda (d-3)\,,
\end{equation}
which is conformal in $d=4$ and $d=6$, but no longer so in other dimensions. In \cite{Didenko:2022qxq}, this mass was shown to be exactly the one that the higher-spin symmetry provides in any dimension. Moreover, its value guaranties that the Kerr-Schild double copy extends to gauge fields of all spins  $s > 2$ , as described by the Fronsdal equations \cite{Fronsdal:1978rb, Metsaev:1997nj}, which include a specific $s$  and  $d$-dependent mass-like term in AdS. This formulation reveals what was referred to as the higher-spin multicopy in \cite{Didenko:2021vui, Didenko:2022qxq}; see also \cite{Didenko:2008va}, where the Kerr-Schild multicopy structure was observed earlier in $d=4$. This signals that higher-spin interactions might reveal the yet-to-be-discovered multicopy structure. {A possible group-theoretical origin of this structure was identified in \cite{Iazeolla:2011cb, Iazeolla:2017vng}. There, the generalized type-D higher-spin black holes, {\it i.e.,} certain Weyl multicopy solutions, were interpreted at the level of $\so(2,3)$ representation theory and related to the so-called twisted projectors, namely Penrose-transformed (in the unfolded sense) projectors onto singleton states, twisted by the star product with the inner Klein operator $\kappa_y$. As found in \cite{Iazeolla:2008ix}, such twisted projectors are dual to the tensor product of singleton and anti-singleton states, which decomposes over irreducible representations of all spins according to the twisted Flato--Fronsdal theorem \cite{Iazeolla:2008ix}, later improved in \cite{Basile:2018dzi}. This composite nature naturally accounts for the simultaneous appearance of fields of all spins, together with the zeroth, single, and double copies, and may well be at the root of the multicopy structure. }

Research on higher-spin multicopy is limited. However, existing findings indicate that the classical double copy is a special case of the more general multicopy, specifically for spins  $s\leq 2$. The literature on classical multicopy includes results at the free level \cite{Didenko:2008va, Didenko:2011ir, Didenko:2021vui, Didenko:2022qxq, Brown:2025xlo}, as well as a few nonlinear generalizations \cite{Didenko:2021vdb, Misuna:2026has}. Additionally, certain exact higher-spin solutions exhibit a clear multicopy structure {\cite{Didenko:2009td, Iazeolla:2007wt, Iazeolla:2011cb}}{; in particular, $Sp(4,\mathbb{R})$-inequivalent Weyl multicopy solutions of the Vasiliev equations associated with symmetry-enhanced $\so(2,3)$ orbits were constructed in \cite{Iazeolla:2011cb}, and some of them were later examined in more detail in \cite{Iazeolla:2020jee}; see also \cite{Aros:2019pgj}}. It is important to note that, despite similarities, there is a distinction between Kerr-Schild double copies and Kerr-Schild multicopies; the latter are not linearly exact.\footnote{\label{foot:HSlocal}{For higher-spin multicopies this statement is more subtle, as it depends on the choice of field variables. It holds provided the locality of the higher-spin equations is imposed. On the other hand, in a non-local frame, the solutions obtained in \cite{Didenko:2009td, Iazeolla:2007wt, Iazeolla:2011cb} provide examples of linearly exact Weyl multicopies.}} For example, the next-to-leading corrections to a planar multicopy associated with the specific orbit have already been calculated in \cite{Didenko:2021vdb}, and the result remains a multicopy.

Considering the applications of higher-spin multicopy at the nonlinear level, we extend our classification of AdS isometry-generated double copies to include bosonic gauge fields of all spins. Specifically, for each orbit $K\in \so(2,3)$ we find the higher-spin Weyl multicopy, thus classifying the corresponding set of solutions. This problem is straightforward to solve using the Penrose transform. We should also note that the problem of constructing exact solutions in higher-spin theory with leftover space-time symmetry from various subalgebras of $\so(2,3)$ was addressed in \cite{Sezgin:2005pv, Aros:2017ror}.

{Since the paper is intended to be self-contained, Sections~\ref{sec:ads}--\ref{sec:class} largely review known material. Section~\ref{sec:ads} collects standard facts about AdS$_4$ and its isometries. {The three-dimensional analogy recalled at the beginning of Section~\ref{sec:class} follows \cite{Banados:1992wn, Banados:1992gq}.} Section~\ref{sec:dc} reviews the construction of the Weyl and Kerr-Schild double copies, the Kerr-Schild vectors, and the metric reconstruction from the AdS symmetry parameter \cite{Didenko:2009tc}, as well as the higher-spin multicopies \cite{Didenko:2008va, Didenko:2021vui, Didenko:2022qxq} and the unfolded form of the AdS Penrose transform \cite{Didenko:2009td, Didenko:2021vui}. Section~\ref{sec:orbits} reproduces the classification of the adjoint $\so(2,3)$ orbits of \cite{Holst:1997tm}. 

The new elements of these sections include comments on the validity and limitations of the constructions reviewed in Sec.~\ref{sec:dc}. Specifically, we discuss cases, where the Penrose transform requires a regularization, which yields Petrov type N solutions. Additionally, in Section ~\ref{sec:class} we show that parameters $K$ from the same adjoint orbit generate equivalent solutions, which reduces the classification of the double copies to that of the adjoint $\so(2,3)$ orbits { (analogous steps were used in \cite{Iazeolla:2011cb} for the symmetry-enhanced orbits in the context of exact multicopy solutions of the Vasiliev equations)}. Finally, we propose an extension of the construction to non-Hermitian parameters $K$ satisfying $[K, K^\dagger]=0$ (Sec.~\ref{miss}), illustrated by the Pleba\'nski-Demia\'nski metric, which is not captured by the classification based on a real parameter $K$ { (complexified parameters $K_{AB}$ were also considered in \cite{Iazeolla:2011cb})}.

The main new results of the paper are contained in Sections 5--9 and summarized in Tables~2 and~5: the case-by-case construction of the double copies for all adjoint orbits, their identification with known spacetimes, the canonical coordinates adapted to their isometries, the symmetry-enhanced and degenerate cases, the unified Carter-Pleba\'nski description of all orbits with two commuting isometries, and the corresponding higher-spin multicopies.

}The paper is organized as follows: First, we introduce the basic conventions and notations. {Table 1 reproduces the classification of the adjoint $\so(2,3)$ orbits of \cite{Holst:1997tm}, while Table 2 summarizes our results on double copies, classifying Einstein spacetimes according to the $\so(2,3)$ parameter. Section~\ref{sec:ads} collects the necessary facts about AdS$_4$.}

Section \ref{sec:dc} {reviews, following \cite{Didenko:2009tc, Didenko:2009td},} how specific Weyl double copies and Kerr-Schild double copies arise from the background isometry, along with a reconstruction of the metric. We also {recall} how these results can be extended to higher-spin multicopies at the linearized order using the Penrose transform{, and comment on the limitations of these constructions}.

In Sec. \ref{sec:class}, {after recalling the analogous classification of solutions of three-dimensional gravity by the adjoint orbits of $\so(2,2)$,} we clarify how $\so(2,3)$ orbits correspond to inequivalent double copies, while also noting that our proposed classification is not exhaustive {and proposing its extension to non-Hermitian parameters}. Section \ref{sec:orbits} {reviews} the standard methods for classifying $\so(2,3)$ orbits.

Sections \ref{sec:I} through \ref{sec:V}  provide a detailed, case-by-case construction of double copies, which constitutes the main new part of the paper. In Sec.~9 we show that all double copies with two commuting isometries are real sections of the Carter-Pleba\'nski metric, with structure functions determined by the characteristic polynomial of $K$. We conclude in Section \ref{sec:conc}.
For further reference, we have included six technical appendices in our paper.   

\subsection{Conventions and summary of the results}
To state our main results, let us introduce the index notation used throughout the paper.
\begin{itemize}
        \item[\textbullet] \textbf{Space-time:} Lowercase Greek letters from the end of the alphabet ($\mu, \nu, \xi, \dots=0\dots 3$) are used for four-dimensional space-time tensor indices.
        \item[\textbullet] \textbf{Fiber space:} Lowercase Latin letters ($a, b, c, \dots=0,\dots, 3$) are used for four-dimensional fiber tensor indices.
    \item[\textbullet] \textbf{Ambient space:} Embedding of AdS into the flat five-dimensional space is described by ambient space tensors that carry uppercase underlined Latin indices ($\underline{A}, \underline{B}, \underline{C}, \dots=0\dots 4$).       \item[\textbullet] \textbf{$\sp(4)$ indices:} As we extensively use the spinorial isomorphism $\sp(4, \mathbb{R})=\so(2,3)$, we denote $\sp(4)$ indices by uppercase Latin letters ($A, B, C, \dots$=1\dots 4).
    \item[\textbullet] \textbf{$\sp(2)$ indices:} For the splitting of $\sp(4)$ indices into two plus two Weyl spinor indices $A=(\alpha, \dot{\alpha})$ we use lowercase Greek letters from the beginning of the alphabet ($\alpha, \dot{\alpha}, \beta, \dot{\beta}\dots=1, 2$).
    \item[\textbullet] \textbf{Metrics and derivatives}: We denote the background AdS metric as $\bar g_{\mu\nu}$ and its covariant derivatives as $\nabla_{\mu}$, $\nabla_{a}$, $\nabla_{\al\dgb}$. The full metric is denoted as $g_{\mu\nu}$, and we use boldface to indicate its covariant derivatives $\boldsymbol{\nabla}_{\mu}, \boldsymbol{\nabla}^2,$ etc.
\end{itemize}

Classical double copies considered in this paper are generated with an AdS symmetry parameter. The general AdS isometry generator $K$ in four dimensions is parameterized by the ten-component antisymmetric matrix $\Omega^{\underline{AB}}=-\Omega^{\underline{BA}}$
\begin{equation}\label{param:vec}
    K=\frac{1}{2}\Omega^{\underline{AB}}J_{\underline{AB}}\,,
\end{equation}
where $J_{\underline{AB}}$ are the $\so(2,3)$ generators, which fulfill the following commutation relations: 
\begin{equation}\label{ads:algebra}
    [J_{\underline{AB}}, J_{\underline{CD}}]= \eta_{\underline{AC}} J_{\underline{DB}} +  \eta_{\underline{AD}} J_{\underline{BC}} +  \eta_{\underline{BC}} J_{\underline{AD}} +  \eta_{\underline{BD}} J_{\underline{CA}}\,.
\end{equation}
{Throughout the paper, the embedding-space directions $\underline{A}=0,1$ are timelike and $\underline{A}=2,3,4$ are spacelike, {\it i.e.,} $\eta_{\underline{AB}}=\mathrm{diag}(-1,-1,+1,+1,+1)$; see Sec.~\ref{sec:ads}. In particular, $J_{01}$ generates a compact rotation in the timelike plane, while, {\it e.g.,} $J_{03}$ generates a boost.} All adjoint orbits of $\so(2,3)$ can be categorized into four major types based on their Jordan blocks, which are parameterized by eigenvalues. This classification is summarized in Table 1, which also includes the values of the Casimir invariants defined as follows:
\begin{equation}\label{Casimirs}
    I_1=\half \Omega_{\underline{AB}}\Omega^{\underline{AB}}\,,\qquad I_2=-\half \Omega^{\underline{AB}}\Omega^{\underline{CD}}\Omega_{\underline{AC}}\Omega_{\underline{BD}}\,.
\end{equation}

\begin{table}[h!]
\label{tab:eigenvalues}
\resizebox{\linewidth}{!}{
\begin{tabular}{|c|c|c|c|c|}
\hline
{\bf Type} & {\bf Eigenvalues} & {\bf Killing Field} & {\bf Casimir} $I_1$ & {\bf Casimir} $I_2$\\
\hline
$I_a$ & $\lambda, -\lambda, \lambda^*, -\lambda^*, 0;\, \lambda=a+ib$ & $b(J_{01} + J_{23}) - a(J_{03} + J_{12})$ & $2b^2 - 2a^2$ & $-2b^4 + 12a^2b^2 - 2a^4$   \\
\hline
$I_b$ & $a_1\,, -a_1\,, a_2\,, -a_2\,, 0$ & $a_1 J_{12} + a_2 J_{03}$ & $-a_1^2 - a_2^2$ & $- a_1^4 - a_2^4$ \\
\hline
$I_c$ & $ib_1\,, -ib_1\,, ib_2\,, -ib_2\,, 0$ &  $b_1 J_{01} + b_2 J_{23}$ & $b_1^2 + b_2^2$ & $- b_1^4 - b_2^4$\\
\hline
$I_d$ & $a\,, -a\,, ib\,, -ib\,, 0$ &  $a J_{03} + b J_{24}$ & $b^2 - a^2$ & $- a^4 - b^4$\\
\hline
$II_a$ & $a\,, -a\,, 0$ &  $a (J_{03} + J_{12}) + J_{01} - J_{02} - J_{13} + J_{23}$ & $-2a^2$ & $-2a^4$\\
\hline
$II_b$ & $ib\,, -ib\,, 0$  & $(b - 1) J_{01} + (b + 1) J_{23} + J_{02} - J_{13}$ & $2b^2$ & $-2b^4$\\
\hline
$III_a$ & $a\,, -a\,, 0$  & $-a J_{14} + J_{23} - J_{03}$ & $-a^2$ & $-a^4$\\
\hline
$III_b$ & $ib\,, -ib\,, 0$  & $-b J_{34} + J_{02} - J_{01}$ & $b^2$ & $-b^4$\\
\hline
$V$ & $0$  & $-J_{01} - J_{03} - J_{12} - J_{14} + J_{23} + J_{34}$ & $0$ & $0$\\
\hline
\end{tabular}
}
\caption{Classification of the adjoint $\so(2,3)$ orbits according to Jordan decomposition and eigenvalues from \cite{Holst:1997tm}. Representatives of a Killing field, as well, as the values of its Casimir invariants are provided for each orbit.}
\end{table}

Each orbit generated by a Killing field defines a non-trivial solution of Einstein's equations with a negative cosmological constant in four dimensions. Their isometries $\xi$ originate from the $\so(2,3)$ subalgebra formed by the centralizer $[\xi, \Omega]=0$. The center is found and classified according to the type of orbit it comes from. The Weyl double copies are found using the Penrose transform, while the Kerr-Schild double copy (if it exists) is obtained from an orbit via the construction of a light-like projector of \cite{Didenko:2009tc}, producing a spacetime of the Petrov type D. In a few degenerate cases, the Petrov type is N.  

{ There is a natural freedom in this classification. The generating parameter $K$ in \eqref{param:vec} can be rescaled by any nonzero constant, $K\to cK$, $c\neq0$. This rescaling changes the Casimir invariants, $I_1\to c^2I_1$ and $I_2\to c^4I_2$, but in the double copy it is compensated by a rescaling of the mass parameter $M$ of the Kerr-Schild metric (for $c<0$, together with time reversal). Up to this rescaling, the families $II_{a,b}$ and $III_{a,b}$ of Table 1 depend on the parameters $a$ and $b$ only through discrete data, since their absolute values can be absorbed into the normalization of $K$. Specifically, the inequivalent types are summarized in Table~\ref{tab:discrete}, and the corresponding conjugations are given in Appendix~{B}.

\begin{table}[h!]
\addtocounter{table}{-1}\renewcommand{\thetable}{1a}
\centering
\resizebox{\linewidth}{!}{
\begin{tabular}{|c|c|c|c|}
\hline
{\bf Type} & {\bf Parameter} & {\bf Representative, up to normalization of $K$} & {\bf Notation}\\
\hline
$II_a$ & $a\neq0$ & $J_{03}+J_{12}+J_{01}-J_{02}-J_{13}+J_{23}$ & $II$\\
\cline{2-4}
 & $a=0$ & $J_{01}-J_{02}-J_{13}+J_{23}$ & $II^0$\\
\hline
$II_b$ & $b=0$ & $-J_{01}+J_{23}+J_{02}-J_{13}$ & $II^0$\\
\cline{2-4}
 & $b>0$ & $2J_{23}+J_{02}-J_{13}$ \ $(b=1)$ & $II^+$\\
\cline{2-4}
 & $b<0$ & $-2J_{01}+J_{02}-J_{13}$ \ $(b=-1)$ & $II^-$\\
\hline
$III_a$ & $a\neq0$ & $-J_{14}+J_{23}-J_{03}$ & $III^+$\\
\cline{2-4}
 & $a=0$ & $J_{23}-J_{03}$ & $III^+_0$\\
\hline
$III_b$ & $b\neq0$ & $-J_{34}+J_{02}-J_{01}$ & $III^-$\\
\cline{2-4}
 & $b=0$ & $J_{02}-J_{01}$ & $III^-_0$\\
\hline
\end{tabular}
}
\caption{Discrete orbits of types $II$ and $III$ up to the normalization of $K$. The two representatives of $II^0$ are conjugate by the reflection $X^0\to-X^0$ (Appendix~{B}), so that $II_{a=0}=II_{b=0}$. All other entries are pairwise non-conjugate, also after any rescaling of $K$.}
\label{tab:discrete}
\end{table}
\renewcommand{\thetable}{\arabic{table}}
}
 
We also define the canonical coordinates that reflect the symmetry of a given orbit, allowing us to provide metrics that are manifestly invariant under the Abelian part of isometries.

The construction of double copies is analyzed for the two-, one-, and zero-parametric families of the resulting Killing fields. In specific cases, the Penrose transform can become degenerate because of a particular degeneracy in $K$ . In these cases, the associated double copy falls under Petrov type N. When this degeneracy occurs, the centralizer of the corresponding orbits has six generators. It appears to have more generators than the actual isometries of the related vacuum spacetimes, which are represented by only four out of these six. A summary of our results can be found in Table 2.
    \begin{table}[h!]
    \centering
    \resizebox{\linewidth}{!}{
    \begin{tabular}{@{}lllll@{}}
        \toprule
        \textbf{Orbit} & \textbf{Petrov type} & \textbf{Isometries$^*$} & \textbf{KS vectors} & \textbf{Spacetime} \\
        \midrule
        $I_a$ & D & $\u(1)\oplus\u(1)$ & $2_{\mathbb{R}}+2_{\mathbb{C}}$ & Generalized Carter-Pleba\'nski \\
        
        $I_b$ & D & $\mathrm{\u(1)} \oplus \mathrm{\u(1)}$ & $2_{\mathbb{R}}+2_{\mathbb{C}}$  & Carter-Pleba\'nski, $t \leftrightarrow i\phi$ \\  
         
        $I_c$ & D & $\mathrm{\u(1)} \oplus \mathrm{\u(1)}$ & $2_{\mathbb{R}}+2_{\mathbb{C}}$  & Kerr; Carter-Pleba\'nski \\
        
        $I_d$ & D & $\mathrm{\u(1)} \oplus \mathrm{\u(1)}$ & $2_{\mathbb{R}}+2_{\mathbb{C}}$  & rotating topological black hole (KMV) $\gga = -a^2$ \\
                 
        $II_{a}$ & D & $\u(1)\oplus \u(1)$ & $2_{\mathbb{R}}+2_{\mathbb{C}}$ & rotating black brane of \cite{Klemm:2014rda} \\
        
        $II_{b<0}$ & D & $\u(1)\oplus \u(1)$ & $2_{\mathbb{R}}+2_{\mathbb{C}}$ & super-entropic black hole \cite{Hennigar:2014cfa} \\

        $II_{b>0}$ & D & $\u(1)\oplus \u(1)$ & $2_{\mathbb{R}}+2_{\mathbb{C}}$ & hyperbolic super-entropic black hole \\

        $III_{a}$ & D & $\u(1)\oplus \u(1)$ & $2_{\mathbb{R}}+2_{\mathbb{C}}$ & KMV at zero rotation ($\gamma\to 0$ contraction) \\

        $III_{b}$ & D & $\u(1)\oplus \u(1)$ & $2_{\mathbb{R}}+2_{\mathbb{C}}$ & KMV at infinite rotation \\

        $V$ & D & $\u(1)\oplus \u(1)$ & $2_{\mathbb{R}}+2_{\mathbb{C}}$ & degenerate Carter-Plebański ($\gep=\gamma = 0$)\\
        \midrule
        {} & {} & \textbf{Symmetry-enhanced cases}\\
        
         \midrule
         $I_c$ $b_2 = 0$  & D & $\mathrm{\u(1)} \oplus \mathrm{\so(3)}$ & $2_{\mathbb{R}}$  & Schwarzschild (AI-metric, $\epsilon_2=1$) \\

         $I_c\Big|_{b_1=0}\sim I_d\Big|_{a=0}$  & D & $\mathrm{\u(1)} \oplus \mathrm{\so(1,2)}$ & $2_{\mathbb{C}}$  & BII-metric, $\epsilon_2=-1$  \\

         $I_b\Big|_{a_2=0}\sim I_b\Big|_{a_1=0}\sim I_d\Big|_{b=0}$  & D & {$\u(1) \oplus\so(1, 2)$}   &  $2_{\mathbb{R}}/2_{\mathbb{C}}^{**}$ &  hyperbolic black hole (AII/BI-metric, $\gep_2=-1$)\\

         $I_b$ $a_1 = \pm a_2$  & D & $\u(1) \oplus \mathrm{\so(2, 1)}$ & $2_{\mathbb{C}}$  &  Wick-rotated hyperbolic Taub-NUT-AdS \\

         $I_c$ $b_1 = \pm b_2$  & D & $\mathrm{\u(1)} \oplus \mathrm{\so(1,2)}$ & $2_{\mathbb{R}}$  & hyperbolic Taub-NUT-AdS, $|n| = \lambda^{-1}/2$ \\

          $II_{a=0}$  & N & $\u(1)\ltimes \h_3$ & 1  & Siklos, $H = Mr^3 u^{-\frac52}$ \\      
        
        $II_{b=0}$  & N & $\u(1)\ltimes \h_3$ & 1  & Siklos, $H=-r^2-x^2+Mr^3$ \\     
        
        $III_{a=0}$  & D & $\mathrm{\u(1)} \oplus \mathrm{\iso(1, 1)} $ & $2_{\mathbb{C}}$  & AdS soliton \\
        
        $III_{b=0}$  & D & $\mathrm{\u(1)} \oplus \mathrm{\iso(2)} $ & $2_{\mathbb{R}}$  & black brane (AIII-metric, $\epsilon_2=0$) \\
          \bottomrule
    \end{tabular}
    }
    \caption{
    Classification of double copies for various orbits: their Petrov types, isometries, the number of independent Kerr-Schild vectors -- labeled by  $\mathbb{R}$ for real vectors and $\mathbb{C}$ for complex vectors -- and the corresponding spacetimes. Types $II_{a=0}$ and $II_{b=0}$ are the same orbit, $II^0$, by Table 1a, but the regularized limits $a\to 0$ and $b\to 0$ provide inequivalent Siklos waves.\\ {\footnotesize{$^*$ $\u(1)$ denotes a one-parameter subgroup, which may be elliptic ($\so(2)$), hyperbolic ($\so(1,1)$) or parabolic (generated by a nilpotent element, with null orbits); we do not distinguish them.\\ 
    $^{**}$  The types of the Kerr-Schild vectors -- whether they are complex conjugates or real --  depend on the form of the AdS background metric; specifically for background classified as either an  $A$-metric or a  $B$-metric. Although both metrics describe AdS patches, they are not connected by real diffeomorphisms.}}} \label{tab:orbits}
    \end{table}
The double copies presented in Table 2 correspond to exact solutions of Einstein's vacuum equations with a negative cosmological constant $(\Lambda < 0)$. These results can be directly extended to Kerr-Schild and Weyl multicopies, which solve the equations of higher-spin gauge fields at the linearized level. The resulting solutions fall under the generalized Petrov types D or N. We naturally interpret these higher-spin solutions as linearized extensions of the corresponding GR vacuum solutions.

\section{Anti-de Sitter spacetime}\label{sec:ads}
{This section collects standard facts about AdS$_4$ and its isometries that are used throughout the paper.}

Anti-de Sitter spacetime $AdS_4$ with the length scale $L$ can be considered as the hyperboloid
\begin{equation}
    -X_0^2 - X_1^2 + X_2^2 + X_3^2 + X_4^2 = - L^2
\end{equation}
embedded in the flat five-dimensional spacetime $M^{2, 3}$
\begin{equation}
    ds^2 = -dX_0^2 - dX_1^2 + dX_2^2 + dX_3^2 + dX_4^2
\end{equation}
with two temporal coordinates $X_0, X_1$. The global symmetry of  $AdS_4$ is represented by a ten-parameter group of isometries $SO(2, 3)$ generated by 
\begin{equation}\label{J: def}
    J_{\underline{AB}}=X_{\underline{A}}\frac{\p}{\p X^{\underline{B}}}-X_{\underline{B}}\frac{\p}{\p X^{\underline{A}}}\,.
\end{equation}
The spacetime satisfies Einstein's equations
\begin{equation}
    R_{\mu \nu} = \Lambda g_{\mu \nu}\,,\quad R = 4 \Lambda\,,\quad \Lambda = -\frac{3}{L^2}=-3\lambda^2\,.
\end{equation}
The maximal compact subgroup of $SO(2,3)$ is $SO(3) \times SO(2) \in SO(2, 3)$, where $SO(3)$ rotates $(X_2, X_3, X_4)$ while $SO(2)$ rotates $(X_0, X_1)$.The set of all solutions to the Killing equation
\begin{equation}
  {\nabla}_\mu \xi_\nu + {\nabla}_\nu \xi_\mu = 0
\end{equation}
generates the isometry Lie algebra $\so(2,3)$ via vector fields.

In the following, we will be interested in the stability algebra $G$ of a given AdS Killing vector: $[G, \xi]=0$. These are typically of the form \(G_2\), \(G_4\), or \(G_6\), where the dimension indicates the number of independent Killing fields. In the case of \(G_2\), there is only one nontrivial commutator; hence, all two-dimensional isometry groups are solvable. If the group $G_2$ is Abelian, then \([\xi_1, \xi_2] = 0\), corresponding to the algebra \(\u(1) \oplus \u(1)\); alternatively, for a non-Abelian \(G_2\), a possible commutator is \([\xi_1, \xi_2] = \pm \xi_1\).

The structure of three-dimensional isometry groups \(G_3\) was originally classified by Bianchi, yielding nine distinct types. In most of our cases,\footnote{In our analysis, algebra $G_3$ appears as $G_4=G_3\oplus\u(1)$.} the relevant types are VIII and IX, with the corresponding algebras \(\sl(2,\mathbb{R}) = \so(1,2)\) and \(\so(3)\), which are semisimple.

A single generator $\xi$ of a transformation group produces a one-parameter subgroup. By choosing a point on its orbit, one can find a coordinate \(x\) such that \(\xi = \partial_x\). Similarly, if there are \(m\) commuting generators \(\xi_{(m)}\), one can introduce \(m\) coordinates \((x^1, \ldots, x^m)\) so that these vector fields are straightened:
\be
\xi_{(n)} = \frac{\partial}{\partial x^{(n)}}.
\ee
This procedure provides a method for selecting appropriate canonical coordinates on each orbit.

The orbits of a group of motions, which have dimension \(d\), may be spacelike, null, or timelike submanifolds, denoted by \(S_d\), \(N_d\), and \(T_d\), respectively, or they may be of mixed type, denoted by \(V_d\). A space on which the group of motions acts transitively is called homogeneous.

\section{Double copies from AdS isometry}\label{sec:dc}
{This section reviews the construction of double copies and multicopies from the AdS global symmetry parameter developed in \cite{Didenko:2008va, Didenko:2009tc, Didenko:2009td, Didenko:2021vui}, which underlies our classification. Apart from the remarks on the validity and limitations of these constructions, which we point out explicitly where they appear, the material of this section reviews the known construction.}

In $d>3$, gravity is not topological. Consider the case of $d=4$ in detail. The AdS spacetime, being a maximally symmetric space, admits a convenient gauge description. Specifically, consider the Cartan equations that describe its frame fields
\begin{subequations}\label{ads:frame}
\begin{align}
    &\dr\mathbf{w}_{ab}+\mathbf{w}_{a}{}^{c}\mathbf{w}_{cb}=\Lambda\mathbf{e}_a\mathbf{e}_b\,,\qquad \Lambda:=-\lambda^2\\
    &\dr\mathbf{e}_a+\mathbf{w}_{a}{}^{b}\mathbf{e}_b=0\,,
\end{align}
\end{subequations}
where $\mathbf{w}_{ab}=-\mathbf{w}_{ba}=\mathbf{w}_{ab,\mu}\dr x^{\mu}$ is the Lorentz connection 1-form and $\mathbf{e}_a=\mathbf{e}_{a, \mu}\dr x^{\mu}$ is the vierbein 1-form. Introducing the covariant differential 
\begin{equation}
    D A_{a}=\dr A_a+\mathbf{w}_{a}{}^{b}A_b\,,\qquad D^2 A_a=\Lambda\mathbf{e}_a\mathbf{e}^bA_b\,,
\end{equation}
one observes the following gauge invariance of \eqref{ads:frame}:
\begin{subequations}\label{ads:gauge}
\begin{align}  
   &\delta\mathbf{w}_{ab}=D\varkappa_{ab}+\Lambda(v_a\mathbf{e}_b-v_b\mathbf{e}_a)\,,\\
   &\delta\mathbf{e}_a=Dv_a-\varkappa_{ab}\mathbf{e}^b\,.
\end{align}
\end{subequations}
Here, $v_a$ and $\varkappa_{ab}=-\varkappa_{ba}$ are arbitrary space-time dependent parameters. The gauge symmetry \eqref{ads:gauge} becomes a global symmetry once one requires $\delta\mathbf{w}_{ab}=\delta\mathbf{e}_a=0$. In that case, these parameters are no longer arbitrary, but satisfy
\begin{subequations}\label{ads:param}
\begin{align}
    &D\varkappa_{ab}=-\Lambda(v_a\mathbf{e}_b-v_b\mathbf{e}_a)\,,\label{CKY}\\
    &Dv_a=\varkappa_{ab}\mathbf{e}^b\,.\label{v:killing}
\end{align}
\end{subequations}
Equation \eqref{v:killing} means that $v_a$ is a Killing vector. Indeed, since $\varkappa_{ab}$ is antisymmetric, we have 
\begin{equation}\label{Kill:eq}
    \nabla_a v_b+\nabla_b v_a=0\,.
\end{equation}
The field $\varkappa_{ab}=-\varkappa_{ba}$ is just a covariant derivative of a Killing vector, but more broadly, Eq. \eqref{CKY} defines the conformal Killing-Yano tensor; see {\it e.g.,} \cite{Krtous:2006qy}, the existence of which in AdS is a consequence of the consistency of \eqref{ads:frame}. 

Equations \eqref{ads:frame} and \eqref{ads:param} become particularly illuminating when written as a flatness condition in AdS isometry algebra. For that matter, we introduce the $\so(2,3)$ valued 0-form  
\begin{equation}
    K=\half \Omega^{\underline{IJ}}J_{\underline{IJ}}\,,
\end{equation}
whose components are the fields $v_a, \varkappa_{ab}$:
\begin{equation}\label{Omega:def}
    \Omega_{\underline{IJ}} = 
        \begin{cases}
        \gl^{-1}\gk_{ab}\,,\qquad (\underline{I},\underline{J})=(a,b)=1,2,3,4 \,,\quad\\
        v_{b}\,,\qquad \underline{I}=0\,,\quad \underline{J}=b=1,2,3,4\,,
    \end{cases}
\end{equation}
where from \eqref{v:killing} it follows 
\begin{equation}
\varkappa_{ab} = \frac{1}{2}(\nabla_{b} v_{a}-\nabla_a v_b)\,.    
\end{equation}
An equivalent form of the system \eqref{ads:param} is 
\begin{equation}\label{embd:param}
    D_0 \Omega_{\underline{IJ}} = 0\,,
\end{equation}
where we introduced the AdS covariant differential $D_0$ with respect to the AdS flat connection $w_{\underline{IJ}}=-w_{\underline{JI}}$ 
\begin{equation}
    w_{\underline{IJ}} =  \begin{cases}
        \w_{ab}\,,\qquad (\underline{I},\underline{J})=(a,b)=1,2,3,4\,,\quad\\
        \gl\e_{b}\,,\qquad \underline{I}=0\,,\quad \underline{J}=b=1,2,3,4
    \end{cases}\,,\qquad D_0 A_{\underline{I}}=\dr A_{\underline{I}}+w_{\underline{I}}{}^{\underline{J}}A_{\underline{J}}\,.
\end{equation}
Equation \eqref{embd:param} is consistent thanks to the AdS zero-curvature condition 
\begin{equation}
    \dr w_{\underline{IJ}}+w_{\underline{I}}{}^{\underline{K}}w_{\underline{KJ}}=0\,,
\end{equation}
the Lorentz component form of which is literally \eqref{ads:frame} and it also entails $D_0^2=0$.

\paragraph{Spinorial representation} 

In four dimensions, spinorial formalism offers great advantages. Not only does it reduce complicated tensor structures into simple symmetric spin-tensors, but it lies in the foundation of the twistor Penrose transform responsible for the manifest construction of classical double copies. We will use it extensively in what follows, while leaving the details of the spinor/vector dictionary to  Appendix A. 

For now, we need the spinorial form of \eqref{ads:param}. In terms of components, the bookkeeping is as follows:
\begin{equation}
    \gk_{ab}\to (\gk_{\al\gb}; \bar\gk_{\dal\dgb})\,,\qquad v_a\to v_{\al\dgb}\,,\qquad \mathbf{e}_a\to \mathbf{e}_{\al\dgb}\,,
\end{equation}
where $\al, \dgb=1,2$ and $\gk_{\al\gb}=\gk_{\gb\al}$ (same for $\bar\gk$). The system \eqref{ads:param} then amounts to
\begin{subequations}\label{ads:param-spin}
\begin{align}
    &D\varkappa_{\al\gb}=\gl^2(\mathbf{e}_{\al}{}^{\dgga}v_{\gb\dgga}+\mathbf{e}_{\gb}{}^{\dgga}v_{\al\dgga})\,,\label{CKY-spin1}\\
    &D\bar\varkappa_{\dal\dgb}=\gl^2(\mathbf{e}^{\gga}{}_{\dal}v_{\gga\dgb}+\mathbf{e}^{\gga}{}_{\dgb}v_{\gga\dal})\,,\label{CKY-spin2}\\
    &Dv_{\al\dgb}=\mathbf{e}_{\al}{}^{\dgga}\bar\varkappa_{\dgb\dgga}+\mathbf{e}^{\gga}{}_{\dgb} \varkappa_{\gga\al}\,.\label{v:killing-spin}
\end{align}
\end{subequations}
The spinorial version of the embedding covariant constancy condition \eqref{embd:param} is reached via the isomorphism $\so(2,3)\sim \sp(4, \mathrm{R})$, which implies 
\begin{equation}
    \Omega_{IJ}\to K_{AB}=K_{BA}\,,\qquad A,B=1,\dots, 4\,,
\end{equation}
where $K$ is a covariant constant 
\begin{equation}\label{param:sp4}
    D_0K_{AB}=0\,,
\end{equation}
and has the following component form: 
\begin{equation}\label{K: comp}
    K_{AB}=\begin{pmatrix}
        \gl^{-1} \gk_{\al\gb} &  v_{\al\dgb}\\
         v_{\gb\dal} & \gl^{-1} \bar\gk_{\dal\dgb}
    \end{pmatrix}\,.
\end{equation}
Its two Casimir invariants \eqref{Casimirs} read
\begin{subequations}\label{Casimir:K}
    \begin{align}
        &C_2=\frac{1}{4} K_{AB}K^{AB}=I_1\,,\\
        &C_4=\frac{1}{4} K_{AB}K_{CD}K^{BC}K^{AD}=3I_1^2 + 2I_2\,.
    \end{align}
\end{subequations}
In terms of the Lorentz components, these take the following form:
\begin{align}
    &2C_2 = \gl^{-2} \varkappa^2 + \gl^{-2} \overline{\varkappa}^2 + 2 v^2\,, \\
    &2C_4 = (\gl^{-2} \varkappa^2 + v^2)^2 + (\gl^{-2} \overline{\varkappa}^2 + v^2)^2 +2 v^2(\varkappa^2 + \overline{\varkappa}^2) + 2 \varkappa_{\ga \ga} \overline{\varkappa}_{\dot{\gb}\dot{\gb}}v^{\ga \dot{\gb}}v^{\ga \dot{\gb}}\,.    
\end{align}

\subsection{Weyl double copy}\label{sec:weyl}

Let us first discuss how the Weyl double copies are generated from \eqref{ads:param-spin}. One effective method to obtain these solutions is outlined in \cite{Didenko:2009tc}, which demonstrates that \eqref{ads:param-spin} produces Petrov type D solutions. In this context, the (anti)self-dual components of the Weyl tensor\footnote{We denote symmetric  spinor indices either with the same letter or by a letter with a number of indices in parenthesis, {\it e.g.,} $A_{\al(2)}:=A_{\al\al}=\frac{1}{2}(A_{\al_1\al_2}+A_{\al_2\al_1})$. } can be expressed as follows:\footnote{Parameters $M$ and $\bar M$ may contain some extra powers of the cosmological parameter $\lambda$ in  order to provide the proper dimension of the Weyl tensor.}

\begin{equation}\label{Weyl:DC}
    C_{\ga(4)} = \frac{M}{\rf^5}\varkappa_{\ga \ga} \varkappa_{\ga \ga}\,,\qquad \overline{C}_{\dot{\ga}(4)} = \frac{\overline{M}}{\overline{\rf}^5}\overline{\varkappa}_{\dot{\ga} \dot{\ga}} \overline{\varkappa}_{\dot{\ga} \dot{\ga}}\,,
\end{equation}
where scalars 
\begin{equation}\label{r: def}
    \quad \rf^2 = -\frac{1}{2 {\gl^4}} \varkappa_{\ga \gb}\varkappa^{\ga \gb},\quad \overline{\rf}^2 = -\frac{1}{2 {\gl^4}}  \overline{\varkappa}_{\dot{\ga} \dot{\gb}}\overline{\varkappa}^{\dot{\ga} \dot{\gb}}\,,   
\end{equation}
\begin{equation}\label{v: def}
    \quad \vf^2 = \frac{1}{2}v_{\al\dgb}v^{\al\dgb}\,,  
\end{equation}
$M$ and its complex conjugate $\bar M$ are arbitrary constants and $\rf$, $\bar \rf$ are defined such that the flat limit $\gl\to 0$ is well-defined. {In \cite{Didenko:2009tc}, it was shown that \eqref{Weyl:DC} satisfies Einstein's equations provided $\gk_{\al\gb}$, $\bar\gk_{\dal\dgb}$, and $v_{\al\dgb}$ obey a consistent deformation of \eqref{ads:param-spin} from AdS to the black-hole background. In the deformed system, Eqs.~\eqref{CKY-spin1} and \eqref{CKY-spin2} keep their form, with the AdS vierbein replaced by the black-hole one, so that $\gk_{ab}$ remains a closed conformal Killing-Yano tensor, while the Killing equation \eqref{v:killing-spin} is deformed by terms whose coefficient depends on $M$, $\bar M$ and on $\gk^2$, $\bar\gk^2$. Equations \eqref{ads:param-spin} themselves single out the AdS isometry parameters, since their consistency condition implies that the Weyl tensor vanishes; they are recovered from the deformed system at $M=\bar M=0$.} The single copy
\begin{equation}\label{Maxwell}
    F_{\al\gb}=\frac{m}{\rf^3}\gk_{\al\gb}\,,\qquad \bar F_{\dal\dgb}= \frac{\overline{m}}{\bar \rf^3}\bar\gk_{\dal\dgb}
\end{equation}
satisfies Maxwell equations 
\begin{equation}
    \nabla_{\gb\dal}F^{\gb}{}_{\al}=\nabla_{\al\dgb}\bar F^{\dgb}{}_{\dal}=0\,,
\end{equation}
while the zeroth copy 
\begin{equation}\label{zero}
    2\phif = \frac{1}{\rf} + \frac{1}{\overline{\rf}}\,.
\end{equation}
satisfies the Klein-Gordon equation
\begin{equation}\label{0:DC}
    \Box\phif=-2 \gl^2 \phif\,.
\end{equation}
as a simple consequence of \eqref{ads:param-spin}. In fact, a stronger statement holds, namely,
\begin{equation}\label{phi}
    \Box\varphif=-2\gl^2\varphif\,,\qquad \varphif:=\frac{1}{\rf}\,,
\end{equation}
where $\varphi$ is generally complex, implying that its real and imaginary parts both satisfy \eqref{0:DC}. This allows us to define a dual zero copy as follows: 
\begin{equation}\label{dual:zero}
2i\tilde\phif=\frac{1}{\rf} - \frac{1}{\overline{\rf}}\,,
\end{equation}
which also satisfies Eq. \eqref{0:DC}
\begin{equation}\label{0:dual-DC}
    \Box\tilde\phif=-2 \gl^2 \tilde\phif\,.
\end{equation}

The equations presented in \eqref{Weyl:DC} do not fully determine the space of nonequivalent solutions they generate. A detailed calculation \cite{Didenko:2009tc} demonstrates that the type D solutions described by \eqref{Weyl:DC} emerge as a smooth deformation with respect to the parameters  $M$  and  $\bar M$  from the global symmetry condition outlined in \eqref{param:sp4}. Notably, the two Casimir constants given in \eqref{Casimir:K} serve to distinguish these solutions, functioning similarly to kinematic parameters, such as the rotational parameter $a$  in the case of a Kerr black hole. 

{Let us note that the field $\gk_{\al\gb}$ (together with $\bar\gk_{\dal\dgb}$), which in AdS defines the closed conformal Killing-Yano tensor \eqref{CKY}, plays a central role in this construction: it remains essentially intact under the deformation, and the Weyl tensor \eqref{Weyl:DC} is built out of it. A related construction was presented by Frolov and Kubiz{\accent7 n}\'ak \cite{Frolov:2008jr} (see also \cite{Krtous:2006qy}), who showed how the Kerr-NUT-(A)dS metric can be generated starting from a principal conformal Killing-Yano tensor. The unfolded formulation used here offers several advantages in this respect. First, the conformal Killing-Yano tensor and the Killing vector are packed into a single covariantly constant $\sp(4)$ parameter $K_{AB}$ \eqref{param:sp4}, so that the input data are purely algebraic. As a result, the classification of the generated solutions reduces to that of the adjoint orbits of $K$, and their isometries are given by its centralizer. Second, the deformation is constructed systematically and yields coordinate-independent expressions for the Weyl tensor, the Kerr-Schild vectors, and the metric. Finally, the same formalism extends directly to the higher-spin multicopies through the Penrose transform.}

To better understand the origin of \eqref{Weyl:DC}, we can refer back to the Penrose transform \cite{Eastwood:1981jy}. A more detailed version of this transform, which explicitly addresses symmetries, has been developed within the context of higher-spin extensions of black hole solutions in \cite{Didenko:2009td}. We will provide the necessary details in our higher-spin section \ref{sec:HS}, while here we offer a brief summary that leads to the desired classification of double copies.

\begin{itemize}
    \item Two solutions \eqref{Weyl:DC} generated by $K_{AB}$ and $K'_{AB}$ are equivalent if there exists $U\in Sp(4, \mathbb{R})$ such that $K'=U^{-1}KU$. Since $\so(2,3)\sim \sp(4, \mathrm{R})$ the statement is reformulated similarly in terms of  $\Omega_{\underline{IJ}}$, \eqref{Omega:def}. Namely, the equivalence holds if and only if the generated elements are within the same conjugacy class $\Omega'=O^{-1}\Omega O$, with $O\in O(2,3)$. Particularly, as already noted, the two invariants \eqref{Casimirs} or \eqref{Casimir:K} label solutions \eqref{Weyl:DC}. More generally, those are as many as adjoint orbits $\so(2,3)$.    
    \item Isometries of the double copy \eqref{Weyl:DC} form an AdS  subalgebra that commutes with the generating $K_{AB}$ (equivalently $\Omega_{\underline{IJ}}$). Therefore, the global symmetry of the so defined double copies is the centralizer\footnote{The isometries coincide with a centralizer, provided the Penrose transform is well-defined. {This caveat, relevant for the degenerate cases, is discussed below in Sec.~\ref{sec:Penrose}.}} of the generating parameter. Particularly, since the parameter commutes with itself, there is always at least one isometry.   
\end{itemize}

\subsection{Kerr-Schild double copy} Type D space-times \eqref{Weyl:DC} are described at the level of the Weyl tensor. The corresponding metric can be recovered in terms of the AdS background symmetry parameter \eqref{ads:param-spin}. In the case of a real $M=\bar M$, the map becomes particularly simple and is  given by the Kerr-Schild representation  \eqref{Kerr-Schild}, where the Kerr-Schild vector $k^\mu$ can be manifestly constructed. 

\subsection{Kerr-Schild vectors} 
Equations \eqref{ads:param-spin} allow us to construct, in general, four Kerr-Schild vectors: two real ones and two mutually complex conjugates. To this end, following \cite{Didenko:2009tc}, let us introduce the rank-1 projectors: 
\begin{subequations}\label{proj:def}
\begin{align}
    &\pi^+_{\al\gb}=\frac{1}{2}\left(\gep_{\al\gb}+\frac{\gk_{\al\gb}}{\lambda\rf}\right)\,,\qquad \bar\pi^+_{\dal\dgb}=\frac{1}{2}\left(\gep_{\dal\dgb}+\frac{\bar\gk_{\dal\dgb}}{\lambda\bar \rf}\right)\,,\\
    &\pi^-_{\al\gb}=\frac{1}{2}\left(\gep_{\al\gb}-\frac{\gk_{\al\gb}}{\lambda\rf}\right)\,,\qquad \bar\pi^-_{\dal\dgb}=\frac{1}{2}\left(\gep_{\dal\dgb}-\frac{\bar\gk_{\dal\dgb}}{\lambda\bar \rf}\right)\,.
\end{align}
\end{subequations}
These are uniquely defined in terms of the fields $\gk_{\al\gb}$ and $\bar\gk_{\dal\dgb}$ from the projector conditions
\begin{equation}\label{proj:prop}
    \pi_{\al}{}^{+\gga}\pi^{+}_{\gga\gb}=\pi^{+}_{\al\gb}\,,\qquad \pi_{\al}{}^{-\gga}\pi^{-}_{\gga\gb}=\pi^{-}_{\al\gb}\,,\qquad \pi_{\al}{}^{+\gga}\pi^{-}_{\gga\gb}=0\,.
\end{equation}
Similar complex conjugate relations for $\bar\pi^{\pm}$ also hold. Note that in checking \eqref{proj:prop} one has to use Schouten's identities for two-component spinors. The projectors allow us to build four light-like vectors $k^{+\mu}$, $k^{-\mu}$, $l^{+\mu}$ and ${l}^{-\mu}$, which are geodesic in addition 
\begin{equation}\label{KS:geo}
k^{\pm\mu}\nabla_{\mu}k^{\pm\nu}=l^{\pm\mu}\nabla_{\mu}l^{\pm\nu}=0
\end{equation}
with respect to the AdS background. They are normalized in such a way, so as to have space-time constant scalar product with the generating Killing vector \eqref{Kill:eq}
\begin{equation}\label{KS:norm}
    k^{\pm\mu} v_{\mu}=l^{\pm\mu} v_{\mu}=-1\,.
\end{equation}
The Kerr-Schild vectors $k^+$ and $k^-$ are real, while $l^+$ and $l^-$ are complex conjugates. Their definition in spinorial form is as follows \cite{Didenko:2009tc}:
\begin{subequations}\label{KS:spin}
\begin{align}   
    &k^{+}_{\al\dal}=\frac{2}{v^+v^-}v^+_{\al\dal}\,,\qquad k^{-}_{\al\dal}=\frac{2}{v^+v^-}v^-_{\al\dal}\,,\label{KS:kn}\\
    &l^{+}_{\al\dal}=\frac{2}{v^{+-}v^{-+}}v^{+-}_{\al\dal}\,,\qquad l^{-}_{\al\dal}=\frac{2}{v^{+-}v^{-+}}v^{-+}_{\al\dal}\,,\label{KS:l}
\end{align}
\end{subequations}
where 
\begin{subequations}\label{v:proj}
    \begin{align}
        &v^-_{\al\dal}=\pi_{\al}^{-\gb}\bar\pi_{\dal}^{-\dgb}v_{\gb\dgb}\,,\qquad v^+_{\al\dal}=\pi_{\al}^{+\gb}\bar\pi_{\dal}^{+\dgb}v_{\gb\dgb}\,, \\
        &v^{-+}_{\al\dal}=\pi_{\al}^{-\gb}\bar\pi_{\dal}^{+\dgb}v_{\gb\dgb}\,,\qquad v^{+-}_{\al\dal}=\pi_{\al}^{+\gb}\bar\pi_{\dal}^{-\dgb}v_{\gb\dgb}\,.
    \end{align}
\end{subequations}
{Here and below, contractions such as $v^+v^-:=\frac12 v^+_{\al\dal}v^{-\al\dal}$, as well as $\gk^2:=\frac12\gk_{\al\gb}\gk^{\al\gb}$ and $\bar\gk^2$, are defined in \eqref{B:notation}; see also \eqref{A:det} in Appendix A.} We will often denote by $k$ one of the two real Kerr-Schild vectors \eqref{KS:kn}, and by $l$ we denote one of the two complex vectors in \eqref{KS:l} (the other one is then $\bar l$). From the above definition, it naturally follows that 
\begin{equation}
    k_{\mu}k^{\mu}=l_{\mu}l^{\mu}=k_{\mu}l^{\mu}=0\,.
\end{equation}
By means of Eqs. \eqref{ads:param-spin} it can be verified that definitions \eqref{KS:spin} indeed entail \eqref{KS:geo} and also \eqref{KS:norm}. The meaning of the construction above is that there are four independent light-like directions onto which a given Killing vector $v$  can be projected using the projectors defined in \eqref{proj:def}. The normalization \eqref{KS:norm} ensures the vectors are geodesic.

Apart from the necessary conditions for the Kerr-Schild definitions being light-like and geodesic \eqref{KS:geo}, Eqs. \eqref{ads:param-spin} give us more. Namely, they generate the following single copy:
\begin{equation}
    \phi_\mu=\phif k_\mu\,,\qquad 
\end{equation}
where $k_{\mu}$ is either of the two real Kerr-Schild vectors \eqref{KS:kn}. Using its definition along with Eqs. \eqref{ads:param-spin}, it can be confirmed \cite{Didenko:2009tc} that the defined single copy satisfies the Maxwell equations in AdS, specifically:
\begin{equation}\label{1:DC}
    \Box\phi_\mu-\nabla^{\nu}\nabla_{\mu} \phi_{\nu} = 0\,.
\end{equation}
As explained in \cite{Didenko:2009tc}, these two single copies are gauge-equivalent,\footnote{More generally, Eqs. \eqref{ads:param-spin} admit discrete symmetry, which results in the interchange $k^{+}_\mu\leftrightarrow k^{-}_\mu$ and $l^+_\mu\leftrightarrow l^-_\mu$.} generating the same Maxwell tensor. Vectors \eqref{KS:l} also generate single copies, however, complex
\begin{equation}\label{1:DC-dual}
    \tilde\phi_{\mu}=\tilde\phif l_{\mu}\,,\qquad \Box\tilde\phi_\mu-\nabla^{\nu}\nabla_{\mu} \tilde\phi_{\nu} = 0\,.
\end{equation}

Additionally, there are double copies, the existence of which is also a consequence of \eqref{ads:param-spin} 
\begin{equation}\label{KS_DC:def}
    \phi_{\mu\nu}=\phif k_{\mu}k_{\nu}\,,\qquad \tilde\phi_{\mu\nu}=\tilde\phif  l_{\mu}l_{\nu}\,.
\end{equation}
Up to a factor, they generate the same Weyl double copy \eqref{Weyl:DC}. Notice, unlike $\phi_{\mu\nu}$, the double copy $\tilde\phi_{\mu\nu}$ is complex. Both satisfy the linearized Einstein equations, {\it e.g.,} 
\begin{equation}\label{2:DC}
    \Box\phi_{\mu\nu}-\nabla_{\rho}\nabla_{\mu}\phi^{\rho}{}_{\nu}-\nabla_{\rho}\nabla_{\nu}\phi^{\rho}{}_{\mu}-6\gl^2\phi_{\mu\nu}=0\,.
\end{equation}
Let us highlight once again that the various copy equations, such as \eqref{0:DC}, \eqref{1:DC} and \eqref{2:DC}, are satisfied as a consequence of \eqref{ads:param-spin}.

\subsection{Metric reconstruction}\label{sec:metric}
Given that the Weyl tensor has the form \eqref{Weyl:DC}, a natural question arises: to what metric does it correspond? The answer is simple for $M=\bar M$. In this case, the Kerr-Schild double copy matches the Weyl double copy, as the metric can be chosen in the Kerr-Schild form \cite{Didenko:2009tc}
\begin{equation}
    g_{\mu\nu}:=\bar g_{\mu\nu}+2M\phif k_{\mu}k_{\nu}\,,
\end{equation}
where $\phif$ is defined in \eqref{zero}, and $k_{\mu}$ is either of the two Kerr-Schild vectors specified in \eqref{KS:kn}. As shown in \cite{Didenko:2009tc}, both choices result in the same Weyl tensor \eqref{Weyl:DC}. {Let us point out a limitation of this construction, which is new.} A subtle issue here is the assumption that \eqref{KS:kn} is well-defined, which turns out not to always be the case for certain generating parameters $K_{AB}$ in \eqref{K: comp}. In these degenerate situations, when real Kerr-Schild vectors do not exist, the corresponding Weyl double copy is not represented by the Kerr-Schild double copy. In practice, the degenerate cases emerge when the isometry of the double copy enhances, corresponding to particular values of Casimir invariants of $K_{AB}$ within types represented in Table 1. The corresponding metric can then be obtained by stepping away from the degenerate point, using the Kerr-Schild representation, and then considering the limit to this point.

The case when $M\neq\bar M$ is more involved. Formally, if one relaxes the reality condition of the metric\footnote{Alternatively, the metric is real but either euclidean or split signature.}, then it can be reached via the double Kerr-Schild ansatz
\begin{equation}\label{Double KS}
    g_{\mu\nu}=\bar g_{\mu\nu}+(M+\bar M)\phif k_{\mu}k_{\nu}+ i(M-\bar M)\tilde\phif l_{\mu}l_{\nu}\,,
\end{equation}
where $\tilde\phif$ is given in \eqref{dual:zero}, while $l_\mu$ is either of the two vectors defined in \eqref{KS:l}. The metric \eqref{Double KS} is complex because $l_{\mu}$ is not real. Additionally, this form assumes $k_{\mu}$ and $l_{\mu}$ are simultaneously well-defined, which is not always so in some symmetry-enhanced cases. 

Interestingly, there is a real representation found in \cite{Didenko:2009tc} of the metric that leads to \eqref{Weyl:DC}. It also uses the four Kerr-Schild vectors. Its form is not as simple as \eqref{Double KS}, and it manifestly contains Casimir invariants $I_1$ and $I_2$ from Table 1. The line element acquires the following form: 
\begin{equation}\label{metric:gen}
    ds^2=\frac{\hat\Delta_{\phif}}{\phif^{-2}+\tilde\phif^{-2}}\hat K^+\hat K^--\frac{\hat\Delta_{\tilde\phif}}{\phif^{-2}+\tilde\phif^{-2}}\hat L\hat{\bar L}\,,
\end{equation}
where 
\begin{subequations}
    \begin{align}
        &\hat K^{\pm}=\frac{\hat\Delta_{\phif}+\Delta_{\phif}}{2\hat{\Delta}_{\phif}}K^{\pm}+\frac{\hat\Delta_{\phif}-\Delta_{\phif}}{2\hat{\Delta}_{\phif}}K^{\mp}\,,\\
        &\hat L^{\pm}=\frac{\hat\Delta_{\tilde\phif}+\Delta_{\tilde\phif}}{2\hat{\Delta}_{\tilde\phif}}L^{\pm}+\frac{\hat\Delta_{\tilde\phif}-\Delta_{\tilde\phif}}{2\hat{\Delta}_{\tilde\phif}}L^{\mp}
    \end{align}
\end{subequations}
and $K^\pm$ and $L^\pm$ are constructed using the Kerr-Schild vectors \eqref{KS:spin} and the background AdS vierbein $\e_a$
\begin{equation}
    K^\pm:=k^{\pm a}\e_{a}\,,\qquad L^\pm:=l^{\pm a}\e_a\,.
\end{equation}
For explicit form of the real functions $\hat\Delta_{\phif}$ and $\hat\Delta_{\tilde\phif}$, which also depend on $M$ and $\bar M$, and the Casimir invariants $I_{1,2}$ with $\Delta:=\hat\Delta|_{M=\bar M=0}$, we refer to \cite{Didenko:2009tc}.        

For our consideration, it is not important whether the parameter $M$ is real or not. However, in what follows, we illustrate our results mostly assuming $M=\bar M$ for simplicity in order to deal with the Kerr-Schild form of the classified solutions.

\subsection{The Kerr-Schild and Weyl multicopies}\label{sec:HS}

A notable feature of the introduced copies is that their sequence of zeroth, single, and double copies does not terminate but rather extends, forming higher-spin multicopies \cite{Didenko:2008va} of any integer spin $s$ in the following simple manner:
\begin{equation}\label{s:DC}
    \phi_{\mu_1\dots\mu_s}=\phif k_{\mu_1}\dots k_{\mu_s}\,.
\end{equation}
The multicopy satisfies 
\begin{equation}\label{Ks:multicopy}
    \Box \phi_{\mu_1\dots\mu_s}-s\nabla_{\nu}\nabla_{(\mu_1}\phi^{\nu}{}_{\mu_2\dots\mu_s)} - 2(s-1)(s+1)\gl^2\phi_{\mu_1\dots\mu_s}+{\text{trace part}}=0\,.
\end{equation}
The trace part contains traces of the Fronsdal field $\phi^{\nu}{}_{\nu\mu_1\dots\mu_{s-2}}$ that vanish on the solutions \eqref{s:DC}. The latter equation is just the zeroth copy \eqref{0:DC} for $s=0$, the single copy \eqref{1:DC} for $s=1$, and the double copy \eqref{2:DC} for $s=2$, while for $s>2$ it is the standard Fronsdal massless spin $s$ equation in AdS \cite{Metsaev:1997nj}. 

The spin $s$  Kerr-Schild multicopy \eqref{s:DC} naturally leads to the corresponding Weyl multicopy. However, this higher-spin Weyl multicopy is linearized, in contrast to the exact version seen in the case of gravity {(see, however, footnote~\ref{foot:HSlocal})}. These linearized higher-spin Weyl tensors (see, {\it e.g.,} \cite{Vasiliev:1999ba} for their definitions) reveal the generalized type D form 
\begin{equation}\label{HS:Weyl DC}
    C_{\al(2s)}=\frac{M_s}{\rf^{2s+1}}\underbrace{\gk_{\al\al}\dots\gk_{\al\al}}_{s}\,,\qquad  \bar C_{\dal(2s)}=\frac{\bar M_s}{\bar \rf^{2s+1}}\underbrace{\bar\gk_{\dal\dal}\dots\bar\gk_{\dal\dal}}_s\,,
\end{equation}
provided $\gk_{\al\gb}$ is nondegenerate. Here $M_s$ and $\bar M_{s}$ are arbitrary spin-dependent constants. The Weyl tensors reduce to \eqref{Maxwell} for $s=1$ and to \eqref{Weyl:DC} for $s=2$.

\subsection{The Penrose transform}
It was realized in \cite{Didenko:2009td} that the zeroth copy \eqref{zero}, the single copy \eqref{Maxwell}, and the double copy \eqref{Weyl:DC}, as well as the higher-spin multicopies \eqref{HS:Weyl DC}, emerge naturally from {a twistor Penrose-like transform in AdS. In flat space, the twistorial origin of the classical double copy was put forward in \cite{White:2020sfn} and developed in a number of subsequent works. These include the derivation of the Weyl double copy from twistor space and its alternative formulations \cite{Chacon:2021wbr, Chacon:2021hfe}, the relation to the Newman--Penrose map \cite{Farnsworth:2021wvs}, the double copy of the multipole expansion \cite{Chacon:2021lox}, the locality of the Weyl double copy in position space \cite{Luna:2022dxo}, non-perturbative aspects of the self-dual double copy \cite{Armstrong-Williams:2022apo}, and the reconstruction of classical spacetimes from the $S$-matrix in twistor space \cite{Guevara:2021yud}. Related constructions include the three-dimensional mini-twistor (Cotton) double copy \cite{Carrillo-Gonzalez:2022ggn}, the double copy of self-dual Yang--Mills theory to self-dual gravity directly on twistor space \cite{Borsten:2023paw}, and the twistor-space form of the KLT kernel \cite{Adamo:2024xpc}; see \cite{White:2024pve} for a review. The equivalence between the twistorial and Kerr--Schild prescriptions for a class of self-dual Kerr--Schild spacetimes was discussed recently in \cite{Albertini:2025ogf}. The Weyl double copy in curved backgrounds with a cosmological constant was studied in \cite{Carrillo-Gonzalez:2017iyj, Han:2022mze, Han:2022ubu, Chawla:2022ogv}. Let us first recall how the Penrose transform works in} the case of a vanishing cosmological constant $\Lambda=0$. 

The standard Penrose transform \cite{Eastwood:1981jy, Adamo:2017qyl} is a certain map from functions of twistors to solutions of free massless field equations. More precisely, a map to solutions of the Maxwell-type equations
\begin{equation}\label{massless:flat}
    \frac{\partial}{\partial x^{\al\dgb}}\bar C^{\dgb}{}_{\dal(2s-1)}=0\,,\qquad  \frac{\partial}{\partial x^{\gb\dal}} C^{\gb}{}_{\al(2s-1)}=0\,.
\end{equation}
Let us define a rank-1 twistor as a pair of spinors $Z^A=(\xi^\al, \bar\xi_{\dal})$ related to each other by the following incidence condition
\begin{equation}\label{incidence}
    \xi^{\al}=ix^{\al\dal}\bar\xi_{\dal}\,,
\end{equation}
which establishes a nonlocal correspondence between points in space-time $x^{\al\dal}$ and points in twistor space $Z^A$. The Penrose transform is then defined as a map from the holomorphic twistor function $f(Z)$ into a solution for the generalized Weyl tensors \eqref{massless:flat} as follows:
\begin{equation}\label{Penrose}
    \bar C_{\dal_1\dots\dal_{2s}}=\oint_{\Gamma} d\bar\xi^{\dgb}\bar\xi_{\dgb}\bar\xi_{\dal_1}\dots\bar\xi_{\dal_{2s}}f(Z)\Big|_{\mathbb{CP}^1}\,,
\end{equation}
where the $\mathbb{CP}^1$ projection implies that the function $f(Z)$ should be calculated on the incidence relation \eqref{incidence}
\begin{equation}
f(Z)\Big|_{\mathbb{CP}^1}=f(ix^{\al\dal}\bar\xi_{\dal}, \bar\xi_{\dgb})\,,  
\end{equation}
while $\Gamma$ is a certain separating pole contour. It is not hard to see that \eqref{massless:flat} does hold after substituting \eqref{Penrose}. For example, type D solution for the massless spin $s$ field can be obtained by choosing, see {{\it e.g.,} \cite{White:2020sfn, Chacon:2021wbr}} 
\begin{equation}\label{exmpl}
    f(Z)=\frac{1}{(Q_{AB}Z^AZ^B)^{s+1}}\,,
\end{equation}
where $Q_{AB}=Q_{BA}$ is some constant matrix. 

In the case of a nonzero cosmological constant, Eqs. \eqref{massless:flat} are modified covariantly,
\begin{equation}\label{massless:ads}
    \nabla_{\al\dgb}\bar C^{\dgb}{}_{\dal(2s-1)}=0\,,\qquad  \nabla_{\gb\dal} C^{\gb}{}_{\al(2s-1)}=0\,,
\end{equation}
where $\nabla_{\al\dgb}$ is a covariant derivative in AdS. {The Penrose transform extends to this case in a well-known way. Since AdS$_4$ is conformally flat, its twistor space is (a region of) the same projective twistor space as that of Minkowski space, and the cosmological constant enters only through the choice of the infinity twistor, which breaks the conformal group $\so(2,4)$ down to $\so(2,3)$. Because the free massless equations are conformally invariant, their solutions on AdS are generated by the same twistor functions, up to the appropriate conformal rescaling. This is classical material, covered in textbook treatments of conformally flat spacetimes \cite{Penrose:1986ca, Huggett:1986fs, Ward:1990vs}. The same framework underlies the twistor description of self-dual Einstein spaces with a cosmological constant \cite{Ward:1980am} and twistor actions for Einstein gravity with $\Lambda\neq0$ \cite{Adamo:2013tja}. AdS twistor methods have also found many modern applications, including holography and higher-spin theory \cite{Adamo:2016rtr, Lang:2024dkt, Neiman:2017mel} and boundary correlators in (A)dS$_4$ \cite{Carrillo-Gonzalez:2025ehb}. Directly relevant to the double copy are the twistor-space double copy of AdS$_4$ boundary correlators \cite{Ansari:2025jdk} and the color/kinematics duality of self-dual gravity in AdS$_4$ \cite{Lipstein:2023pih}. Rather than the standard contour-integral form, in what follows we use the unfolded version of the AdS Penrose transform \cite{Didenko:2009td, Gelfond:2008td, Neiman:2017mel}, which makes the global symmetries of the solutions manifest.} 

The standard Penrose transform keeps no track of global symmetries of solutions for massless equations. In particular, in the example above \eqref{exmpl}, there is no immediate answer to the question of what the leftover symmetries of the solution \eqref{Penrose} are for a given set of constants $Q_{AB}$. This occurs because the Penrose transform maps twistors to solutions for primaries (Weyl tensors), rather than to solutions of the entire module, which also includes descendants. In the latter case, if the symmetries acting on the module are under control, it becomes possible to identify the global symmetries. This can indeed be accomplished within the so-called unfolded approach to higher spins \cite{Vasiliev:1988sa}; see also \cite{Misuna:2024ccj, Misuna:2024dlx, Iazeolla:2025btr, Misuna:2026bhy} for recent applications. 

\subsection{The Penrose transform and global symmetries}\label{sec:Penrose}

Let us have a look at the free massless field equations \eqref{massless:ads} in AdS. It is  convenient to pack the generalized (anti)holomorphic Weyl tensors $C_{\al(2s)}$ and $\bar{C}_{\dal(2s)}$ into a generating function by introducing commuting variables $y_{\al}$ and $\bar y_{\dal}$, following \cite{Vasiliev:1999ba}
\begin{equation}\label{C:gener}
    C(y,\bar y|x)=\sum_{m,n=0}^{\infty}\frac{1}{m!n!}C_{\al(m), \dal(n)}\underbrace{y^{\al}\dots y^{\al}}_{m}\underbrace{\bar y^{\dal}\dots\bar y^{\dal}}_n\,.
\end{equation}
Notice that the generating function \eqref{C:gener} contains more fields  than needed. Indeed, apart from (anti)holomorphic spin $s$ Weyl tensors, it also contains mixed fields $C_{\al(m), \dal(n)}$ with $m\cdot n\neq 0$. These components are auxiliary and will be expressible via covariant AdS derivatives of $C_{\al(2s)}$ and $\bar C_{\dal(2s)}$ on-shell.  
Equations \eqref{massless:ads} acquire the following form in terms of the generating function \eqref{C:gener}:
\begin{equation}\label{C:twadj}
    \dr C+W_0*C-C*\pi(W_0)=0\,,
\end{equation}
where 
\begin{equation}\label{W0}
    W_0=-\frac{1}{4}(\go_{\al\gb}y^{\al}y^{\gb}+\bar\go_{\dal\dgb}\bar y^{\dal}\bar y^{\dgb}+2\gl\,\mathbf{e}_{\al\dgb}y^{\al}\bar y^{\dgb})
\end{equation}
is the generating function that encodes AdS frame fields being the (anti)holomorphic Lorentz connections $\go$ and $\bar\go$, and the vierbein $\mathbf{e}$. Here, the star-product operation $*$ corresponding to the symmetric Weyl ordering of symbols $y (\bar y)$ is introduced 
\begin{equation}\label{star}
    f(y, \bar y)*g(y, \bar y)=f(y, \bar y) 
    e^{i\gep^{\al\gb}\overleftarrow\partial_{\al}\overrightarrow\partial_\gb+ i\gep^{\dal\dgb}\overleftarrow{\partial}_{\dal}\overrightarrow{\partial}_{\dgb}}
    g(y, \bar y)\,.
\end{equation}
Its integral representation formula is more useful in practice 
\begin{equation}\label{star:int}
    f(y, \bar y)*g(y, \bar y)=\frac{1}{(2\pi)^4}\int du dv d\bar u d\bar v\,f(y+u, \bar y+\bar u)g(y+v, \bar y+\bar v)e^{i u_{\al}v^{\al}+i\bar u_{\dal}\bar v^{\dal}}\,.
\end{equation}
The introduced star product generates the following multiplication rules:
\begin{subequations}
    \begin{align}
        &y*=y+i\partial^y\,,\qquad *y=y-i\partial^y\,,\\
        &\bar y*=\bar y+i\bar\partial^{\bar y}\,,\qquad *\bar y=\bar y-i\bar\partial^{\bar y}\,.
    \end{align}
\end{subequations}
There is an important automorphism $\pi$ in \eqref{C:twadj} that properly flips the sign 
\begin{equation}
    \pi f(y, \bar y)=f(-y, \bar y)
\end{equation}
making the representation \eqref{C:twadj} twisted-adjoint.\footnote{\label{foot} The adjoint case with $\pi=Id$ is also consistent but does not describe propagating massless fields; rather, it describes a collection of Killing tensors packed in $C(y, \bar y)$. In the higher-spin context, such fields are called topological \cite{Vasiliev:1999ba}; see also \cite{Kirakosiants:2026spd} for a recent account.} It is this twist that leads to the dynamical content \eqref{massless:ads} for the spin $s$ Weyl tensors and to \eqref{0:DC} for a scalar \begin{equation}\label{HS:scalar}
    \phi:=C(0, 0|x)\,.
\end{equation}  
The consistency of \eqref{C:twadj} forces the connection $W_0$ to be flat
\begin{equation}\label{HS:flat}
    \dr W_0+W_0*W_0=0
\end{equation}
that with the prescription \eqref{W0} makes fields in $W_0$ the frame fields of AdS. Now, it is a matter of straightforward calculation using \eqref{W0} and \eqref{star} to check that \eqref{C:twadj} is equivalent to \eqref{massless:ads}, along with infinitely many consequences that simply express fields $C_{\al(m), \dot\alpha(n)}$ with $m\cdot n\neq 0$ via the primary components $C_{\al(m)}$ and $C_{\dal(n)}$. For $s=0$, the dynamical content differs from \eqref{massless:ads}, resulting in \eqref{0:DC} for \eqref{HS:scalar}.  

A nice feature of Eq. \eqref{C:twadj} is that, unlike \eqref{massless:ads}, it makes global symmetries manifest. Indeed, suppose one is given a certain solution $C_0(y, \bar y|x)$ of \eqref{C:twadj} and $W_0$ of \eqref{HS:flat}. From the fact that the transformation 
\begin{align}
    &\gd C=\xi*C-C*\pi (\xi)\,,\\
    &\gd W=\dr \xi+[W,\xi]_*\,,
\end{align}
where $\xi(y, \bar y|x)$ is an arbitrary local parameter, leaves \eqref{C:twadj}, \eqref{HS:flat} invariant it follows that the parameter $\xi$ satisfying  
\begin{align}
    &\xi*C_0-C_0*\pi (\xi)=0\,,\label{global:C}\\
    &\dr\xi+[W_0, \xi]_*=0\label{global:W}
\end{align}
generates the leftover global symmetry of the solution $C_0$. The condition \eqref{global:W} is the unfolded way of saying that the parameter $\xi_0$ satisfies the equation for Killing tensors in AdS; see also footnote \ref{foot}. Indeed, taking, for example,
\begin{equation}
    \xi=K_{AB}Y^AY^B\,,\qquad Y^A:=(y^{\alpha}, \bar y^{\dal})\,,
\end{equation}
where $K_{AB}$ is given in \eqref{K: comp}, Eq. \eqref{global:W} amounts to \eqref{ads:param-spin} in components.

Notice that the difference between \eqref{global:W} and \eqref{C:twadj} is the twist $\pi$. This is where the Penrose transform comes about. To get at it, we need the following useful observation \cite{Didenko:2009td}. The star-product element 
\begin{equation}
    \gk_y:=2\pi \gd^{(2)}(y)
\end{equation}
can be shown to generate a sign flip 
\begin{equation}
    f(y, \bar y)*\gk_y=\gk_y*f(-y, \bar y)\,,\qquad \gk_y*\gk_y=1\,.
\end{equation}
Therefore, the action of $\pi$ can be arranged in the following form:
\begin{equation}\label{Klein}
    \pi f(y, \bar y)=\gk_y*f(y, \bar y)*\gk_y\,.
\end{equation}
We call a map from solution of \eqref{global:W} to a solution of \eqref{C:twadj} the Penrose transform. Its explicit realization is as follows. Suppose we are given a function $\xi(y, \bar y|x)$ that satisfies \eqref{global:W} then \eqref{Klein} entails that
\begin{equation}\label{Penrose:unfl}
    C(y, \bar y|x):=\xi*\gk_y
\end{equation}
satisfies \eqref{C:twadj}. The star product with $\gd$-function in \eqref{Penrose:unfl} generates the Fourier transformation with respect to the variable $y$
\begin{equation}\label{Penrose:ads}
    C=\xi*\gk_y=\int d^2u\,\xi(u, \bar y)e^{iy_{\al}u^{\al}}\,.
\end{equation}
This fact can be seen using \eqref{star:int}.
Notice that, unlike the standard form of the Penrose transform given by a contour integral \eqref{Penrose}, Eq. \eqref{Penrose:ads} features integration along a two-dimensional plane. The equivalence of these two representations is discussed in \cite{Didenko:2021vui}. Loosely speaking, the Penrose transform is a map between solutions of dynamical equations \eqref{massless:ads} and a space of Killing tensors encoded in \eqref{global:W}.  

Now, suppose we have a solution of \eqref{massless:ads} that has the form \eqref{Penrose:unfl}. In this case, its global symmetries are easy to identify using \eqref{global:C}, \eqref{global:W}, and \eqref{Klein}. The corresponding parameter $\xi_0$ enjoys
\begin{equation}\label{center}
    [\xi_0, \xi]_*=0\,,
\end{equation}
which means that $\xi_0$ belongs to the centralizer of $\xi$. 

The Weyl double copies \eqref{Weyl:DC} along with their higher-spin cousin multicopies in AdS can be constructed in this way \cite{Didenko:2009td}. The analog of type D solutions \eqref{exmpl} in the case of a negative cosmological constant can be generated using the AdS parameter \eqref{param:sp4}. Indeed, the unfolded Penrose map \eqref{Penrose:unfl} can be chosen in the form 
\begin{equation}\label{C:Dtype}
    C=\xi*\gk_y\,,\qquad \xi:=f(K_{AB}Y^AY^B)
\end{equation}
where $f$ is an unspecified function. From \eqref{global:W}, it follows that 
\begin{equation}\label{K:eq}
\dr(K_{AB}Y^AY^B)+[W_0, K_{AB}Y^AY^B]_*=0\,,    
\end{equation}
which is equivalent to \eqref{param:sp4}

Notice that $f$ cannot be a polynomial, since in this case, the Penrose transform results in a distribution with respect to the variables $Y$. Carrying out integration using \eqref{star:int} and extracting Weyl tensors from the final result, we arrive at \eqref{HS:Weyl DC} (see \cite{Didenko:2021vui} for more detail). Constants $M_s$ in \eqref{HS:Weyl DC} depend on the particular function $f$ chosen. The remaining global symmetries of \eqref{C:Dtype}, \eqref{HS:Weyl DC} are given by \eqref{center}. Focusing on the isometries, which are parameterized by the bilinear parameters
\begin{equation}
    \xi_0=\varepsilon_{AB}Y^AY^B\,,
\end{equation}
where the Killing vector is associated with the component $\varepsilon_{\al\dgb}$, we find that these parameters $\varepsilon_{AB}$ must be centralizers of $K_{AB}$
\begin{equation}\label{center:isom}
    [\varepsilon, K]_{AB}=\varepsilon_{A}{}^C K_{CB}-K_{A}{}^C\varepsilon_{CB}=0\,.
\end{equation}  

Let us clarify an important point regarding Eq. \eqref{center}, which is related to Eq. \eqref{center:isom}{; this remark on the applicability of the construction is new}. This counting of isometries is valid only when the decomposition in Eq. \eqref{C:Dtype} is applicable. Specifically, this requires that $\det\gk_{\al\gb}\neq 0$, a condition that is not satisfied for certain cases of $K_{AB}$. In such situations, the Penrose transform can still be applied by regularizing $\det\gk_{\al\gb} = \gep \to 0$, leading to Petrov type N solutions instead of type D. However, the global symmetries in this case may not align with the center of $K_{AB}$ because Eq. \eqref{center} was derived under the assumption of a factorized form given by Eq. \eqref{Penrose:ads}, which does not hold as $\gep$ approaches zero. Therefore, in these cases that we encounter in our analysis, the corresponding global symmetries need to be re-evaluated.

Another comment highlights the difference between a double copy and a multicopy. The double copy, as specified in \eqref{Weyl:DC}, is derived from the Penrose transform and is designed to solve the linearized Einstein equations \eqref{massless:ads} for $s=2$. Notably, these solutions are also exact solutions of the Einstein equations. In contrast, for multicopies with $s>2$, this property no longer holds {(see, however, footnote~\ref{foot:HSlocal})}. As demonstrated in \cite{Didenko:2021vdb}, quadratic corrections to the linearized solutions \eqref{massless:ads} are generally nonzero.

\section{Classifying copies}\label{sec:class}
{Unlike Sections~\ref{sec:ads}--\ref{sec:dc}, this section contains new material: the argument relating inequivalent solutions to adjoint orbits, and the discussion of the solutions not captured by a real parameter $K$, together with the proposed extension to non-Hermitian parameters (Sec.~\ref{miss}). Section~\ref{sec:orbits} reviews the classification of the adjoint orbits \cite{Holst:1997tm}. The three-dimensional analogy recalled below is also known \cite{Banados:1992wn, Banados:1992gq}.}

To clarify the classification of multicopies and double copies, we focus on the solutions obtained through the Penrose transform, which is generated by the AdS global symmetry parameter $K_{AB}$, as given in \eqref{C:Dtype}. These solutions correspond to Weyl tensors of the form given in \eqref{Weyl:DC} and their higher-spin extensions \eqref{HS:Weyl DC}. The key question is: how many inequivalent solutions can be generated by different parameters $K_{AB}$? 

{\paragraph{Analogy with three dimensions.} This question has a well-known counterpart in three dimensions, which motivates our analysis. Pure three-dimensional gravity with $\Lambda<0$ is topological: all its solutions are locally AdS$_3$. As noted by Witten (see a comment in \cite{Banados:1992wn}), the BTZ black hole \cite{Banados:1992wn} is obtained from AdS$_3$ by identifying points along a Killing vector field $\xi\in\so(2,2)$. More generally, the elements of $\so(2,2)$ were classified in \cite{Banados:1992gq} up to the adjoint action $\Omega\to U^{-1}\Omega U$, $U\in SO(2,2)$, and the inequivalent identifications, and hence the resulting solutions (black holes, deficit angles, naked singularities, and others), correspond one-to-one to the adjoint orbits listed in Table~\ref{tab:3d}. Two features of this construction carry over to four dimensions. First, the solutions are labeled by adjoint orbits, which are distinguished in part by the Casimir invariants. Second, the isometries of a solution are the background Killing vectors that commute with the generating element, {\it i.e.,} they form its centralizer. The difference is that in $d=4$ gravity is not topological, so the solutions cannot be obtained by identifications. Instead, the generating element $K\in\so(2,3)$ produces them through the Penrose transform, and the classification of solutions reduces to that of the adjoint orbits of $K$, as we now show.

\begin{table}[h]
    \centering
    \renewcommand{\arraystretch}{1.4}
    \begin{tabular}{c c c c}
        \toprule
        Type & Killing field & $\frac{1}{4} I_1$ & $\frac{1}{4} I_2$ \\
        \midrule
        $I_a$ & $b(J_{01} + J_{23}) - a(J_{03} + J_{12})$ & $b^2 - a^2$ & $b^2 + a^2$ \\
        $I_b$ & $\lambda_1 J_{12} + \lambda_2 J_{03}$ & $-\frac{1}{2} (\lambda_1^2 + \lambda_2^2)$ & $\lambda_1 \lambda_2$ \\
        $I_c$ & $b_1 J_{01} + b_2 J_{23}$ & $\frac{1}{2} (b_1^2 + b_2^2)$ & $b_1 b_2$ \\
        $II_a$ & $\lambda (J_{03} + J_{12}) + J_{01} - J_{02} - J_{13} + J_{23}$ & $-\lambda^2$ & $\lambda^2$ \\
        & or $\lambda (-J_{03} + J_{12}) - J_{13} + J_{23}$ ($\lambda \neq 0$) & $-\lambda^2$ & $-\lambda^2$ \\
        $II_b$ & $(b - 1) J_{01} + (b - 1) J_{23} + J_{02} - J_{13}$ & $b^2$ & $b^2$ \\
        $III^+$ & $-J_{13} + J_{23}$ & $0$ & $0$ \\
        $III^-$ & $-J_{01} + J_{02}$ & $0$ & $0$ \\
        \bottomrule
    \end{tabular}
    \caption{Classification of one-parameter subgroups of $SO(2,2)$ from \cite{Banados:1992gq}.}\label{tab:3d}
\end{table}

}
To address this, we need to examine the defining equations that underlie the Penrose transform. The AdS background is characterized by the condition \eqref{HS:flat}, which we can solve in a pure gauge form
\begin{equation}\label{W0:sol}
    W_0=g^{-1}(y,\bar y|x)*\dr g(y, \bar y|x)\,.
\end{equation}
where $g$ is such that $W_0$ is bilinear \eqref{W0} and $g(y,\bar y|x_0)=1$ at a certain space-time point $x_0$. Eq. \eqref{K:eq} is solved as follows
\begin{equation}\label{K:gauge}
    K=g^{-1}*K_0*g\,,\qquad K:=K_{AB}Y^AY^B\,,\quad K_{0}:=K_{AB}(x_0)Y^AY^B\,,
\end{equation}
where $K_0$ is the value of $K(x)$ at $x=x_0$. Now, notice that \eqref{W0:sol} is invariant under 
\begin{equation}\label{inv:left}
    g\to h(y, \bar y)*g\,,
\end{equation}
where $h$ does not depend on spacetime coordinates. The invariance \eqref{inv:left} when substituted into \eqref{K:gauge} implies the following equivalence
\begin{equation}\label{K0:equiv}
    K_0\sim h^{-1}*K_0*h\,,
\end{equation}
where we recall that $K_0$ must remain bilinear. The equivalence \eqref{K0:equiv} represents the equivalence under the group adjoint action. Given $K_{AB}\in \sp(4)$, we have the following equivalence relation
\begin{equation}\label{K:equiv}
    K_{AB}\sim (U^{-1}KU)_{AB}\,,\qquad U\in Sp(4)\,,
\end{equation}
where $U$ is a group element, the action of which is represented by the element $h$ in \eqref{K:equiv}. From this straightforward analysis, we conclude that the two parameters  $K_{AB}$  and  $K'_{AB}$ , which belong to the same orbit generated by the adjoint group action, lead to equivalent Penrose transforms up to similarity transformation. Consequently,  $K_{AB}$  can be divided into various $\sp(4)$ orbits that we need to classify. In particular, the two solutions driven by the Penrose map using $K$ and $K'$ differ if the Casimir invariants \eqref{Casimir:K} of $K$ differ from those of $K'$. However, the Casimir invariants do not entirely fix the adjoint $\sp(4)$ orbit.

{ Let us note that some steps of the orbit classification described above were performed in \cite{Iazeolla:2011cb} to single out $Sp(4,\mathbb{R})$-inequivalent Weyl multicopy solutions of the Vasiliev equations, where complexifications of the global symmetry parameter $K_{AB}$ were also considered. For technical reasons related to the construction of exact solutions of the Vasiliev system, that analysis was restricted to a few concrete orbits (type $I$ in our classification). The corresponding Weyl multicopies were found as exact solutions of the Vasiliev equations in Sections 5.2 and 5.3 there. Their residual symmetries were also discussed and summarized in \cite{Iazeolla:2011cb}. This analysis was further extended in \cite{Iazeolla:2020jee}, and analogous methods were used in \cite{Aros:2019pgj} to analyze fluctuations over a four-dimensional BTZ black hole. In context of multicopies our analysis extends these results to all adjoint orbits of $\so(2,3)$.}  

\subsection{Missing solutions}\label{miss}
It may seem that all double copies \eqref{Weyl:DC} and their higher-spin extensions \eqref{HS:Weyl DC} are well captured by the AdS global symmetry parameter $K_{AB}$ in \eqref{K: comp}, but there is a subtlety that makes this statement false. The isomorphism $\so(2,3)\sim\sp(4)$ implies that the $\sp(4)$ element is Hermitian; see Appendix A
\begin{equation}\label{K:Herm}
    K_{AB}=(K_{AB})^\dagger\quad\Rightarrow\quad v_{\al\dgb}=(v_{\al\dgb})^\dagger\,,\qquad (\gk_{\al\gb})^\dagger=\bar\gk_{\dal\dgb}\,.
\end{equation}
From the perspective of the Weyl double copy in \eqref{Weyl:DC}, the field  $\gk_{\al\gb}$  is inherently complex. Therefore, there is no requirement for the parameter  $K_{AB}$  to be Hermitian. Relaxing the condition \eqref{K:Herm} effectively means that the seed Killing vector $v$  is also complex
\begin{equation}\label{K:nHerm}
    K_{AB}\neq(K_{AB})^\dagger\quad\Rightarrow\quad v_{\al\dgb}\neq v_{\dgb\al}\,,\qquad (\gk_{\al\gb})^\dagger=\bar\gk_{\dal\dgb}\,,
\end{equation}
which means that one deals with two real AdS Killing vectors: $v_1:=\text{Re}\,v$ and $v_2:=\text{Im}\,v$. At the linearized level, the Weyl double copy generated using a complex parameter $K$ via the Penrose transform works well. {Complexifications of $K_{AB}$ were also considered in \cite{Iazeolla:2011cb} in the context of exact higher-spin solutions.} However, it is unlikely that the results obtained this way in \eqref{Weyl:DC} will hold true at the full GR level.

The proof that the solution in equation \eqref{Weyl:DC} is linearly exact, as presented in \cite{Didenko:2009tc}, utilized integrating flow equations that depend on the mass parameters  $M$  and  $\bar M$ . In this context, the reality conditions for $K$  played a crucial role. However, we can argue that if the condition
\begin{equation}\label{com:conj}
    [K, K^\dagger]_{AB}=0\,,
\end{equation}
holds true -- indicating that the two Killing vectors $v_1$ and $v_2$  commute -- then Eq. \eqref{Weyl:DC} remains linearly exact. To support this assertion, we refer to observations made in \cite{Didenko:2009tc} (see also \cite{Easson:2022zoh}), where it was shown that starting with the following seed AdS metric in the coordinates  $x^{\mu}=(\tau, \psi, p, q)$:
\begin{equation}
    ds^2=\frac{1}{(1-qp)^2}\left(\frac{Q_0(q)(d\tau-p^2d\psi)^2}{q^2+p^2}-\frac{P_0(p)(d\tau+q^2d\psi)^2}{q^2+p^2}-\frac{q^2+p^2}{Q_0(q)}dq^2-\frac{q^2+p^2}{P_0(p)}dp^2\right)\,,
\end{equation}
where 
\begin{subequations}
    \begin{align}
        &Q_0(q):=\gl^2/2+\gamma+\varepsilon q^2-(\gl^2/2-\gamma)q^4\,,\\
        &P_0(p):=\gl^2/2+\gamma-\varepsilon p^2-(\gl^2/2-\gamma)p^4
    \end{align}
\end{subequations}
and $\gamma$, $\varepsilon$ are some arbitrary constants, and by utilizing  
a complex Killing vector $v^{\mu}=(1,i, 0, 0)$, which corresponds to two real vector fields $\p_{\tau}$ and $\p_{\psi}$, one can derive the so-called Pleba\'{n}ski-Demia\'{n}ski metric \cite{Plebanski:1976gy}, which is not captured by construction by any real AdS global symmetry parameter.

Consequently, the rich classification provided in this paper based on the real global symmetry parameter is still not exhaustive. The interplay of the complex structure arguably allows for many more solutions being a feature of four dimensions; see also \cite{Krasnov:2024qyk} for a closely related discussion. 

If the condition \eqref{com:conj} is the only constraint for the complex $K_{AB}$,
we anticipate that each real  $\sp(4)$  orbit of $K_{AB}=K_{AB}^\dagger$  can generate additional solutions through the parameter 
\begin{equation}
    K'_{AB}=K_{AB}+i\gga G_{AB}\,,
\end{equation}
where $G=G^\dagger$ belongs to the center of $K$ and $\gga$ is an arbitrary constant\footnote{In the context of the Pleba\'nski-Demia\'nski solution, this constant serves as an acceleration parameter.}. We will not consider this option in this paper and restrict ourselves to the Hermitian case \eqref{K:Herm}.

\subsection{Adjoint orbits}\label{sec:orbits} 

Bearing in mind higher-dimensional generalization, we will look for the adjoint $\sp(4)$ orbits \eqref{K:equiv} using the isomorphism $\so(2,3)\sim \sp(4)$. So, instead of classifying $\sp(4)$ symmetric matrices $K_{AB}=K_{BA}$, we classify antisymmetric $\so(2,3)$ matrices $\Omega_{\underline{AB}}=-\Omega_{\underline{BA}}$; see \eqref{param:vec}. Specifically, we are interested in the classification  of antisymmetric matrices $\Omega_{\underline{IJ}}$ up to an equivalence relation:
\begin{equation}
    \Omega'^{\underline{I}}{}_{\underline{J}} = (O^{-1})^{\underline{I}}{}_{\underline{K}} \Omega^{\underline{K}}{}_{\underline{L}} O^{\underline{L}}{}_{\underline{J}},\qquad O \in O(2, 3)\,.
\end{equation}
This problem has been solved in {\it e.g.,} \cite{Holst:1997tm}. Notice that while symmetric matrices are classified by their eigenvalues, the matrix $\Omega_{\underline{AB}}$ is not diagonalizable; instead, it can be reduced to the normal Jordan form $J$: $\Omega = O^{-1} J O$.
\begin{equation}
   J =  \begin{pmatrix}
       \gl_1 & a_1 & 0 & 0 & 0\\
       0 & \gl_2 & a_2 & 0 & 0\\
       0 & 0 & \gl_3 & a_3 & 0\\
       0 & 0 & 0 & \gl_4 & a_4\\
       0 & 0 & 0 & 0 & \gl_5\\
   \end{pmatrix},\qquad a_i \in \{0, 1\}.
\end{equation}
If $a_k = 1$, then $\gl_k = \gl_{k + 1}$. For example, consider the case $a_1 = 1$ and $a_2 = a_3 = a_4 = 0$, which leads to the existence of one eigenvector $\boldsymbol{x}$ with eigenvalue $\gl_1 = \gl_2 = \gl$: $M \boldsymbol{x} = \gl \boldsymbol{x}$, and a vector $\boldsymbol{y}$ associated with it: $M \boldsymbol{y} = \gl \boldsymbol{y} + \boldsymbol{x}$. When the operator $M$ acts on any vector from the linear span formed from the vectors $\{\boldsymbol{x}, \boldsymbol{y}\}$, it remains in the linear span. Thus, the vectors $\{\boldsymbol{x}, \boldsymbol{y}\}$ form an invariant subspace with the corresponding eigenvalue $\gl$. Consequently, the matrix $\Omega$ is completely described by the eigenvalues and the dimension of the invariant subspace. 

The linear operator $\Omega$ can be conveniently decomposed into the sum of a semisimple operator $S$ and a nilpotent operator $N$ according to the Jordan-Chevalley decomposition:
\begin{equation}\label{W=S+N}
    \Omega = S + N,\qquad [S, N] = 0,\qquad N^q = 0, \qquad q \in \mathbb{Z}
\end{equation}
where $S$ is a diagonalizable matrix with complex eigenvalues: $\Lambda = O S O^{-1}$. 

The explicit derivation of all algebraically non-equivalent cases can be found in \cite{Holst:1997tm}, with the results summarized in the introduction; see Table 1, which basically reproduces the result of \cite{Holst:1997tm}.

The proposed classification, based on the root space of the AdS global symmetry parameter $K$, groups various Einstein spacetimes along with linearized higher-spin multicopies according to the eigenvalues and dimensions of the invariant spaces of the symmetry parameter. 

We explore the parametric family of Killing fields for each orbit, analyzing the exact Einstein solution they induce via the Penrose transform and identifying the corresponding Weyl and Kerr-Schild (when they exist) double copies. Additionally, we look at the specific values of the orbit parameters that lead to degeneracy or symmetry enhancement of the respective solutions. These cases will be studied in detail.  

In all parametrically non-degenerate cases, the stabilizer algebra coincides with the Cartan subalgebra for $\mathfrak{so}(2, 3)$, having only two generators. However, there are special points in the orbit parameter space where the centralizer enhances and may contain up to six generators.  

In classifying solutions, we will begin with type \rom{1} $(N = 0)$, which corresponds to semisimple orbits. Recall that semisimple elements -- elements for which the adjoint action is diagonalizable -- are divided into three qualitative types, often named in analogy with the isometries of hyperbolic space (or M{\"o}bius transformations): elliptic, hyperbolic, loxodromic, depending on the isometries formed by compact, non-compact, or mixed operators; see \cite{fresse2021approximationnilpotentorbitssimple}, and \cite{Parikh:2011aa}. For example, in the given classification, the orbit type $\rom{1}_c$ with imaginary eigenvalues is elliptic and is described by two compact rotations; type $\rom{1}_b $ with real eigenvalues is hyperbolic and is described by non-compact boosts; and type $\rom{1}_d $ with one real and one imaginary eigenvalue is loxodromic, which represents a combination of boost and rotation, all in orthogonal planes. There exist infinitely many distinct semisimple orbits \cite{McGovern1993NilpotentOrbits}. Each orbit is characterized by its own unique values of parameters $a$ or $b$ (see Table 1) or Casimir invariants.

The construction of the Kerr-Schild space, where the symmetry of two commuting Killing functions is explicitly realized, qualitatively differs in the structure of the defining generators. When one of the parameters $a$ or $b$ is zero, we have only three non-trivial possibilities: when the generating Killing field describes rotation in the timelike $(--)$ or spatial $(++)$ planes, and when it is a boost $(-+)$. Another special case is the reduction of the parameters $a=\pm b$. The dimension of the stabilizer algebra in these special cases is four.

Pure nilpotent adjoint orbits in $\so(2,3)$ 
obtained by degenerating the parameters $a, b = 0$ for the mixed orbits $\rom{2}$ and $\rom{3}$ are in one-to-one correspondence with certain signed Young diagrams of size five (see \cite[Theorem~9.3.4]{McGovern1993NilpotentOrbits}). In fact, there are four distinct cases of real non-trivial nilpotent orbits in the semisimple Lie algebra $\so(2,3)$; see \cite{fresse2021approximationnilpotentorbitssimple} and \cite{McGovern1993NilpotentOrbits}: {Up to conjugation in $O(2,3)$, these are the minimal orbit $II^0$ (it splits into two orbits of the identity component, exchanged by the reflection $X^0\to-X^0$; see Table~\ref{tab:discrete}), the two subregular orbits $III^+_0$ and $III^-_0$, and the principal nilpotent orbit of type $V$.} 

Below, we describe each type, presenting for each a representative $K \in \so(2, 3)$, Casimir operators,  the character of the corresponding isometry, and the stabilizer/dimension of the orbit.

\section{TYPE $I$}\label{sec:I}
This case is quite rich and includes four general types: $I_a$, $I_b$, $I_c$, and $I_d$. Additionally, certain eigenvalues lead to sub-cases that exhibit enhanced global symmetries in their corresponding double copies. Some of these cases are clearly isomorphic. For instance, the following straightforward isomorphisms occur (see Table 1):

\begin{equation}
    I_c\Big|_{b_1=0}\sim I_d\Big|_{a=0}\,,\qquad I_b\Big|_{a_2=0}\sim I_b\Big|_{a_1=0}\sim I_d\Big|_{b=0}\,.
\end{equation}
Our approach in the following section is to focus on the type $I_c$, as it includes physically significant solutions such as black holes and the Carter-Pleba\'nski metric \cite{Carter:1968rr, Plebanski:1975xfb}. We will carefully examine and identify each symmetry-enhanced sub-case within $I_c$. In addition, we will analyze types $I_b$ and $I_d$ in terms of their general eigenvalues, demonstrating that they can be derived from $I_c$ through specific coordinate Wick rotations. Due to this relationship and the previously mentioned isomorphisms, we will not explore their symmetry-enhanced sub-cases in detail.

{ Types $I_b$, $I_c$ and $I_d$ are related to each other by Wick rotations, while type $I_a$ looks different at first sight: the eigenvalues of $K$ form a complex quartet $\pm(b\pm ia)$, and its double copy, depending on two real kinematical parameters and a mass parameter, does not appear in the literature in this form. We refer to it as the generalized Carter-Pleba\'nski spacetime. As shown in Sec.~\ref{sec:Ia}, it is nevertheless a member of the same family. In Carter-Pleba\'nski coordinates its metric takes the standard form fixed by the characteristic polynomial of $K$. Type $I_a$ turns out to be a real section of the complexified Kerr-AdS metric with the complex  rotation parameter $a_K$ on the circle $|a_K|=\gl^{-1}$, see \eqref{Ia:aK}. All four semisimple type $I$ orbits are then described by the single formula \eqref{I:uniform}. The solution belongs to the class identified in \cite{Klemm:2014rda} as rotating generalizations of AdS black branes, although, to our knowledge, this particular region of parameters associated with type $I_a$ has not been analyzed before.}

We will begin our discussion with the most symmetric cases before moving on to the general ones.

\subsection{Symmetry-enhanced cases} As we discussed, the isometry algebra of the double copy in question is equal to the centralizer of its generating parameters $K_{AB}$. This algebra is maximal for 
\begin{align}
    &I_c:\qquad b_1b_2=0\,,\quad b_{1}=\pm b_2\,,\\
    &I_{b}:\qquad a_1a_2=0\,,\quad a_{1}=\pm a_2\,,\\
    &I_{d}:\qquad ab=0\,.
\end{align}
Below we review some of these cases with an emphasis on type $I_c$.  

\subsection{$I_c$,  $b_2 = 0$: Schwarzschild}
According to Table 1, type $I_c$ is parameterized by the two constants $b_{1,2}$, which enter the generating parameter $K=b_1 J_{01}+b_2 J_{23}$. In the degenerate case $b_2=0$, we can set $b_1=1$ for convenience. The generators commuting with the defining Killing field 
\begin{equation}
K = J_{01}\,,    
\end{equation}
where $ KK < 0$, read  
\begin{equation}
G_1 = J_{23}\,, \quad G_2 = J_{24}\,, \quad G_3 = J_{34}\,.
\end{equation}
Their commutation relations are:  
\[
[K, G_i] = 0\,, \quad [G_1, G_2] =  -G_3\,, \quad [G_2, G_3] = -G_1\,, \quad [G_3, G_1] = -G_2\,.
\]   
Together with \( K \), these generators form the maximally compact subalgebra  
\[
\u(1) \oplus \so(3) \subset \so(2,3),
\]  
where \( \u(1) \) is an Abelian part generated by $K$ itself. The algebra \( \so(3) \) corresponds to type  $\rom{9}$ in the Bianchi classification of the three-dimensional Lie algebras; see \cite{Bergshoeff:2003ri}. 

Choosing the global AdS coordinates
\begin{subequations}\label{coord: global}
\begin{align}
    &X^0 = \gl^{-1}\sqrt{1+\gl^2r^2}\cos{\gl t}\,, \\
    &X^1 = \gl^{-1}\sqrt{1+\gl^2r^2}\sin{\gl t}\,,\\
    &X^2 = r\sin{\theta}\sin{\varphi}\,,\\
    &X^3 = r\sin{\theta}\cos{\varphi}\,,\\
    &X^4 = r\cos{\theta}
\end{align}
\end{subequations}
leads us to the following representation of the centralizer of $K$: 
\begin{subequations}
\begin{align}
    &J_{01} = -\gl^{-1} \partial_t\,,\qquad J_{23} =- \partial_{\phi}\,,\label{comm: global}\\
    &J_{34}=-\cos{\phi}\partial_{\theta} + \cot{\theta}\sin{\phi} \partial_{\phi}\,,\\
    &J_{24}=-\sin{\phi}\partial_{\theta} - \cot{\theta}\cos{\phi} \partial_{\phi}\,.
\end{align}
\end{subequations}
In a given coordinate system, the AdS metric takes the form
\begin{equation}\label{AdS:AI}
    d\s^2 = - (1+\gl^2r^2)dt^2 + \frac{d r^2}{1+\gl^2 r^2} + r^2( d\theta^2 + \sin^2{\theta}  d\varphi^2)\,, 
\end{equation}
while the vierbein \( \e^m = \e^m_\mu d x^\mu \) can be chosen to be
\begin{equation}\label{vierbein: gl}
    \e^0 = \sqrt{1+\gl^2r^2}dt\,, \quad \e^1 = \frac{dr}{\sqrt{1+\gl^2r^2}}\,, \quad \e^2 = r d\theta\,, \quad \e^3 = r\sin{\theta} d\varphi\,.
\end{equation}
From the point of view of the formalism used, the convenience of the global coordinates $(t, r, \theta, \phi)$ is justified by the fact that they straighten the commuting Killing vectors \eqref{comm: global}. This is how \eqref{comm: global} can be derived by imposing the following partial differential equations for variables $(t, r, \theta, \phi)$ as functions of the embedding coordinates $X_{\underline{A}}$:
\begin{subequations}
\begin{align}
    &\gl J_{01} t=-1\,,\qquad J_{23}\phi=-1\,,\\
    &J_{01}r=J_{01}\theta=J_{01}\phi=0\,,\\
    &J_{23}r=J_{23}\theta=J_{23}t=0\,,
\end{align}
\end{subequations}
where we recall that $J_{\underline{AB}}$ is defined in \eqref{J: def}. {It can be verified that the coordinates \eqref{coord: global} satisfy the equations above, which yields the representation \eqref{comm: global} of the centralizer.} Notice, however, that the solution \eqref{coord: global} is not unique. For example, there remains reparameterization freedom $r\to r'(r,\theta)$ and $\theta\to\theta'(r, \theta)$.

The Killing vector field $K = -\gl^{-1}\partial_t$ corresponds to the Killing vector
\begin{equation}
v^{\mu}=(-1, 0, 0, 0)    
\end{equation}
defined in the global coordinates. It provides us with the following spinorial representation for the components of the global symmetry parameter $K^{AB}$ (see Eq. \eqref{K: comp}):  
\begin{equation}
    v_{\ga \dot{\gb}} = -\sqrt{1+\gl^2r^2}
        \begin{pmatrix}
             1 &  0 \\
            0 & 1 \\ 
        \end{pmatrix},\quad  
    \varkappa_{\alpha \beta} = \gl^2r
        \begin{pmatrix}
         1 &  0 \\
        0 & -1 \\ 
        \end{pmatrix}
\end{equation}
with the Casimir invariants \eqref{Casimir:K} given by
\begin{equation}
    C_2=C_4=1\,.
\end{equation}
The parameter $K^{AB}$ generates two real Kerr-Schild vectors \eqref{KS:kn}. Their manifest form is
\begin{equation}\label{Ic(b2=0):KS}
    k_\mu^{\pm} = \e^{\ a}_{\mu}k_a^{\pm} =\left(1, \pm \frac{1}{1+ \gl^2 r^2}, 0, 0\right)\,,
\end{equation}
while the complex Kerr-Schild vectors $l^{\pm}$ from \eqref{KS:l} appear to be divergent and, consequently, do not exist.   The double and zeroth copies follow from \eqref{KS_DC:def} and \eqref{zero} 
\begin{equation}
     h_{\mu\nu} = {2}M\phif k_{\mu} k_{\nu}\,,\qquad \phif = \frac{1}{r}\,,
\end{equation}
where $k_{\mu}$ is one of the two Kerr-Schild vectors $k^{\pm}_{\mu}$ in \eqref{Ic(b2=0):KS}. 
The Weyl tensor, computed for some test spinor $\xi^{\al}$, is given by 
\begin{equation}
    C_{\al(4)} \xi^{\al(4)} = M \frac{(\xi_1^2 - \xi_2^2)^2}{r^3}\,.
\end{equation}
The considered case defines the AdS Schwarzschild black hole of mass $M$ in the Kerr-Schild form through \eqref{Kerr-Schild}. Rewriting it in the conventional form results in the so called $AI$-metric; see \cite{Podolsk__2017}, with the cosmological constant $\Lambda$, where the respective variables  (see Appendix {E}) are chosen as
\begin{equation}
\epsilon_0 = \epsilon_2 = 1\,,\quad p = r\,,\quad  q = \cos{\theta}\,,\quad M \rightarrow {-M}\,.
\end{equation}

\subsection{$I_c$, $b_1 = 0$:  $BII$-space}

There is another symmetry-enhanced case where $b_1=0$ and $b_2\neq 0$. Without loss of generality, we can set $b_2=1$. This gives us the following defining Killing vector field: 
\begin{equation}
K = J_{23}\,,    
\end{equation}
where $KK > 0$. The algebra, originating from the generators that commute with $K$, defines the global symmetry of the respective double copy. Specifically, its generators are     
\begin{equation}
G_1 = J_{01}\,,\quad G_2 = J_{04}\,,\quad G_3 = J_{14}\,, \quad [K, G_i] = 0\,,
\end{equation}
and 
\begin{equation}
    \quad [G_1, G_2] = G_3\,, \quad [G_2, G_3] = -G_1\,, \quad [G_3, G_1] = G_2\,.
\end{equation}
Together with \( K \), they form the subalgebra  
\[
\u(1) \oplus \so(1, 2) \subset \so(2,3)\,.
\]
Similarly to the previous case with $b_2=0$, it is convenient to use the special coordinates: 
\begin{subequations}\label{coord: BII}
\begin{align}
    &X^2 = \gl^{-1}\sqrt{\gl^2r^2 - 1}\cos{\gl t}\,, \\
    &X^3 = \gl^{-1}\sqrt{\gl^2r^2 - 1}\sin{\gl t}\,,\\
    &X^0 = r\cosh{\chi}\sin{\varphi}\,,\\
    &X^1 = r\cosh{\chi}\cos{\varphi}\,,\\
    &X^4 = r\sinh{\chi}
\end{align}
\end{subequations}
which allows us to straighten Killing fields of the Abelian subalgebra spanned by $K$ and $G_1$ and in which $(\varphi, r, \chi, t)$ AdS metric is realized as:
\begin{equation}\label{AdS:BII}
    d\s^2 = (\gl^2r^2 - 1)dt^2 + \frac{d r^2}{\gl^2 r^2 - 1} + r^2(d\chi^2 - \cosh{\chi}^2  d\varphi^2)\,, 
\end{equation}
The vector field $K$ corresponds to the Killing vector $v^\mu = (0, 0, 0, 1)$. Using the vierbein:
\begin{equation}\label{vierbein:BII}
    \e^0 = r\cosh{\chi} d\varphi\,, \quad \e^1 = \frac{dr}{\sqrt{\gl^2r^2-1}}\,, \quad \e^2 = r d\chi\,, \quad \e^3 = \sqrt{\gl^2r^2-1}dt\,.
\end{equation}
we obtain the following components of the global symmetry parameter: 
\begin{subequations}
\begin{align}   
    v_{\ga \dot{\gb}} = \sqrt{\gl^2 r^2 - 1}
    \begin{pmatrix}
     1 &  0 \\
    0 & -1 \\ 
    \end{pmatrix}\,,\quad
    \varkappa_{\ga \gb} = \gl^2 r \begin{pmatrix}
     1 &  0 \\
    0 &  1 \\ 
    \end{pmatrix}\,.
 \end{align}  
\end{subequations}
with the Casimir invariants 
\begin{equation}
   C_2=C_4=1\,.
\end{equation}
Using \eqref{KS:kn}, one finds that the real Kerr-Schild vectors are ill-defined, while the complex-conjugate pair \eqref{KS:l} exists
\begin{equation}\label{BII:KS}
    l_\mu^{\pm} dx^{\mu} = -dt \pm \frac{i}{-1+ \gl^2 r^2}dr\,, 
\end{equation}
Similarly, the real zeroth copy \eqref{zero} vanishes, while the purely imaginary \eqref{dual:zero} remains
\begin{equation}
    \Tilde{\phif} = \frac{1}{r} 
\end{equation}
The Weyl tensor obtained from the metric $g_{\mu \nu} = \overline{g}_{\mu \nu} + {2}M \Tilde{\phi} l_{\mu} l_{\nu} $ with real $M$ is
\begin{equation}
C_{\alpha(4)}\xi^{\alpha(4)} = M \frac{\left(\xi_1^2 + \xi_2^2\right)^2}{r^3}\,,
\end{equation}
After recasting the full metric into canonical form, one obtains the $BII$-metric of Eq.~\eqref{B-metrics}, with the parameters given by:
\begin{equation}
    \epsilon_0 = 1\,,\quad  \epsilon_2 = -1\,,\quad \varphi \leftrightarrow t\,,\quad p = r\,,\quad q=\sinh{\chi}\,.
\end{equation}
The two sub-cases with $b_1=0$ and $b_2\neq 0$ or $b_2=0$ and $b_1\neq 0$ just considered  are related through a Wick rotation with respect to the coordinates $X_{2,3} \to iX_{1,0}$. Notice that the representative $J_{23}$ is of elliptic type: a semisimple element conjugate to a compact rotation. This isometry is a spatial rotation spanned by $(X^2, X^3)$, while the generator $J_{01}$ rotates the timelike 2-plane spanned by $(X^0, X^1)$.  The corresponding spacetime is described by the $BII$-metric \cite{Podolsk__2017}; see Eq. \eqref{B-metrics}.

\subsection{$I_c$,  $b_1= \pm  b_2$: {hyperbolic Taub-NUT}}\label{b1=b2}

This is also a symmetry-enhanced case. Conveniently normalizing the Killing field, such that $1=b_1=\pm b_2$, one identifies three generators that commute with the defining generator
\begin{equation}\label{K:b1=pm b2}
  K =  J_{01} \pm J_{23}\,.   
\end{equation}
Specifically, these are 
    \begin{equation}\label{Ic_b1=pm b2:glob}
         G_1 = J_{01} \mp J_{23},\quad  G_2 = \pm J_{30} - J_{12},\quad  G_3 = \pm J_{13} - J_{02} 
    \end{equation}
with the commutation relations
    \begin{equation}
        [K, G_i] = 0,\quad  [G_1, G_2] = -2G_3,\quad [G_3, G_1] = -2G_2, \quad  [G_2, G_3] = 2G_1\,. 
    \end{equation}
Thus, in this case, $I_c$ has a total of four Killing vector fields. The three generators \( G_i \) together with \( K \) form algebra \( \u(1) \oplus \so(2, 1) \subset \so(2, 3) \), where $K$ in \eqref{K:b1=pm b2} represents a compact operator given by a sum of two rotations. 

In what follows, we set 
\begin{equation}
    b_1=b_2=1\,.
\end{equation}
The opposite case of $b_1=-b_2=1$ corresponds to a flipping of signs for certain coordinates, as we comment below. To construct a convenient AdS internal coordinate system, denoted as $(t,r,\theta, \phi)$, we choose to straighten the two commuting generators
\begin{equation}\label{Ic_b1=b2:coord}
   J_{01} t=\gl^{-1}\,,\qquad J_{23}\phi=-1\,. 
\end{equation}
Setting $X^4:=r$ and requiring the rest of $X^{\underline{A}}$ to depend on the internal coordinates in a factorized fashion, after solving Eqs. \eqref{Ic_b1=b2:coord}, we eventually find 
\begin{subequations}\label{coord: ex.Kerr} 
\begin{align}
    &X^0 = \gl^{-1} \sqrt{(1+\gl^2 r^2)(1+\theta^2)}\cos{\gl t}\,, \\
    &X^1 = \gl^{-1} \sqrt{(1+\gl^2 r^2)(1+\theta^2)}\sin{\gl t}\,, \\
    &X^2 = \gl^{-1} \theta \sqrt{1+\gl^2 r^2} \sin{\varphi}\,, \\
    &X^3 = \gl^{-1} \theta \sqrt{1+\gl^2 r^2} \cos{\varphi}\,, \\
    &X^4 = r\,.
\end{align}
\end{subequations}
Notice how the variable $\theta$ enters \eqref{coord: ex.Kerr}, compared to the AdS global coordinates given in \eqref{coord: global}. The action of the generators that form the centralizer of $K$ reads
\begin{subequations}   
\begin{align}
    &J_{01} = -\gl^{-1} \partial_t\,,\qquad J_{23} =- \partial_{\varphi}\,,\label{comm: critical_velocity}\\
    &J_{30} - J_{12} = \gl^{-1} \theta \frac{\sin{(\varphi - \gl t)}}{\sqrt{1 + \theta^2}}\partial_t + \sqrt{1 + \theta^2}\cos{(\varphi - \gl t)}\partial_{\theta} - \frac{\sqrt{1 + \theta^2}}{\theta}\sin{(\varphi - \gl t)} \partial_{\varphi}\,,\\
    &J_{13} - J_{02} = -\gl^{-1} \theta \frac{\cos{(\varphi - \gl t)}}{\sqrt{1 + \theta^2}}\partial_t + \sqrt{1 + \theta^2}\sin{(\varphi - \gl t)}\partial_{\theta} + \frac{\sqrt{1 + \theta^2}}{\theta}\cos{(\varphi - \gl t)} \partial_{\varphi}\,.
\end{align}
\end{subequations}
Notice that the Killing field has a negative norm:  $K K = -\gl^2 - (X^4)^2 < 0$.
A sign flip in the generators $K, G_1, G_2, G_3$, as specified above in \eqref{Ic_b1=pm b2:glob}, translates into the flipping in $\theta \rightarrow \pm\theta$ and $\varphi \rightarrow \pm\varphi$ in the internal coordinates. In these coordinates ($t, r, \theta, \varphi$), the AdS spacetime is expressed as
\begin{equation}\label{AdS:critical}
    d\overline{s}_{AdS}^2 = -(1+\gl^2 r^2)(1 + \theta^2)\, dt^2 + \frac{dr^2}{1+\gl^2 r^2} + \gl^{-2}(1+\gl^2 r^2) \left(\frac{d\theta^2}{1 + \theta^2} + \theta^2 \, d\varphi^2 \right) \,.
\end{equation}
In the standard diagonal frame, the vierbein \( \e^m = \e^m_\mu dx^\mu \)  is given by
\begin{subequations}\label{vierbein: gl2}
\begin{align}
    &\e^0 = \sqrt{(1+\gl^2r^2)(1 + \theta^2)}dt\,, \quad \e^1 = \frac{dr}{\sqrt{1+\gl^2r^2}}\,, \\ &\e^2 = \gl^{-1} \sqrt{\frac{1+\gl^2r^2}{1 + \theta^2}} d\theta\,, \quad \e^3 = \gl^{-1} \theta \sqrt{1+\gl^2r^2} d\varphi\,.
\end{align}
\end{subequations}
The Killing vector $v^{\mu}= (-1, 0, 0, -\gl)$ induces the following components of the global symmetry parameter:
\begin{align}
    &v_{\ga \dot{\gb}} =
    \sqrt{1+\gl^2 r^2}
    \begin{pmatrix}
        \sqrt{1+\theta^2} + \theta & 0 \\
        0 & \sqrt{1+\theta^2} - \theta \\
        \end{pmatrix}\,,\\
    &\varkappa_{\ga \gb} = 
        \gl (\gl r - i)
        \begin{pmatrix}
         \sqrt{1+\theta^2} + \theta & 0 \\
        0 & -\sqrt{1+\theta^2} + \theta \\
        \end{pmatrix}\,.\label{kappa:exKerr}
\end{align}
The Casimir invariants acquire the following values:
\begin{equation}
    C_2=2\,,\qquad C_4=8\,.
\end{equation}
To evaluate the zeroth copy, we use  \eqref{kappa:exKerr} and its complex conjugate. Substituting these into Eq. \eqref{zero}, we arrive at 
\begin{equation}
    \phif = \frac{\gl^2 r}{1 + \gl^2 r^2}\,.
\end{equation}
There exist only two real Kerr-Schild vectors that one can construct using \eqref{KS:kn}
\begin{align*}
    k_\mu^{\pm} =\left( -1 - \theta^2, \pm \frac{1}{1 + \gl^2 r^2}, 0, \gl^{-1} \theta^2 \right)\,,
\end{align*}
while the complex ones from \eqref{KS:l} appear to be divergent. The self-dual part of the Weyl tensor is expressed as:
\begin{equation}\label{Weyl:exKerr}
    C_{\al(4)}(\xi^{\al})^4 = M \frac{ \left( (\sqrt{1+\theta^2} + \theta)^2 \xi_1^2 -  (\sqrt{1+\theta^2} - \theta)^2 \xi_2^2 \right )^2}{(\gl r - i)^3}\,.
\end{equation}
The expression for the Kerr–Schild vectors corresponding to the field $K = J_{01} - J_{23}$ can be obtained by performing the redefinition $\theta \rightarrow -\theta$ and $\varphi \rightarrow -\varphi$, while the corresponding Weyl tensor  \eqref{Weyl:exKerr} requires, in addition, to replace $\gl r-i$ with $\gl r+i$. 

{Let us show that the metric $g_{\mu\nu}=\bar{g}_{\mu\nu}+2M\phif k_{\mu}k_{\nu}$ describes the hyperbolic Taub--NUT--AdS solution with the special value $|n|=\gl^{-1}/2$ of the NUT parameter. In the language of \cite{Hennigar:2015wxa}, it is the four-dimensional rotating black hyperboloid membrane. Indeed, by rescaling $t\to t/2$ and $r\to2r$, together with $\theta=\sinh(\sigma/2)$ and $\varphi\to\varphi+\gl t/2$, the AdS metric \eqref{AdS:critical} takes the form
\begin{equation}\label{crit:NUT}
    ds^2 = -F(r)\big(dt + 2n(\cosh\sigma - 1)d\varphi\big)^2 + \frac{dr^2}{F(r)} + (r^2+n^2)\big(d\sigma^2 + \sinh^2\sigma\, d\varphi^2\big)\,,
\end{equation}
with $n=\mp\gl^{-1}/2$ and
\begin{equation}
    F(r) = \gl^2(r^2+n^2) - \frac{2mr}{r^2+n^2}\,.
\end{equation}
For $m=0$, \eqref{crit:NUT} reduces to the AdS metric \eqref{AdS:critical}. For $n\neq0$ the massless hyperbolic Taub--NUT--AdS metric is locally AdS only when $n^2=\gl^{-2}/4$, which is why the NUT parameter is fixed. The Kerr--Schild deformation, after removing the $dr$ cross terms, gives \eqref{crit:NUT} with $m = M/8$ in the present normalization. This agrees with the Weyl tensor \eqref{Weyl:exKerr}, since after the rescaling $(\gl r - i)^3\to(2\gl r-i)^3 = 8\gl^3(r - i\gl^{-1}/2)^3$ exhibits the characteristic NUT dependence on $r\pm in$. The isometry algebra of \eqref{crit:NUT}, consisting of $\partial_t$ and the $\so(2,1)$ of the hyperbolic plane, is the $\u(1)\oplus\so(2,1)$ found above. This solution can be obtained from the Kerr--AdS solution in the critical limit $\gl a\to\pm1$ combined with zooming near the axis, $\theta\to\theta\sqrt{1-\gl^2a^2}$. This is the hyperboloid-membrane ultraspinning limit of \cite{Hennigar:2015wxa}, which will be shown explicitly in Section~\ref{sec: Kerr}.}

\subsection{$I_b$, $a_1 = 0$: hyperbolic black hole}\label{static-hyp-bh}
This case, which also arises from  $\rom{1}_d$ with $b = 0$, is generated by a Killing field that we choose in the form
\begin{equation}
  K = J_{03}\,.  
\end{equation}
It represents a hyperbolic type: a semisimple element conjugate to a non-compact boost. It has real eigenvalues and generates a Lorentz boost in the plane $(X^0, X^3)$. The generators commuting with the Killing field \( K \) are given by
\begin{equation}
    G_1 = J_{24}\,,\quad G_2 = J_{14}\,,\quad G_3 = J_{12}\,.
\end{equation}
Their commutation relations are: 
\begin{equation}
    [K, G_i] = 0\,, \quad [G_1, G_2] = G_3\,, \quad [G_2, G_3] = -G_1\,, \quad [G_3, G_1] = G_2\,.
\end{equation}
Together with \( K \), they form the subalgebra  
\[
\so(1, 1) \oplus \so(1, 2) \subset \so(2,3)\,.
\]
The algebra \( \so(1, 2) \) corresponds to type $\rom{8}$ in the Bianchi classification of three-dimensional Lie algebras.
It is convenient to use the hyperbolic coordinates: 
\begin{subequations}   
\begin{align}
    &X^0 = \gl^{-1}\sqrt{\gl^2r^2 - 1}\sinh{\gl t}\,, \\
    &X^3 = \gl^{-1}\sqrt{\gl^2r^2 - 1}\cosh{\gl t}\,, \\
    &X^2 = r\sinh{\chi}\sin{\varphi}\,,\\
    &X^4 = r\sinh{\chi}\cos{\varphi}\,, \\
    &X^1 = r\cosh{\chi}\,.
\end{align}
\end{subequations}
In these coordinates, the AdS metric is expressed as
\begin{equation}\label{AdS:AII}
    d\s^2 = - (\gl^2r^2 - 1)dt^2 + \frac{d r^2}{\gl^2 r^2 - 1} + r^2( d\chi^2 + \sinh{\chi}^2  d\varphi^2)\,, 
\end{equation}
while the global symmetry algebra generators become
\begin{subequations}
\begin{align}
    &J_{03} = -\gl^{-1}\partial_t\,,\qquad J_{24} = -\partial_{\varphi}\,,\\
    &J_{14}=-\cos{\phi}\partial_{\chi} + \coth{\chi}\sin{\phi} \partial_{\phi}\,,\\
    &J_{12}=-\sin{\phi}\partial_{\chi} -\coth{\chi}\cos{\phi} \partial_{\phi}\,.
\end{align}
\end{subequations}
Notice that the Killing vector $\partial_t$ is non-compact, unlike in the case of the Schwarzschild solution represented using global coordinates. It corresponds to a boost (hyperbolic rotation) acting in the  $(X^0, X^3)$ plane. The remaining isometries act on slices of hyperbolic space $\mathbb{H}^2$, parameterized by the coordinates $X^1$, $X^2$, and $X^4$. They generate the isometry group $SO(1,2)$. Specifically, $\partial_\phi$ generates rotations in the $(X^2, X^4)$ plane, while the remaining transformations generate boosts along the direction $\chi$. Globally, these coordinates are defined only within the so-called \emph{Rindler wedge}, which in our notation corresponds to the domain $\lambda^2 r^2 > 1$.

Consider the following choice of vierbein \( \e^m = \e^m_\mu dx^\mu \):  
\begin{equation}
    \e^0 = \sqrt{\gl^2r^2 - 1}dt\,, \quad \e^1 = \frac{dr}{\sqrt{\gl^2r^2-1}}\,, \quad \e^2 = r d\chi, \quad \e^3 = r\sinh{\chi} d\varphi\,.
\end{equation}
The Killing vector field $K$ provides the following spinorial representation for the components of the global symmetry parameter:  
\begin{equation}
    v_{\ga \dot{\gb}} = -\sqrt{\gl^2r^2 - 1}
        \begin{pmatrix}
             1 &  0 \\
            0 & 1 \\ 
        \end{pmatrix}\,,\quad  
    \varkappa_{\alpha \beta} = \gl^2r
        \begin{pmatrix}
         1 &  0 \\
        0 & -1 \\ 
        \end{pmatrix}\,.
\end{equation}
corresponding to the following Casimir invariants:
\begin{equation}
    C_2 = -1\,,\quad C_4 = 1\,.
\end{equation}
The parameter $K$ generates two real Kerr-Schild vectors 
\begin{equation}\label{Ia:KS}
    k_\mu^{\pm} = \e^{\ a}_{\mu}k_a^{\pm} =\left(1, \pm \frac{1}{\gl^2 r^2 - 1}, 0, 0\right)\,,
\end{equation}
while the complex $l^{\pm}$ are ill-defined in this case. The corresponding double and zeroth copies are:
\begin{equation}
     h_{\mu\nu} = {2}M\phif k_{\mu} k_{\nu}\,,\quad \phif = \frac{1}{r}\,.
\end{equation}
Its Weyl tensor computed for some test spinor $\xi^{\al}$ is 
\begin{equation}
    C_{\al(4)} \xi^{\al(4)} = M \frac{(\xi_1^2 - \xi_2^2)^2}{r^3}\,.
\end{equation}
This metric corresponds to a black hole with a hyperbolic horizon of mass $M$ in the Kerr-Schild form. Rewriting it in the conventional form; see Appendix~{E}, one arrives at the so called $AII$-metric with a cosmological constant $\Lambda$, where the parameters are identified as $AII$-metric; see Eq. \eqref{A-metrics} 
\begin{equation}
\epsilon_0 = \epsilon_2 = -1\,,\qquad p = r\,,\qquad q = \cosh{\chi}\,,\qquad M \rightarrow {-M}\,.
\end{equation}

\paragraph{$BI$-metric} The type of the metric, previously  identified as $AII$, in fact depends on the patch of the background AdS spacetime used in constructing the double copy.  We demonstrate here that the same type $I_b$ with $a_1=0$ generated by $J_{03}$ admits another, inequivalent realization corresponding to the \(BI\)-metric. 

To this end, we begin with the global AdS metric \eqref{AdS:AI} and make the following formal substitution: $t \to it$, $\theta \to \frac{\pi}{2} + i \chi$, which entails the following changes in metric:
\begin{equation}\label{AdS:BI}
    d\s^2 = (1+\gl^2r^2)dt^2 + \frac{d r^2}{1+\gl^2 r^2} + r^2( -d\chi^2 + \cosh^2{\chi}  d\varphi^2)\,.
\end{equation}
In terms of the ambient coordinates, this form can be reached by redefining \eqref{coord: global} 
\begin{equation}
    X^1 \to iX^4\,,\qquad X^3 \to - i X^1\,,\qquad X^4 \to X^3
\end{equation}
that induces the following embedding:
\begin{subequations}\label{coord: BI}
\begin{align}
    &X^0 = \gl^{-1}\sqrt{1+\gl^2r^2}\cosh{\gl t}\,, \\
    &X^3 = \gl^{-1}\sqrt{1+\gl^2r^2}\sinh{\gl t}\,,\\
    &X^2 = r\cosh{\chi}\sin{\varphi}\,,\\
    &X^4 = r\cosh{\chi}\cos{\varphi}\,,\\
    &X^1 = r\sinh{\chi}\,.
\end{align}
\end{subequations}
Next, we fix the vierbein \( \e^m = \e^m_\mu dx^\mu \) in the form:  
\begin{equation}
    \e^0 = r d\chi\,, \quad \e^1 = \frac{dr}{\sqrt{1 + \gl^2r^2}}\,, \quad 
    \e^2 = \sqrt{1 + \gl^2r^2}dt , \quad 
    \e^3 = r\cosh{\chi}d\varphi\,.
\end{equation}
In the coordinate chart $(\chi, r, t, \phi)$, the Killing generator under consideration is represented by the Killing vector $v^{\mu} = (0, 0, 1, 0)$. The corresponding global symmetry parameter then takes the following spinorial form:
\begin{equation}
    v_{\ga \dot{\gb}} = i \sqrt{1 + \gl^2r^2}
        \begin{pmatrix}
             0 &  -1 \\
            1 & 0 \\ 
        \end{pmatrix}\,,\quad  
    \varkappa_{\alpha \beta} = i\gl^2r
        \begin{pmatrix}
         0 &  1 \\
        1 & 0 \\ 
        \end{pmatrix}
\end{equation}
with the expected Casimir invariants
\begin{equation}
    C_2 = -1,\qquad C_4 = 1\,.
\end{equation}
The important difference between the  $BI$-patch in  \eqref{AdS:BI} and the  $AII$-metric presented in  \eqref{AdS:AII} is that the $BI$  patch yields a complex pair of Kerr-Schild vectors through \eqref{KS:spin} instead of the real vectors in \eqref{Ia:KS}   
\begin{equation}\label{BI:KS}
    l_\mu^{\pm} dx^{\mu} = -dt \pm \frac{i}{1+ \gl^2 r^2}dr\,. 
\end{equation}
The zeroth copy is also purely imaginary
\begin{equation}
    \Tilde{\phif} = \frac{1}{r} \,. 
\end{equation}
The Weyl tensor obtained from the metric $g_{\mu \nu} = \overline{g}_{\mu \nu} + {2}M \Tilde{\phif} l_{\mu} l_{\nu} $ with real $M$ is
\begin{equation}
C_{\alpha(4)}\xi^{\alpha(4)} = M \frac{\left(\xi_1^2 + \xi_2^2\right)^2}{r^3}\,,
\end{equation}
Expressing the full metric in canonical form, we recover the $BI$-metric, cf. Eq.~\eqref{B-metrics}, corresponding to the parameter values:
\begin{equation}
    \epsilon_0 = -1\,,\quad  \epsilon_2 = 1\,,\quad \varphi \leftrightarrow t\,,\quad p = r\,,\quad q=\sinh{\chi}\,.
\end{equation}
which is direct counterpart of the Schwarzschild solution \cite{Podolsk__2017}.
There are quite a few works discussing the physical interpretation and the relation between $A$ and $B$ metrics. For example, both the A\rom{2} and B\rom{1} metrics form a complete gravitational field of the tachyon moving in superluminal speed in AdS spacetime with the special boundary called Mach-Cherenkov shockwave $\cite{Hruska:2018djo}$.

\subsection{General cases}
The minimal isometry algebra of type $I$ is $\u(1)\oplus\u(1)$ that occurs for general eigenvalues. Below, we specify these cases. 

\subsection{$I_c:$ Kerr metric}\label{sec: Kerr}
In the general case of type $I_c$, where $a$ and  $b$ are nonzero constants from Table 1, we can normalize one of the two parameters in the Killing field $K$ to unity. The other parameter will remain arbitrary, which we denote as $a$ . As we will explain further below, this remaining parameter $a$  represents the rotation in the AdS-Kerr geometry of the resulting double copy.
Specifically, we have 
\begin{equation}\label{K:Kerr}
    K = J_{10} + \gl a J_{23}\,,\quad G = J_{10} - \gl a J_{23}\,, \quad [K, G] = 0\,.
\end{equation}
where if $|\gl a| < 1$, then the norm of the Killing field is negative,  $KK < 0$.
The two generators $K$ and $G$ form an Abelian subalgebra, which is the Cartan subalgebra of $\so(2, 3)$:
\[
\u(1) \oplus \u(1) \subset \so(2,3)\,.
\]
To exhibit the commutative structure explicitly, we construct such coordinates $(t, r, \theta, \varphi)$ in which $K$ and its Abelian centralizer $G$ are straightened 
\begin{align}
    J_{01} = -\gl^{-1} \partial_t\,,\qquad J_{23} = - \partial_{\varphi}\,.
\end{align}  
In addition, we would like to enforce  the factorized dependence of embedding $X^{\underline{A}}$ on the AdS internal coordinates. Since there is no manifest $X^4$ in the definitions \eqref{K:Kerr}, we keep $X^4:=r\cos{\theta}$ as it happens with the global coordinates \eqref{coord: global}. The solution one finds in this way is captured by the Boyer-Lindquist spheroidal coordinates    
\begin{subequations}\label{coord:BL}  
\begin{align}
    &X^0 = \gl^{-1} \sqrt{\frac{(1+\gl^2 r^2)(1-\gl^2 a^2\cos^2{\theta})}{1-\gl^2 a^2}}\cos{\gl t}\,,\\
    &X^1 = \gl^{-1} \sqrt{\frac{(1+\gl^2 r^2)(1-\gl^2 a^2\cos^2{\theta})}{1-\gl^2 a^2}}\sin{\gl t}\,, \\
    &X^2 =\sqrt{\frac{a^2+r^2}{1-\gl^2 a^2}}\sin{\theta}\sin{\varphi}\,, \\
    &X^3 = \sqrt{\frac{a^2+r^2}{1-\gl^2 a^2}}\sin{\theta}\cos{\varphi}\,,\\
    &X^4 = r \cos{\theta}\,,
\end{align} 
\end{subequations}
where the parameter $a$ in \eqref{K:Kerr} could be arbitrary. Then the AdS line element takes the following form:
\begin{equation}\label{AdS:BL}
    d\overline{s}_{AdS}^2 = -\frac{(1+\gl^2 r^2)\Delta_{\theta}\, dt^2}{1-\gl^2 a^2} + \frac{\rho^2\, dr^2}{(1+\gl^2 r^2)(r^2+a^2)} + \frac{\rho^2\, d\theta^2}{\Delta_{\theta}} + \frac{(r^2 + a^2)\sin^2{\theta}\, d\varphi^2}{1-\gl^2 a^2}\,,
\end{equation}
where 
\begin{equation}
    \rho^2 := r^2 + a^2\cos^2{\theta}\, ,\qquad
    \Delta_{\theta} := 1-\gl^2 a^2\cos^2{\theta}\, .
\end{equation}
We now choose the following vierbein $\e^m = \e^m_{\mu} dx^{\mu}$:
\begin{align}
    &\e^0 = \sqrt{\frac{1+\gl^2 r^2}{1-\gl^2 a^2}}\sqrt{\Delta_{\theta}} dt\, , \qquad     
    \e^1 = \frac{\rho}{\sqrt{(r^2+a^2)(1+\gl^2 r^2)}} dr\,, \\
    &\e^2 = \frac{\rho}{\sqrt{\Delta_{\theta}}}d\theta\, ,
    \qquad    \e^3 = \sqrt{\frac{r^2 + a^2}{1-\gl^2 a^2}}\sin{\theta} d\varphi\, .
\end{align}
 The Killing vector in the chosen spacetime coordinates $v^{\mu}= (-1, 0, 0, -\gl^2 a)$ generates the global symmetry parameter $K_{AB}$ with the following Lorentz components:
\begin{align}
    v_{\ga \dot{\gb}}  &=  
    \frac{1}{\sqrt{1-\gl^2 a^2}}
    \begin{pmatrix}
        \sqrt{1+\gl^2 r^2}\sqrt{\Delta_{\theta}} + \gl^2 a\sqrt{r^2+a^2}\sin{\theta}&  0 \\
        0 & \sqrt{1+\gl^2 r^2}\sqrt{\Delta_{\theta}} - \gl^2 a\sqrt{r^2+a^2}\sin{\theta} \\ 
    \end{pmatrix}\label{v:Kerr}\\  
    \varkappa_{\ga \gb} &= 
    \frac{\gl^2 (r-ia\cos{\theta})}{\rho \sqrt{1-\gl^2 a^2}}
    \begin{pmatrix}    \sqrt{\Delta_{\theta}}\sqrt{a^2+r^2} + a\sin{\theta}\sqrt{1+\gl^2r^2} &  0 \\
        0 &  -\sqrt{\Delta_{\theta}} \sqrt{a^2+r^2} + a\sin{\theta}\sqrt{1+\gl^2 r^2} 
\end{pmatrix}\,.\label{gk:Kerr}
\end{align}
From here, we find that
\begin{equation}
     \sqrt{-\varkappa^2}:=\sqrt{-\det\gk_{\al\gb}} = \gl^2 (r - ia\cos{\theta})
\end{equation}
{where the notation $\varkappa^2$ is defined in \eqref{B:notation},}
and so for the zeroth copy, according to \eqref{zero}, we arrive at
\begin{equation}
    \phif  =  \frac{r}{r^2 + a^2\cos^2{\theta}}\,.
\end{equation}
The Casimir invariants are  
\begin{equation}
    C_2 = 1 + \gl^2 a^2, \quad 
    C_4 = 1 + 6 \gl^2 a^2 + \gl^4 a^4\,.
\end{equation}
Using \eqref{v:Kerr} and \eqref{gk:Kerr}, we extract the Kerr-Schild vectors from \eqref{KS:spin}, resulting in a final outcome that matches the known one from the literature, see {\it e.g.,}  \cite{Gibbons:2004uw}, up to the definition of the cosmological constant $\gl^2$
\begin{equation}\label{Kerr}
    k^{\pm}_{\mu}dx^{\mu} = \frac{1-\gl^2 a^2\cos^2{\theta}}{1-\gl^2a^2}dt \pm \frac{r^2+a^2\cos^2{\theta}}{(1+\gl^2 r^2)(a^2 + r^2)}dr - \frac{a\sin^2{\theta}}{1-\gl^2 a^2}d\varphi\,, 
\end{equation}
\begin{equation}
    l^{\pm}_{\mu}dx^{\mu} = \frac{1+\gl^2 r^2}{1-\gl^2a^2}dt \pm \frac{i}{a \sin{\theta}} \frac{r^2+a^2\cos^2{\theta}}{(1-\gl^2 a^2\cos^2{\theta})}d\theta - \frac{a^2+r^2}{a(1-\gl^2 a^2)}d\varphi\,. 
\end{equation}
The conditions on the Kerr-Schild vectors \eqref{KS:geo} are satisfied, while the space described by 
\begin{equation}\label{Kerr:metric}
g_{\mu \nu} = \bar{g}_{\mu \nu} + 2M\phif k_\mu k_\nu\,,    
\end{equation}
where $M$ is the mass of a black hole, satisfies the Einstein equations. Recall that the field $\phi$ is a conformal scalar on AdS \eqref{0:DC}. This is not the case in the Kerr geometry, as one can find that \cite{Didenko:2008va}
\begin{equation}
    \boldsymbol\nabla^2 \phif + 2\gl^2\phif + 2M \boldsymbol\nabla_{\gamma}(\phif^{\beta \gamma} \boldsymbol\nabla_{\beta} \phi) = 0\,,
\end{equation}
where $\phif^{\beta \gamma} = \phif k^{\beta}k^{ \gamma}$ and the covariant derivatives are defined with respect to the Kerr metric \eqref{Kerr:metric}. The self-dual part of the Weyl tensor is 
\begin{equation}\label{Weyl:BL}
    C_{\alpha(4)}(\xi^{\alpha})^4 = M \frac{\left( a\sin{\theta}\sqrt{1+\gl^2 r^2}(\xi_1^2 + \xi_2^2) + \sqrt{(a^2 + r^2)\Delta_{\theta}}(\xi_1^2 - \xi_2^2)\right )^2}{\rho^2 (r-ia\cos{\theta})^3 (1-\gl^2 a^2)}\,.
\end{equation}

{\subsubsection{Three inequivalent limits $\gl a\to\pm1$.}\label{Sec:3lims} 

It is important to note that the limit $\gl a=\pm 1$ is ill-defined in the background metric given by \eqref{AdS:BL}. However, the Weyl tensor behaves well under three types of rescalings discussed below. No pathology in the definition of $K$ in \eqref{K:Kerr} arises too. One of those limits corresponds to a black hole rotating at a critical velocity; see \cite{Hawking:1998kw}, \cite{Berman:1999mh}, which is related to the so-called super-entropic black hole of \cite{Hennigar:2014cfa} and which reproduces solution generated by orbit $II_b$. Another limit results naturally in the earlier considered symmetry-enhanced case $I_c$ with $b_1=\pm b_2$ of the Taub-NUT solution. Notice, this limit does not change the type of the orbit and thus offers a nontrivial isomorphism of the critical Kerr and the Taub-NUT-AdS with a fine-tuned NUT-parameter. Notably, such a limit exhibits four isometries instead of the usual two found in typical rotating states. Finally, the third limit involves the rescaling of the mass parameter, resulting in the Petrov type N solution described by the particular Siklos space-time. We are not aware of this limit in the literature. In all these cases, the Boyer-Lindquist coordinates do not adequately capture the limit, as they exhibit a coordinate singularity.

Consider these limits in detail. Let 
\begin{equation}
 \Xi:=1-\gl^2a^2   \,.
\end{equation}
As $\Xi\to0$, the components $g_{tt}$, $g_{t\varphi}$, $g_{\varphi\varphi}$ of both the background \eqref{AdS:BL} and the Kerr--Schild term in \eqref{Kerr:metric} diverge, as $\Xi^{-1}$ and $M\Xi^{-2}$ respectively. A limit is therefore specified by a scaling of $M$ together with a rescaling of coordinates. There are three inequivalent choices:
\begin{itemize}
    \item[A.] $M\to M\Xi$, with $(t,\varphi)\to\sqrt\Xi\,(t,\varphi)$ and $(r,\theta)$ fixed.
    \item[B.] $M$ fixed, with $\varphi=\Xi\psi+\gl^2a\,t$ and $(t,r,\theta)$ fixed.
    \item[C.] $M$ fixed, with $\sin\theta=\sqrt\Xi\,\sinh(\sigma/2)$ and $(t,r,\varphi)$ fixed.
\end{itemize}
See Table~\ref{tab:limits} for a summary.
\begin{table}[h]
\centering

\renewcommand{\arraystretch}{1.3}
\begin{tabular}{@{}lllll@{}}
\toprule
Limit & Mass & Coordinates & Petrov type & Space-time\\
\midrule
A & $M\to M\Xi$ & $t,\varphi\to\sqrt\Xi\,(t,\varphi)$ & N & Siklos AdS wave, 2 Killing vectors\\
B & fixed & $\varphi=\Xi\psi+\gl^2a\,t$ & D & super-entropic black hole \cite{Hennigar:2014cfa}\\
C & fixed & $\sin\theta=\sqrt\Xi\sinh(\sigma/2)$ & D & hyperbolic Taub--NUT--AdS, $n=\mp\gl^{-1}/2$ \cite{Hennigar:2015wxa}\\
\bottomrule
\end{tabular}
\caption{The three inequivalent limits $\gl a\to\pm1$ of the Kerr--AdS double copy \eqref{Kerr:metric}, with $\Xi=1-\gl^2a^2$.}\label{tab:limits}
\end{table}
The rescaling of $(t,\varphi)$ in (A) only serves to give the metric finite components. It does not affect the frame components \eqref{Weyl:BL}.

{\bf Limit A} is the one suggested by the $1/\Xi$ growth of \eqref{Weyl:BL}. This growth is, however, a property of the static vierbein, which becomes infinitely null-rotated with respect to the principal null frame. At $\gl a=\pm1$ the two coefficients in the numerator of \eqref{Weyl:BL} coincide, so that \eqref{Weyl:BL} reduces to
\begin{equation}
C_{\al(4)}\xi^{\al(4)}=\frac{4M(1+\gl^2r^2)\sin^2\theta}{\gl^2\rho^2(r\mp i\gl^{-1}\cos\theta)^3}\,\xi_1^4\,.
\end{equation}
This is Petrov type N: the Weyl tensor does not vanish, although all its invariants do. In the coordinates
\begin{equation}
u=\gl^{-1}\varphi-t\,,\qquad v=\tfrac12(t+\gl^{-1}\varphi)\,,\qquad z=\frac{1}{\gl\sqrt{1+\gl^2r^2}\,\sin\theta}\,,\qquad x=\frac{r\cos\theta}{\sqrt{1+\gl^2r^2}\,\sin\theta}\,,
\end{equation}
the limiting metric is a Siklos wave \cite{Siklos},
\begin{equation}\label{limA:Siklos}
ds^2=\frac{1}{\gl^2z^2}\big(2du\,dv+dx^2+dz^2+H\,du^2\big)\,,\qquad H=\frac{2Mr\sin^2\theta}{(1+\gl^2r^2)(r^2+\gl^{-2}\cos^2\theta)}\,,
\end{equation}
with a $u$-independent profile. A direct analysis of the Killing equations shows that \eqref{limA:Siklos} has only two isometries, $\partial_u$ and $\partial_v$. It is therefore inequivalent both to two regularizations of type $II_{b=0}=
II_{a=0}$ Siklos solutions \eqref{II0:H} and \eqref{IIa:Siklos} to be discussed later, which have four isometries, and to the Kaigorodov space-time \cite{Kaigorodov}, which has five. The metric \eqref{limA:Siklos} is thus generated by no orbits classified in Table 1.  

Physically, after $m\to m\Xi$ the charges $E=m\Xi^{-2}$ and $J=am\Xi^{-2}$ \cite{Gibbons:2004ai} diverge as $\Xi^{-1}$, while $E-\gl J=m/(1+\gl a)$ stays finite. It is the light-speed-rotation analogue of the infinite-momentum-frame limit producing the Kaigorodov space-time \cite{Cvetic:1998jf}, and of the shock-wave limits \cite{PodolskyGriffiths:1997,Horowitz:1999gf}. 

{\bf Limit B} is the ultraspinning limit of \cite{Hennigar:2014cfa}. The divergent $dt^2$ terms cancel in \eqref{AdS:BL} after introducing the coordinate $\psi$ (see Table 4), and one obtains the type D super-entropic black hole with mass parameter $M$ and a non-compact horizon of finite area once $\psi$ is compactified. As we will see later, this solution is reproduced by type $II_b$.

{\bf Limit C} is the zoom
\begin{equation}
    \theta\to\theta\sqrt{1-\gl^2a^2}
\end{equation}
described above for type $I_c$ with $b_1=\pm b_2$, which is the hyperboloid-membrane limit of \cite{Hennigar:2015wxa}. With $M$ unchanged, $\Delta_\theta\to\Xi(1+\theta^2)$ and the factor $\Xi$ cancels between the numerator and the denominator of \eqref{Weyl:BL}, which tends to \eqref{Weyl:exKerr}. The result is the hyperbolic Taub--NUT--AdS solution \eqref{crit:NUT} with $n=\mp\gl^{-1}/2$ and $m=M/8$ in the normalization \eqref{Kerr:metric}, i.e., the solution of Section~\ref{b1=b2}. }

\subsubsection{The Carter-Pleba\'{n}ski form} The Kerr solution is captured by the more general class known due to Carter \cite{Carter:1968rr} and Pleba\'nski, who rediscovered this metric independently in  \cite{Plebanski:1975xfb}. So, it is useful to have a map between the Boyer-Lindquist coordinates and the Carter ones. For AdS, the map is established via the following change of variables:
\begin{align}
    \tau = \frac{t - a\varphi}{1 - \gl^2 a^2},\quad \psi = \frac{\varphi / a - \gl^2 t}{1 - \gl^2 a^2},\quad y = a \cos{\theta}.
\end{align}
Under this transformation, the AdS metric takes the Carter form:
\begin{multline}    
    d\overline{s}_{AdS}^2 = -\frac{\Delta_r}{r^2+y^2}(d\tau + y^2 d\psi)^2  +\frac{\Delta_y}{r^2+y^2}(d\tau - r^2 d\psi)^2 + \frac{(r^2 + y^2)dr^2}{\Delta_r} + \frac{(r^2 + y^2)dy^2}{\Delta_y}\,,
\end{multline}
with the structure functions given by 
\begin{equation}
\Delta_r = (1+\gl^2 r^2)(r^2 + a^2)\,,\quad  \Delta_y = (1-\gl^2 y^2)(a^2 - y^2)\,.     
\end{equation}
These functions match those appearing in the Carter-Pleba\'nski metric for the parameter choice
\begin{equation}
\gga = a^2\,,\qquad \epsilon = 1 + \gl^2 a^2\,.    
\end{equation}
So, that 
\begin{equation}
    \Delta_r = r^2(\gl^2r^2 + \epsilon) + \gga\,,\qquad 
    \Delta_y = y^2(\gl^2y^2 - \epsilon) + \gga\,.
\end{equation}


\subsection{$I_c:$ Carter-Pleba\'{n}ski spacetime}\label{CP}
The Kerr case can be reconsidered to capture the various previously discussed symmetry-enhanced cases. To achieve this, we must keep both parameters  $b_1$ and $b_2$ arbitrary in our $I_c$  case. In general, there are only two independent, commuting Killing vector fields: the generating 
\begin{equation}
  K = b_1 J_{01} + b_2 J_{23}\,,  
\end{equation}
and its centralizer
\begin{equation}
    G = b_2 J_{01} + b_1 J_{23}\,, \qquad [K, G] = 0\,.    
\end{equation}
We are now no longer confined to $b_1 \cdot b_2 \neq 0$.
The appropriate AdS coordinates  ($\tau, \psi, r, y$) are chosen in such a way as to bring the system into the so-called Carter-Pleba\'{n}ski form with the kinematical parameters $a$ and $\epsilon$; see \cite{Didenko:2009tc}
\begin{subequations}\label{CPembs}    
\begin{align}
    &X^0 = \sqrt{\frac{\kappa^2(a^2 + \gl^2 r^2 y^2) + \kappa(y^2 -r^2 - \gl^{-2} \epsilon )}{\gl^2 a^2 \kappa^2 - 1}} \cos\left( \frac{\gl}{\sqrt{\kappa}} \tau + \gl a^2 \sqrt{\kappa} \psi \right)\,, \\
    &X^1 = \sqrt{\frac{\kappa^2(a^2 + \gl^2 r^2 y^2) + \kappa(y^2 -r^2 - \gl^{-2} \epsilon )}{\gl^2 a^2 \kappa^2 - 1}}  \sin\left( \frac{\gl}{\sqrt{\kappa}} \tau + \gl a^2 \sqrt{\kappa} \psi \right)\,, \\
    &X^2 = \sqrt{\frac{a^2 + \gl^2 r^2 y^2 + \gl^2 a^2\kappa(y^2 -r^2 - \gl^{-2} \epsilon)}{\gl^{4} a^4 \kappa^2 - \gl^{2} a^2}} \cos\left(\gl^{2} a \sqrt{\kappa} \tau + \frac{a}{\sqrt{\kappa}} \psi \right)\,, \\
    &X^3 = \sqrt{\frac{a^2 + \gl^2 r^2 y^2 + \gl^2 a^2\kappa(y^2 -r^2 - \gl^{-2} \epsilon)}{\gl^{4} a^4 \kappa^2 - \gl^{2} a^2}} \sin\left(\gl^{2} a \sqrt{\kappa} \tau + \frac{a}{\sqrt{\kappa}} \psi \right)\,, \\
    &X^4 = \frac{ry}{a}\,,
\end{align}
\end{subequations}
where 
\begin{equation}\label{I_c_gen:b12}
    b_1 = \frac{1}{\sqrt{\kappa}}\,,\qquad b_2 = - \gl a \sqrt{\kappa}\,,
\end{equation}
and the parameter $\kappa$ satisfies
\begin{equation}\label{Ic_gen:kappa_eq}
   \gl^2 a^2 \kappa^2 - \epsilon \kappa + 1 = 0\,. 
\end{equation}
The symmetry generators take the following form in the internal coordinates:
\begin{equation}
    K=-\gl^{-1} \partial_\tau = \frac{1}{\sqrt{\kappa}}J_{01} - \gl a \sqrt{\kappa} J_{23}\,,\qquad G=-a^{-1} \partial_\psi = \gl a\sqrt{\kappa}J_{01} - \frac{1}{\sqrt{\kappa}} J_{23}\,. 
\end{equation}
One finds the eigenvalues of $K$ to be purely imaginary, as expected 
\begin{equation}
\lambda_{1, 2} = \pm  \frac{i}{\sqrt{\kappa}}\,,\qquad \gl_{3,4}=\pm i \gl a \sqrt{\kappa}\,.    
\end{equation}
The corresponding Casimir invariants  \eqref{Casimirs} of $\so(2, 3)$ take the following form:
\begin{equation}\label{Ic:I12}
    I_1 = b_1^2 + b_2^2 = \epsilon>0\,,\qquad I_2 =  -b_1^4 - b_2^4  = 2\gl^2a^2 - \epsilon^2<0\,.
\end{equation}
The AdS spacetime in the Carter-Pleba\'{n}ski form reads
\begin{equation}    d\overline{s}_{AdS}^2 = -\frac{\Delta_r}{r^2+y^2}(d\tau + y^2 d\psi)^2 + \frac{\Delta_y}{r^2+y^2}(d\tau - r^2 d\psi)^2 + \frac{(r^2 + y^2)dr^2}{\Delta_r} + \frac{(r^2 + y^2)dy^2}{\Delta_y}\,,
\end{equation}
where
\begin{equation}
\Delta_r = r^2(\gl^2 r^2 + \epsilon) + a^2,\qquad 
\Delta_y = y^2(\gl^2 y^2 - \epsilon) + a^2\,.
\end{equation}
The associated with $K$ space-time Killing vector $v^{\mu} = (-1, 0, 0, 0)$ delivers the corresponding global symmetry parameter
\begin{equation}
v_{\ga \dot{\gb}} = -\frac{1}{\sqrt{r^2 + y^2}}
\begin{pmatrix}
    \sqrt{\Delta_r}&  \sqrt{\Delta_y} \\
    \sqrt{\Delta_y} & \sqrt{\Delta_r}
\end{pmatrix},\quad
\varkappa_{\ga \gb} =    
\gl^2
\begin{pmatrix}
    y-ir &  0 \\
    0 &   y-ir
\end{pmatrix}.
\end{equation}
with 
\begin{equation}
    C_2 = \epsilon\,, \qquad C_4 = \epsilon^2 + 4\gl^2 a^2\,.
\end{equation}
The coordinate choice generates, through \eqref{KS:spin}, the following Kerr-Schild fields: 
\begin{equation}
    k^{\pm}_{\mu}dx^{\mu}=d\tau + y^2 d\psi \pm \frac{r^2+y^2}{\gl^2 r^4 +\epsilon r^2 + a^2}dr\,,
\end{equation}
\begin{equation}
    l_{\mu}^{\pm}dx^{\mu}=d\tau - r^2 d\psi \pm i \frac{r^2+y^2}{\gl^2 y^4 -\epsilon y^2 + a^2}dy
\end{equation}
with the zeroth copy and curvature of the form
\begin{equation}
    \phif = \frac{r}{r^2 + y^2}\,,\qquad C_{\alpha(4)}\xi^{\alpha(4)} = M \frac{(\xi_1^2 + \xi_2^2)^2}{(r + iy)^3}\,.
\end{equation}
The corresponding spacetime belongs to the class of Carter and Pleba\'nski \cite{Carter:1968rr, Plebanski:1975xfb}.

The degenerate cases, where the global symmetry of the double copies enhances, arise for $b_1b_2=0$ and $b_1=\pm b_2$. All of these have four isometries for type $I_c$. For example, the case $b_1=\pm b_2$ yields 
\begin{equation}
    \gl a=\mp 1\,,\qquad \kappa=1\,,\qquad \gep=2\,.
\end{equation}
It describes the previously considered critical cases.

\subsection{Type $I_b$}
This case, parameterized by the two real constants $a_1$ and $a_2$, can be fully recovered from type $I_c$ by the following formal exchange of the embedding coordinates:
\begin{equation}
    X^0 \leftrightarrow iX^2\,.
\end{equation}
Upon the above substitution, the generating field $K$ and its centralizer become  
\begin{equation}
    K = a_1 J_{12} + a_2 J_{03} =|_{X^0 \leftrightarrow i X^2} = -ia_1 J_{01} - ia_2 J_{23},\quad G = K|_{a_1 \leftrightarrow a_2}.
\end{equation}
They form the Cartan (split) subalgebra  $\mathfrak{h} = \mathfrak{so}(1,1) \oplus \mathfrak{so}(1,1)$ of $\mathfrak{so}(2,3)$.
Comparing with the previous case $I_c$, the Killing field then yields the formal relations $a_1 = ib_1$ and $a_2 = ib_2$. Now, to reproduce the general $I_b$ case with $a_1a_2\neq 0$, we can consider coordinates of the Carter-Pleba\'{n}ski type for the embedding \eqref{CPembs}. Assuming the eigenvalues $a_i, b_i$ to be real, the embedding coordinates for type $I_b$ follow from the change of variables in \eqref{CPembs} 
\begin{equation}
    \kappa\to-\kappa\,,\qquad r\to y\,,\qquad y\to r\,,\qquad \gep\to-\gep\,.
\end{equation}
As a result, the relevant embedding coordinates for type $I_b$ read
\begin{subequations}\label{Ib_gen:embd} 
\begin{align}
    &X^2 = \sqrt{\frac{\kappa^2(a^2 + \gl^2 r^2 y^2) + \kappa(y^2 -r^2 - \gl^{-2} \epsilon )}{1 - \gl^2 a^2 \kappa^2}} \cosh\left( \frac{\gl}{\sqrt{\kappa}} \tau + \gl a^2 \sqrt{\kappa} \psi \right)\,,  \\
   &X^1 = \sqrt{\frac{\kappa^2(a^2 + \gl^2 r^2 y^2) + \kappa(y^2 -r^2 - \gl^{-2} \epsilon )}{1 - \gl^2 a^2 \kappa^2}} \sinh\left( \frac{\gl}{\sqrt{\kappa}} \tau + \gl a^2 \sqrt{\kappa} \psi \right)\,, \\
    &X^0 = \sqrt{\frac{a^2 + \gl^2 r^2 y^2 + \gl^2 a^2\kappa(y^2 -r^2 - \gl^{-2} \epsilon)}{\gl^{2} a^2 - \gl^{4} a^4 \kappa^2}}   \cosh\left( \gl^{2} a \sqrt{\kappa} \tau + \frac{a}{\sqrt{\kappa}} \psi \right)\,,  \\
    &X^3 = \sqrt{\frac{a^2 + \gl^2 r^2 y^2 + \gl^2 a^2\kappa(y^2 -r^2 - \gl^{-2} \epsilon)}{\gl^{2} a^2 - \gl^{4} a^4 \kappa^2}}  \sinh\left( \gl^{2} a \sqrt{\kappa} \tau + \frac{a}{\sqrt{\kappa}} \psi \right)\,,  \\
    &X^4 = \frac{ry}{a}\,,
\end{align}    
\end{subequations}    
which gives us
\begin{equation}
    -\gl^{-1}\partial_{\tau} = \gl a \sqrt{\kappa}J_{03} + \frac{1}{\sqrt{\kappa}} J_{12}\,,\qquad -a^{-1}\partial_{\psi} = \frac{1}{\sqrt{\kappa}} J_{03} + \gl a \sqrt{\kappa} J_{12}\,.
\end{equation}
The respective Killing fields in this chart are $K = -\gl^{-1} \partial_{\tau}$ and $G = -a^{-1} \partial_{\psi}$. The numbers $a_{1,2}$ characterizing type $I_b$ are parameterized via
\begin{equation}
    a_1 = \pm \frac{1}{\sqrt{\kappa}}\,,\qquad a_2 = \pm \gl a \sqrt{\kappa}\,,
\end{equation}
where we recall that $\kappa$ represents any two of roots of Eq. \eqref{Ic_gen:kappa_eq}.
The $\so(2, 3)$ algebra invariants are
\begin{equation}
    I_1 = -a_1^2 - a_2^2 = -\epsilon\,,\qquad I_2 =  -a_1^4 - a_2^4  = 2\gl^2a^2 - \epsilon^2\,.
\end{equation}
The AdS spacetime in the Wick-rotated Carter-Pleba\'nski coordinates \eqref{Ib_gen:embd} has the form:
\begin{equation}    
d\overline{s}_{AdS}^2 =  - \frac{\Delta_y}{r^2+y^2}(d\tau - r^2 d\psi)^2 +\frac{\Delta_r}{r^2+y^2}(d\tau + y^2 d\psi)^2 + \frac{(r^2 + y^2)dr^2}{\Delta_r} + \frac{(r^2 + y^2)dy^2}{\Delta_y}\,.
\end{equation}
Ordering our coordinates as $(\psi, \tau, r, y)$, the Killing field $K$ corresponds to the Killing vector  $v^\mu = (0, -1, 0, 0)$, which generates the components of the global symmetry parameter $K_{AB}$ via
\begin{equation}
v_{\ga \dot{\gb}} = -\frac{1}{\sqrt{r^2 + y^2}}
\begin{pmatrix}
    -\sqrt{\Delta_y}&  \sqrt{\Delta_r} \\
    \sqrt{\Delta_r} & -\sqrt{\Delta_y}
\end{pmatrix}\,,\quad
\varkappa_{\ga \gb} =    
-\gl^2\begin{pmatrix}
    0 &  y-ir \\
    y-ir &   0
\end{pmatrix}\,.
\end{equation}
with the Casimir invariants
\begin{equation}
    C_2 = -\epsilon\,, \qquad C_4 = \epsilon^2 + 4\gl^2 a^2\,,
\end{equation}
and eigenvalues:
\begin{equation}
    K_A^{\ B} \xi_B = \Tilde{\gl} \xi_A\,,\qquad \Tilde{\gl} = \pm \sqrt{\epsilon \pm 2 \gl a}\,.
\end{equation}
Using \eqref{KS:spin}, this choice generates the following Kerr-Schild vectors:
\begin{equation}
    k^{\pm}_{\mu}dx^{\mu}=- r^2 d\psi +d\tau \pm \frac{r^2+y^2}{\gl^2 y^4 -\epsilon y^2 + a^2}dy\,,
\end{equation}
\begin{equation}
    l^{\pm}_{\mu}dx^{\mu}= y^2 d\psi + d\tau \mp i\frac{r^2+y^2}{\gl^2 r^4 +\epsilon r^2 + a^2}dr
\end{equation}
with the zeroth copy and the Weyl curvature being
\begin{equation}
    \phif = \frac{y}{r^2 + y^2}\,,\qquad C_{\alpha(4)}\xi^{\alpha(4)} = M \frac{\xi_1^2 \xi_2^2}{(r + iy)^3}\,.
\end{equation}

\subsection{Type $I_d$}
Similarly, type $I_d$ can be analytically recovered from type $I_c$ by Wick-rotating 
\begin{equation}\label{I_d_gen:isom}
    X^0 \leftrightarrow i X^2\,,\qquad  X^1 \leftrightarrow i X^4\,.
\end{equation}
Indeed, the canonical $I_d$--Killing field $K$ (see Table 1) transforms as follows 
\begin{equation}
    K = (a_1 J_{03} + a_2 J_{24})\Big|_{X^0 \to i X^2,\, X^4 \to -i X^1 } = -a_2 J_{01} - ia_1 J_{23}\,,
\end{equation}
establishing a correspondence between cases $\rom{1}_c$ and $\rom{1}_d$. The centralizer of $K$ for general $a_1a_2\neq 0$ is one-dimensional 
\begin{equation}
   G = K|_{a_1 \leftrightarrow -a_2}\,,\qquad [K, G]=0\,. 
\end{equation}
Indeed, let us rewrite the canonical Killing field by replacing the following embedding coordinates $X^0 \leftrightarrow i X^2$ and  $X^1 \leftrightarrow i X^4$:
\begin{equation}
    K = a_1 J_{03} + a_2 J_{24} = -a_2 J_{01} - ia_1 J_{23}\,,\qquad G = K|_{a_1 \leftrightarrow -a_2}\,.
\end{equation}
The two generators form the Abelian subalgebra $\mathfrak{h} = \mathfrak{so}(1,1) \oplus \mathfrak{so}(2)$ of $\mathfrak{so}(2,3)$.
Comparing with the case $\rom{1}_c$, we find the formal parameter relations
\begin{equation}
    a_1 = ib_2\,,\qquad a_2 = -b_1\,.    
\end{equation}
Assuming that each parameter pair remains real in its respective case, one must also perform the substitution $a \rightarrow ia$ (see Eq. \eqref{I_c_gen:b12}), which induces the corresponding adjustments in the real embedding coordinates: 
\begin{subequations} 
\begin{align}
    &X^2 = \sqrt{\frac{\kappa^2(-a^2 + \gl^2 r^2 y^2) \pm \kappa (y^2 -r^2 - \gl^{-2} \epsilon)}{1 + \gl^2 a^2 \kappa^2}} \cos\left( \frac{\gl}{\sqrt{\kappa}} \tau - a^2 \gl \sqrt{\kappa} \psi \right)\,,  \\
    &X^4 = \sqrt{\frac{\kappa^2(-a^2 + \gl^2 r^2 y^2) \pm \kappa (y^2 -r^2 - \gl^{-2} \epsilon)}{1 + \gl^2 a^2 \kappa^2}}  \sin\left( \frac{\gl}{\sqrt{\kappa}} \tau - a^2 \gl \sqrt{\kappa} \psi \right)\,, \\
    &X^0 = \sqrt{\frac{a^2 - \gl^2 r^2 y^2 \pm \gl^2 a^2\kappa(y^2 -r^2 - \gl^{-2} \epsilon)}{\gl^2 a^2 + \gl^4 a^4 \kappa^2}}  \cosh\left( \gl^2 a \sqrt{\kappa} \tau + \frac{a}{\sqrt{\kappa}} \psi \right)\,, \\
    &X^3 = \sqrt{\frac{a^2 - \gl^2 r^2 y^2 \pm \gl^{2} a^2\kappa(y^2 -r^2 - \gl^{-2} \epsilon)}{\gl^2 a^2 + \gl^4 a^4 \kappa^2}} \sinh\left( \gl^2 a \sqrt{\kappa} \tau + \frac{a}{\sqrt{\kappa}} \psi \right)\,, \\
    &X^1 = \frac{ry}{a}.
\end{align}
\end{subequations} 
In this coordinate system, the corresponding Killing fields are given by
\begin{equation}
    K=-\gl^{-1}\partial_{\tau} = \gl a \sqrt{\kappa} J_{03} - \frac{1}{\sqrt{\kappa}} J_{24}\,,\qquad G=-a^{-1}\partial_{\psi} = \frac{1}{\sqrt{\kappa}} J_{03} + \gl a\sqrt{\kappa} J_{24}
\end{equation}
with two real and two imaginary eigenvalues of the Killing field $K$
\begin{equation}
    a_1 = \pm \gl \sqrt{\kappa}a\,,\qquad ia_2 = \pm \frac{i}{\sqrt{\kappa}}\,,
\end{equation}
where $\kappa$ is a positive root of the equation (cf. \eqref{Ic_gen:kappa_eq})
\begin{equation}
    -\gl^2 a^2 \kappa^2 - \epsilon \kappa + 1 = 0\,.
\end{equation}
In this parameterization, the $\so(2, 3)$ invariants are
\begin{equation}
    I_1 = \epsilon\,,\qquad I_2  = -2\gl^2 a^2 - \epsilon^2\,.
\end{equation}
The resulting metric retains the Carter-Pleba\'{n}ski form, but with the parameter replacement $a \rightarrow ia$ in the chart $(\tau, \psi, r, y)$:

\begin{equation}    d\overline{s}_{AdS}^2 = -\frac{\Delta_r}{r^2+y^2}(d\tau + y^2 d\psi)^2 + \frac{\Delta_y}{r^2+y^2}(d\tau - r^2 d\psi)^2 + \frac{(r^2 + y^2)dr^2}{\Delta_r} + \frac{(r^2 + y^2)dy^2}{\Delta_y}\,,
\end{equation}
where 
\begin{equation}
    \Delta_r = \gl^2 r^4 + \epsilon r^2 - a^2,\qquad  \Delta_y = \gl^2 y^4 - \epsilon y^2 - a^2\,.
\end{equation}

As in type $I_c$, $K$ defines the Killing vector $v^\mu = (-1, 0, 0, 0)$, while the spinorial components of the global symmetry parameter take the following form:
\begin{equation}\label{Id:K_AB}
v_{\ga \dot{\gb}} = -\frac{1}{\sqrt{r^2 + y^2}}
\begin{pmatrix}
    \sqrt{\Delta_r}&  \sqrt{\Delta_y} \\
    \sqrt{\Delta_y} & \sqrt{\Delta_r}
\end{pmatrix}\,,\quad
\varkappa_{\ga \gb} =    
\gl^2 \begin{pmatrix}
    y-ir &  0 \\
    0 &   y-ir
\end{pmatrix}\,.
\end{equation}
They yield the Casimir invariants 
\begin{equation}
    C_2 = \epsilon, \quad C_4 = \epsilon^2 - 4\gl^2 a^2\,.
\end{equation}
Using \eqref{Id:K_AB}, one calculates the following Kerr-Schild fields:
\begin{equation}
    k^{\pm}_{\mu}dx^{\mu}=d\tau + y^2 d\psi \pm \frac{r^2+y^2}{\gl^2 r^4 +\epsilon r^2 - a^2}dr\,,
\end{equation}
\begin{equation}
    l^{\pm}_{\mu}dx^{\mu}=d\tau - r^2 d\psi \pm i\frac{r^2+y^2}{\gl^2 y^4 -\epsilon y^2 - a^2}dy
\end{equation}
along with the corresponding zero-copy and the curvature Weyl tensor 
\begin{equation}
    \phif =\frac{r}{r^2 + y^2}\,,\qquad C_{\alpha(4)}\xi^{\alpha(4)} = M \frac{(\xi_1^2 + \xi_2^2)^2}{(r + iy)^3}\,.
\end{equation}
Formally, we conclude that the two orbits $I_c$ and $I_d$ are connected by analytical continuation 
\begin{equation}
    \rom{1}_d = {\rom{1}_c}\Big|_{a \rightarrow ia}\,.
\end{equation}
It is worth noting that both arise from the Pleba\'{n}ski and Demia\'{n}ski family of type D solutions from \cite{Plebanski:1976gy} with different signs of the kinematical parameter, denoted as $\gga$ in \cite{Plebanski:1976gy}, which introduces rotation $\gga = a^2$ for the case $\rom{1}_c$ and $\gga = -a^2$ for its topological version $\rom{1}_d$. More details on this type are provided in Appendix {D}.

\subsection{Type $I_a$}\label{sec:Ia}


{
The Killing field $K$ generating this case is driven by the orbit parameterized by the two numbers $a$ and $b$; see Table 1. For nonzero such parameters, $K$ and its centralizer $G$ form $\u(1) \oplus \u(1)$ symmetry algebra:  
\begin{subequations}\label{KG:Ia}
\begin{align}
    &K = (aX^3 + bX^1)\partial_0 + (aX^2 - bX^0)\partial_1 + (aX^1 - bX^3)\partial_2 + (aX^0 + bX^2)\partial_3\,,\\
    &G = K|_{b \rightarrow -b}\,.
\end{align}    
\end{subequations}
To identify the corresponding solution, we now introduce the canonical $(\tau, \psi, r, y)$ coordinates -- as discussed earlier -- that make this symmetry manifest. We select them to adapt the coordinates to the eigen-structure of $K$ inspired by the Carter--Pleba\'nski form of the AdS metric i.e., the non-accelerating, NUT-free member of the Pleba\'nski--Demia\'nski family \cite{Plebanski:1975xfb,Plebanski:1976gy}. This puts type $I_a$ on the same footing as the case of $I_c$ studied in Section~\ref{CP}. 

Let
\begin{equation}\label{Ia:beta}
    \beta = b + i a\,,
\end{equation}
so that the eigenvalues of $K^{\underline{A}}{}_{\underline{B}}$ are $0$, $\pm i\beta$, $\pm i\bar\beta$ (cf. Table 1). We introduce the Killing coordinates $(\tau,\psi)$ through
\begin{equation}\label{Ia:taupsi}
    K = -\gl^{-1}\partial_\tau\,,\qquad G = \frac{\gl}{|\beta|^2}\,\partial_\psi\,,
\end{equation} and 
\begin{equation}\label{Ia:ry}
     K \psi=K r=K y=G \tau=G r=G y = 0\,,
\end{equation}
The embedding coordinates that solve \eqref{Ia:taupsi} and \eqref{Ia:ry} then take the form
\begin{subequations}\label{Ia:emb}
\begin{align}
    &X^0 + i X^2 = \sqrt2\,\rho\cos\varphi\,,\qquad X^1 - i X^3 = \sqrt2\,\rho\sin\varphi\,,\qquad X^4 = \frac{\gl\, r y}{|\beta|^2}\,,\\
    &\rho^2 = \frac{(\gl^2 r^2 + \beta^2)(\beta^2 - \gl^2 y^2)}{\gl^2\beta^2(\beta^2 - \bar\beta^2)}\,,\qquad
    \varphi = \gl\beta\,\tau + \frac{\beta\bar\beta^2}{\gl}\,\psi\,.
\end{align}
\end{subequations}
The embedding \eqref{Ia:emb} requires $ab\neq 0$ through the factor $\beta^2-\bar\beta^2$. Its pullback gives the AdS metric in the Carter--Pleba\'nski form
\begin{equation}\label{Ia:CP}
    d\overline{s}^2_{AdS} = -\frac{\Delta_r}{r^2+y^2}(d\tau + y^2 d\psi)^2 + \frac{\Delta_y}{r^2+y^2}(d\tau - r^2 d\psi)^2 + \frac{(r^2+y^2)dr^2}{\Delta_r} + \frac{(r^2+y^2)dy^2}{\Delta_y}\,,
\end{equation}
with the structure functions
\begin{subequations}\label{Ia:Delta}
\begin{align}
    &\Delta_r = \frac{|\gl^2 r^2 + \beta^2|^2}{\gl^2} = r^2(\gl^2 r^2 + \epsilon) + \gga\,,\qquad
    \Delta_y = \frac{|\gl^2 y^2 - \beta^2|^2}{\gl^2} = y^2(\gl^2 y^2 - \epsilon) + \gga\,,\\
    &\epsilon = 2(b^2 - a^2)\,,\qquad \gga = \frac{(a^2+b^2)^2}{\gl^2}\,.
\end{align}
\end{subequations}
Two features distinguish this case from Kerr--AdS. Both $\Delta_r$ and $\Delta_y$ are strictly positive for $ab\neq0$, so $r$ and $y$ range over the whole real line. Also, the compact element of the stabilizer, $K - G = 2b(J_{01} + J_{23})$, has no fixed points on the hyperboloid (these would require $X^0=\dots=X^3=0$), so there is no axis of rotation. This agrees with the absence of real zeros of $\Delta_y$.

Interestingly, the structure functions are fixed by the orbit in a manifestly invariant way:
\begin{equation}\label{Ia:charpoly}
    \Delta_r(r) = \frac{\det(\gl r - K)}{\gl^3 r}\,,\qquad \Delta_y(y) = \Delta_r(iy)\,,
\end{equation}
where $K$ is understood as the $5\times5$ matrix $K^{\underline{A}}{}_{\underline{B}}$. Equivalently, $\epsilon = I_1$ and $\gl^2\gga = \half(I_1^2 + I_2)$ in terms of the Casimir invariants of Table 1. The second relation in \eqref{Ia:charpoly} reflects the fact that the Carter--Pleba\'nski metric depends on the complex combination $r+iy$ only through a single quartic. The Casimir invariants resulting from \eqref{Casimir:K} are
\begin{equation}
    C_2 = \epsilon = 2(b^2-a^2)\,,\qquad C_4 = \epsilon^2 + 4\gl^2\gga = 8(a^4 + b^4)>0\,,
\end{equation}
which agrees with the $I_c$ relations $C_2=\epsilon$, $C_4=\epsilon^2+4\gl^2a^2$ of Section~\ref{CP} upon $a^2\to\gga$.

\paragraph{Double copy.} The metric \eqref{Ia:CP} is formally identical to that of Section~\ref{CP}, and the Killing vector is again $v^\mu = (-1,0,0,0)$, i.e., $v=\gl K$. Therefore, in the same vierbein, the spinorial components of the global symmetry parameter are
\begin{equation}\label{Ia:vk}
v_{\ga \dot{\gb}} = -\frac{1}{\sqrt{r^2 + y^2}}
\begin{pmatrix}
    \sqrt{\Delta_r}&  \sqrt{\Delta_y} \\
    \sqrt{\Delta_y} & \sqrt{\Delta_r}
\end{pmatrix}\,,\qquad
\varkappa_{\ga \gb} = \gl^2
\begin{pmatrix}
    y-ir &  0 \\
    0 &   y-ir
\end{pmatrix}\,,
\end{equation}
where the square roots are now real for all $r,y$. With these components, \eqref{KS:spin} gives the four Kerr–Schild vectors. 
\begin{subequations}\label{Ia:KS:CP}
\begin{align}
    &k^{\pm}_{\mu}dx^{\mu} = d\tau + y^2 d\psi \pm \frac{r^2+y^2}{\Delta_r}\,dr\,,\\
    &l^{\pm}_{\mu}dx^{\mu} = d\tau - r^2 d\psi \pm i\,\frac{r^2+y^2}{\Delta_y}\,dy\,.
\end{align}
\end{subequations}
All four are null and geodesic with respect to \eqref{Ia:CP}. The zeroth copy \eqref{zero} and the Weyl double copy are
\begin{equation}\label{Ia:zero-Weyl}
    \phif = \frac{r}{r^2+y^2}\,,\qquad C_{\alpha(4)}\xi^{\alpha(4)} = M\,\frac{(\xi_1^2+\xi_2^2)^2}{(r+iy)^3}\,.
\end{equation}
The Kerr--Schild metric \eqref{Kerr-Schild}, $g_{\mu\nu} = \bar g_{\mu\nu} + {2}M\phif k^{\pm}_\mu k^{\pm}_\nu$, is of Petrov type D for $M\neq0$. After the standard removal of the $dr$ cross terms
\begin{equation}
    d\tau\to d\tau\mp r^2\zeta\,dr\,,\qquad d\psi\to d\psi\mp\zeta\,dr\,,\qquad \zeta=\frac{{2}Mr}{\Delta_r(\Delta_r-{2}Mr)}\,,
\end{equation}
it is again of the Carter--Pleba\'nski form \eqref{Ia:CP} with $\Delta_r\to\Delta_r - {2}Mr$ and $\Delta_y$ unchanged. 

\paragraph{Complexified rotation parameter.} It is of interest to observe that type $I_a$ can be regarded as Kerr with a ``complex rotation''. Indeed, the Carter--Pleba\'nski data $(\epsilon,\gga)$ are defined up to the rescaling $(\tau,\psi,r,y,M)\to(s^{-1}\tau,s^{-3}\psi,sr,sy,s^3M)$, under which $(\epsilon,\gga)\to(s^2\epsilon,s^4\gga)$. This is the rescaling $K\to sK$, $\beta\to s\beta$, discussed after Table 1. The only invariant is
\begin{equation}
    \mathcal{I}:=\frac{\epsilon^2}{\gl^2\gga} = \frac{(\beta^2+\bar\beta^2)^2}{|\beta|^4} = 4\cos^2(2\arg\beta)\,.
\end{equation}
Kerr--AdS in the Carter--Pleba\'nski form corresponds to $\epsilon = 1 + \gl^2 a_K^2$ and $\gga = a_K^2$ (Section~\ref{sec: Kerr}), for which $\mathcal{I}\geq 4$ when $a_K$ is real. Type $I_a$ instead fills the complementary region $\epsilon^2 < 4\gl^2\gga$. Matching the invariant gives
\begin{equation}\label{Ia:aK}
    \gl a_K = \frac{\bar\beta}{\beta} = e^{-2i\arg\beta}\,,\qquad \lambda |a_K| = 1\,.
\end{equation}
Indeed, the complex rescaling $s=\beta^{-1}$ brings \eqref{Ia:Delta} exactly to $\epsilon = 1+\gl^2a_K^2$, $\gga=a_K^2$. Type $I_a$ is therefore a real section of the complexified Kerr--AdS metric whose rotation parameter lies on the circle $|a_K|=\gl^{-1}$ in the complex plane. The reality of the $I_a$ section is restored by the complex coordinate change $s=\beta^{-1}$, which is equivalent to the complex combinations $X^0+iX^2$ and $X^1-iX^3$ in \eqref{Ia:emb}. The circle meets the real axis at $\gl a_K=\pm1$, i.e., at $ab=0$. These are precisely the critical values of Section~\ref{b1=b2}. 

Summarizing, one can regard the real line $|\gl a_K|\neq 1$ as ordinary Kerr (type $I_c$) or Wick rotated Kerr (type $I_b$), depending on whether the rescaling $s$ is real or imaginary. Similarly, the imaginary line $|\gl a_K|\neq 1$ can be shown to match the $I_d$ case. The two points $\gl a_K = \pm 1$ generate the boundary with the ratio $\mathcal{I}=4$, corresponding to the critical case, while the circle $|\lambda a_K| = 1$ off the real axis generates $I_a$.

\paragraph{All type $I$ orbits from type $I_a$.} Naturally, type $I_a$ can be used to generate all other types $I$. Indeed, Eqs. \eqref{Ia:emb}--\eqref{Ia:aK} are holomorphic in $\beta$ and $\bar\beta$, i.e., they nowhere imply that $\beta$ and $\bar\beta$ are complex conjugates. Treating them as two independent parameters, $\beta\to\beta_1$ and $\bar\beta\to\beta_2$, yields a single description of all semisimple orbits with eigenvalues $0,\pm i\beta_1,\pm i\beta_2$:
\begin{equation}\label{I:uniform}
    \Delta_r = \frac{(\gl^2r^2+\beta_1^2)(\gl^2r^2+\beta_2^2)}{\gl^2}\,,\quad \Delta_y(y)=\Delta_r(iy)\,,\quad
    \epsilon = \beta_1^2+\beta_2^2\,,\quad \gl^2\gga = \beta_1^2\beta_2^2\,,\quad \gl a_K = \frac{\beta_2}{\beta_1}\,.
\end{equation}
The Kerr--Schild vectors, zeroth copy and Weyl tensor \eqref{Ia:KS:CP}--\eqref{Ia:zero-Weyl} are unchanged. Reality of $\Delta_{r,y}$ requires the set $\{\beta_1^2,\beta_2^2\}$ to be closed under complex conjugation, which leaves exactly four possibilities:
\begin{itemize}
    \item $I_a$: $\beta_2 = \bar\beta_1 = b - ia$. Then $\epsilon^2 < 4\gl^2\gga$ and $|\gl a_K|=1$.
    \item $I_c$: $\beta_{1,2} = b_{1,2}$ real. This is Kerr--AdS with real $\gl a_K = b_2/b_1$. With \eqref{I_c_gen:b12} one finds $a_K = a\kappa$, which reduces to $a$ in the normalization $\kappa=1$, $\epsilon = 1+\gl^2a^2$. In this case \eqref{Ia:emb} becomes $X^0+iX^1 = |\rho_1| e^{i\varphi_1}$, $X^2+iX^3=|\rho_2|e^{i\varphi_2}$, with $\rho_j$, $\varphi_j$ obtained from $\rho$, $\varphi$ by $(\beta,\bar\beta)\to(\beta_j,\beta_{k\neq j})$.
    \item $I_b$: $\beta_{1,2} = i a_{1,2}$. Then $\epsilon = -(a_1^2+a_2^2)<0$ and $\gl a_K = a_2/a_1$ is real. However, the normalization $s=\beta_1^{-1}$ is imaginary, so the phases $\varphi_j$ become real rapidities ($\cos\to\cosh$). This is the Wick rotation $X^0\leftrightarrow iX^2$ of the type $I_b$ subsection; cf. the exchange $r\leftrightarrow y$, $\epsilon\to-\epsilon$ used for \eqref{Ib_gen:embd}.
    \item $I_d$: $\beta_1 = b$ real and $\beta_2 = ia$ imaginary. Then $\epsilon = b^2-a^2$, $\gga = -a^2b^2/\gl^2 < 0$, and $a_K$ is imaginary. This is the substitution $a\to ia$ of the type $I_d$ subsection, with the topological sign $\gga<0$ noted there.
\end{itemize}
Thus the four type $I$ families correspond to the four real forms of one holomorphic Carter--Pleba\'nski metric, distinguished by the location of $\beta_{1,2}^2$ in the complex plane: complex-conjugate pair, both positive, both negative, and opposite signs. In all four cases the discriminant is $\epsilon^2-4\gl^2\gga = (\beta_1^2-\beta_2^2)^2$, which is negative only for $I_a$.

\paragraph{Degenerate limits.} The derivation requires $ab\neq0$, since $\rho^2$ in \eqref{Ia:emb} contains the factor $(\beta^2-\bar\beta^2)^{-1}$. For $a=0$ the generator $K=b(J_{01}+J_{23})$ is the symmetry-enhanced case $I_c$ with $b_1=\pm b_2$, the critical solution with four isometries. For $b=0$ one obtains the symmetry-enhanced topological case $I_b$ with $a_1=a_2$, which also has four isometries. In both limits, the discriminant vanishes and the structure functions develop double roots, $\Delta_r = (\gl^2r^2+b^2)^2/\gl^2$ and $\Delta_r = (\gl^2r^2-a^2)^2/\gl^2$, respectively. The same local functions appear for types $II_b$ and $II_a$. Type $II_b$ is the ultraspinning (super-entropic) Kerr--AdS solution \cite{Hennigar:2014cfa}, reached as $a_K\to\gl^{-1}$ on the circle \eqref{Ia:aK} together with a singular rescaling of the azimuthal coordinate. Because \eqref{Ia:emb} breaks down at $ab=0$, which of these orbits is reached depends on how the limit is taken. Reaching type $II$ requires a contraction $K\to S+N$ with $N\neq0$, as in Appendix~{B}, rather than simply setting $a$ or $b$ to zero.
}

\section{TYPE $II$}\label{sec:II}
According to Table 1, Type $II$ includes two subtypes: Type $II_a$ and Type $II_b$, which are characterized by the kinematical parameters  $a$  and  $b$, respectively. These parameters are, in fact, discrete, as shown in Table 1a. When these parameters are nonzero, the centralizer of the generating global symmetry parameter  $K_{AB}$  yields two Killing vectors for the double copies. Using canonical coordinates, we identify the zeroth copy, the Kerr-Schild single copy,  and the Weyl double copy for these types. Type $II^-=II_{b<0}$ describes the so-called super-entropic black hole considered in \cite{Hennigar:2014cfa}. Type $II^+=II_{b>0}$ describes its Wick-rotated version being the hyperbolic super-entropic solution. Type $II=II_{a\neq 0}$ generates rotating black membrane of \cite{Klemm:2014rda}.

The case with $a=0$ or $b=0$ corresponding to orbit {$II^0$}$=II_{a=0}=II_{b=0}$ is degenerate: it leads to a global symmetry enhancement of the corresponding double copies. The centralizer of the orbit includes six Lie generators (including $K_{AB}$ itself). One might expect that the corresponding double copy would exhibit six isometries in accordance with the general analysis from Sec. \ref{sec:Penrose}. However, this expectation is not met because the Penrose transform, which generates multicopies through $K_{AB}$ using \eqref{C:Dtype}, becomes ill-defined when $a=0$ or $b=0$  (specifically, $(\det\gk_{\al\gb}=\det\bar\gk_{\dal\dgb}=0)$).This is contrasted with type $I$, where the symmetry-enhanced cases are perfectly regular. 

Despite this complication, it is still possible to derive solutions in this case. This can be done by regularizing the limit $a\to 0$ or $b\to 0$. Surprisingly, the two different limits reproduce different solutions despite corresponding to the same orbit. Thus, type {$II^0$} characterizes the solution up to the regularization prescription. The resulting double copies are classified as Petrov type N due to the degeneracy of $\gk_{\al\gb}$; see \eqref{Weyl:DC}. The general analysis of global symmetries presented in Sec. \ref{sec:Penrose} does not hold in these degenerate situations too. Therefore, to determine the appropriate isometries, we investigate the Lie equation   $\mathcal{L}_\xi g_{\mu\nu}=0$ and identify four Killing vectors for each degenerate case instead of six. We recognize the resulting metrics as particular representatives of AdS plane-wave solutions, known as Siklos metrics \cite{Siklos}; see also \cite{Podolsky:1997ik} for their interpretation.  

It is noteworthy that by using the orbit with $a=0$ or $b=0$, we can create double copies that feature five or six isometries. The resulting solutions are of type N, but they differ from the Einstein $\Lambda$-vacuums because they may have a nonzero stress tensor. The solutions are not derived through the regularization of the Penrose transform; instead, we obtain them using the ansatze based on the generating parameter $K_{AB}$. The corresponding metrics are referred to as Kaigorodov vacuum spacetime in the case of five isometries \cite{Kaigorodov} and the Defrise spacetime \cite{Defrise1969} with six isometries, which form the center of the $II_{a,b=0}$ generator. The latter yields a nonzero stress tensor. Since these solutions are not connected to the solution-generating scheme developed in this paper, we present them separately as a curious case.

\subsection{Type $\rom{2}_b$}\label{sec:IIb}
In the general case $b \neq 0$, this orbit provides one with two commuting Killing fields:
\begin{subequations}\label{KG:IIa}
\begin{align}
    &K = (b-1)J_{01} + J_{02} -J_{13} +(b+1)J_{23}\,,\\
    &G = b(J_{23} + J_{01})\,. 
\end{align}
\end{subequations}
{
For all $b$, the difference
\begin{equation}\label{IIb:N}
    N:=K-G=-J_{01}+J_{02}-J_{13}+J_{23}\,,\qquad N^2=0\,,\qquad [G,N]=0\,,
\end{equation}
is the same minimal nilpotent element (orbit $II^0$; see Table 1a). Hence \eqref{KG:IIa} is the Jordan decomposition $K=G+N$, whose semisimple part $G=b(J_{01}+J_{23})$ is compact.

\subsubsection{Type $II^-$: super-entropic black hole} This orbit is generated by a negative $b$. A particular choice of $b$ is irrelevant, as it represents a gauge choice. We set $b=-1$ for convenience, so that 
\begin{equation}
   K=-2J_{01}+J_{02}-J_{13}\,,\qquad  G=-(J_{01}+J_{23})
\end{equation}
By Table~\ref{tab:discrete}, this represents the discrete orbit $II^-$. To identify the inner coordinates, we straighten $K$ together with the nilpotent part $N$ of its centralizer. A convenient embedding that does this is
\begin{subequations}\label{IIb:emb}
\begin{align}
    &X^0+iX^1=\frac{e^{i\gl t}}{2\sqrt2}\Big[\chi+2R-2i\chi\,(2\varphi-\gl t)\Big]\,,\\
    &X^3+iX^2=\frac{e^{i\gl t}}{2\sqrt2}\Big[\chi-2R+2i\chi\,(2\varphi-\gl t)\Big]\,,\\
    &X^4=r\cos\theta\,,
\end{align}
\end{subequations}
where
\begin{equation}
    \chi=\sin\theta\sqrt{r^2+\gl^{-2}}\,,\qquad R=\frac{\gl^{-2}+r^2\cos^2\theta}{\chi}\,.
\end{equation}
The hyperboloid condition holds identically: 
\begin{equation}
 -|X^0+iX^1|^2+|X^3+iX^2|^2+(X^4)^2=-\chi R+r^2\cos^2\theta=-\gl^{-2}\,.  
\end{equation}
In these coordinates
\begin{equation}\label{IIb:straight}
    \partial_t=-\gl K\,,\qquad \partial_\varphi=2N\,,
\end{equation}
and
\begin{equation}
  K r=K\theta=K\varphi= Nt=Nr=N\theta=0\,. 
\end{equation}
The inner coordinates are recovered from the embedding coordinates as
\begin{subequations}\label{IIb:inner}
\begin{align}
    &\gl t=\arctan\frac{X^1+X^2}{X^0+X^3}\,,\qquad \chi^2=2\big[(X^1+X^2)^2+(X^0+X^3)^2\big]\,,\qquad r\cos\theta=X^4\,,\\
    &\varphi=\frac{\gl t}{2}+\frac{X^0X^2-X^1X^3}{2\big[(X^1+X^2)^2+(X^0+X^3)^2\big]}\,.
\end{align}
\end{subequations}
Here $X^4$ and $(X^1+X^2)^2+(X^0+X^3)^2$ are the two joint invariants of $K$ and $G$; they determine $r$ and $\theta$. The AdS metric takes the form
\begin{equation}\label{IIb:AdS:g}
    d\overline{s}^2=-\frac{\Delta}{\rho^2}\big(dt-\gl^{-1}\sin^2\theta\,d\varphi\big)^2+\frac{\rho^2dr^2}{\Delta}+\frac{\rho^2d\theta^2}{\sin^2\theta}+\frac{\sin^4\theta}{\rho^2}\big(\gl^{-1}dt-(r^2+\gl^{-2})d\varphi\big)^2\,,
\end{equation}
with
\begin{equation}
    \Delta=\frac{(1+\gl^2r^2)^2}{\gl^2}\,,\qquad \rho^2=r^2+\gl^{-2}\cos^2\theta\,.
\end{equation}
This is the background of the super-entropic black hole \cite{Hennigar:2014cfa}, i.e., the limit (B) of Table~\ref{tab:limits}. It is a Carter--Pleba\'nski metric with the double roots $\Delta_r=(1+\gl^2r^2)^2/\gl^2$ and $\Delta_y=(1-\gl^2y^2)^2/\gl^2$, where $y=\gl^{-1}\cos\theta$. Note that $g_{\varphi\varphi}=0$: the generator $\partial_\varphi=2N$ is a null Killing vector. Compactifying $\varphi$, as required for the super-entropic black hole, is therefore an identification of AdS along the parabolic element $N$.

We choose the vierbein
\begin{subequations}\label{IIb:SE:e}
\begin{align}
    &\e^0=\frac{\sqrt\Delta}{\rho}\big(dt-\gl^{-1}\sin^2\theta\,d\varphi\big)\,,\qquad \e^1=\frac{\rho\,dr}{\sqrt\Delta}\,,\\
    &\e^2=\frac{\rho\,d\theta}{\sin\theta}\,,\qquad \e^3=\frac{\sin^2\theta}{\rho}\big(\gl^{-1}dt-(r^2+\gl^{-2})d\varphi\big)\,.
\end{align}
\end{subequations}
The Killing vector $v=\gl K=-\partial_t$, i.e., $v^\mu=(-1,0,0,0)$, is the limit of the Kerr--AdS vector $v^\mu=(-1,0,0,-\gl^2a)$ of Section~\ref{sec: Kerr} under (B). With the conventions of \eqref{v:Kerr}--\eqref{gk:Kerr}, one finds the diagonal components
\begin{equation}\label{IIb:SE:vk}
    v_{\ga\dgb}=\frac{1}{\gl\rho}\begin{pmatrix}1+\gl^2r^2+\sin^2\theta & 0\\ 0 & \gl^2\rho^2\end{pmatrix}\,,\qquad
    \gk_{\ga\gb}=\gl^2\big(r+i\gl^{-1}\cos\theta\big)\begin{pmatrix}1&0\\0&-1\end{pmatrix}\,.
\end{equation}
Hence $\rf=r+i\gl^{-1}\cos\theta$, and the Casimir invariants \eqref{Casimir:K} are $C_2=2$ and $C_4=8$. The Kerr--Schild vectors \eqref{KS:spin}, normalized by \eqref{KS:norm}, are
\begin{subequations}\label{IIb:SE:KS}
\begin{align}
    &k^\pm_\mu dx^\mu=dt-\gl^{-1}\sin^2\theta\,d\varphi\mp\frac{\rho^2}{\Delta}\,dr\,,\\
    &l^\pm_\mu dx^\mu=dt-\gl(r^2+\gl^{-2})\,d\varphi\mp i\,\frac{\gl\rho^2}{\sin^3\theta}\,d\theta\,.
\end{align}
\end{subequations}
The zeroth copy \eqref{zero} and the Weyl tensor \eqref{Weyl:DC} are
\begin{equation}\label{IIb:SE:Weyl}
    \phif=\frac{r}{r^2+\gl^{-2}\cos^2\theta}\,,\qquad C_{\ga(4)}\xi^{\ga(4)}=M\gl^4\frac{(\xi_1^2-\xi_2^2)^2}{(r+i\gl^{-1}\cos\theta)^3}\,.
\end{equation}
The change of coordinates
\begin{equation}
  dt\to dt+(r^2+\gl^{-2})\zeta\,dr\,,\quad d\varphi\to d\varphi+\gl\zeta\,dr\,,\quad  \zeta=\mp2Mr/(\Delta(\Delta-2Mr)) 
\end{equation}
removes the $dr$ cross terms in the metric. It brings the Kerr--Schild metric $g=\bar g+2M\phif kk$ to \eqref{IIb:AdS:g} with $\Delta\to\Delta-2Mr$. This is precisely the super-entropic black hole of \cite{Hennigar:2014cfa}, with mass parameter $m=M$ in the normalization \eqref{Kerr:metric}. The type $II^-$ double copy is therefore the super-entropic black hole.

\subsubsection{Type $II^+$: hyperbolic super-entropic black hole} The orbit $II_{b>0}$ can be taken with any $b>0$. All such choices are equivalent. Let us pick $b=+1$. In this case, Eqs.~\eqref{KG:IIa} give
\begin{equation}
    K=J_{02}-J_{13}+2J_{23}\,,\qquad G=J_{01}+J_{23}\,,
\end{equation}
which represents the orbit $II^+$ of Table~\ref{tab:discrete}. In the chart \eqref{IIb:emb}, the corresponding Killing vector is $v=\partial_t+\gl\partial_\varphi$ (see \eqref{IIb:vb} below), which is not static. A static chart is obtained by the analytic continuation $\theta\to i\eta$ of \eqref{IIb:emb}, with $X^{1,2}\to-X^{1,2}$:
\begin{subequations}\label{IIb:emb+}
\begin{align}
    &X^0-iX^1=\frac{e^{i\gl t}}{2\sqrt2}\Big[\tilde\chi+2\tilde R+2i\tilde\chi\,(2\varphi-\gl t)\Big]\,,\\
    &X^3-iX^2=\frac{e^{i\gl t}}{2\sqrt2}\Big[\tilde\chi-2\tilde R-2i\tilde\chi\,(2\varphi-\gl t)\Big]\,,\\
    &X^4=r\cosh\eta\,,
\end{align}
\end{subequations}
where
\begin{equation}
    \tilde\chi=\sinh\eta\sqrt{r^2+\gl^{-2}}\,,\qquad \tilde R=\frac{\gl^{-2}+r^2\cosh^2\eta}{\tilde\chi}\,.
\end{equation}
In this chart, again $\partial_t=-\gl K$ and $\partial_\varphi=2N$. The AdS metric \eqref{IIb:AdS:g} is modified by $\sin\theta\to i\sinh\eta$ and $\cos\theta\to\cosh\eta$:
\begin{equation}\label{IIb:AdS:g+}
    d\overline{s}^2=-\frac{\Delta}{\tilde\rho^2}\big(dt+\gl^{-1}\sinh^2\eta\,d\varphi\big)^2+\frac{\tilde\rho^2dr^2}{\Delta}+\frac{\tilde\rho^2d\eta^2}{\sinh^2\eta}+\frac{\sinh^4\eta}{\tilde\rho^2}\big(\gl^{-1}dt-(r^2+\gl^{-2})d\varphi\big)^2\,,
\end{equation}
with the same $\Delta$ and
\begin{equation}
  \tilde\rho^2=r^2+\gl^{-2}\cosh^2\eta\,.  
\end{equation}
In the Carter--Pleba\'nski form, \eqref{IIb:AdS:g} and \eqref{IIb:AdS:g+} are the same metric, with $\Delta_r=(1+\gl^2r^2)^2/\gl^2$ and $\Delta_y=(1-\gl^2y^2)^2/\gl^2$. They describe the two sides of the double root of $\Delta_y$: 
\begin{subequations}
    \begin{align}
       &|y|=\gl^{-1}|\cos\theta|<\gl^{-1}\quad  \text{for}\quad  b=-1\,,\\ 
       &|y|=\gl^{-1}\cosh\eta>\gl^{-1}\quad \text{for}\quad  b=+1\,.
    \end{align}
\end{subequations}
The orbit of the static Killing vector $\partial_t$ changes from $II^-$ to $II^+$ across $|y|=\gl^{-1}$, as follows from the invariant of Appendix~{B}.

In the vierbein
\begin{subequations}
\begin{align}
    &\e^0=\frac{\sqrt\Delta}{\tilde\rho}\big(dt+\gl^{-1}\sinh^2\eta\,d\varphi\big)\,,\qquad \e^1=\frac{\tilde\rho\,dr}{\sqrt\Delta}\,,\\ &\e^2=\frac{\tilde\rho\,d\eta}{\sinh\eta}\,,\qquad \e^3=\frac{\sinh^2\eta}{\tilde\rho}\big(\gl^{-1}dt-(r^2+\gl^{-2})d\varphi\big)\,,
\end{align}   
\end{subequations}
the Killing vector $v=\gl K=-\partial_t$ gives
\begin{equation}
    v_{\ga\dgb}=\frac{1}{\gl\tilde\rho}\begin{pmatrix}\gl^2\tilde\rho^2 & 0\\ 0 & 1+\gl^2r^2-\sinh^2\eta\end{pmatrix}\,,\qquad
    \gk_{\ga\gb}=\gl^2\big(r-i\gl^{-1}\cosh\eta\big)\begin{pmatrix}1&0\\0&-1\end{pmatrix}\,,
\end{equation}
with $C_2=2$ and $C_4=8$. The Kerr--Schild vectors, the zeroth copy and the Weyl tensor are
\begin{subequations}\label{IIb:KS+}
\begin{align}
    &k^\pm_\mu dx^\mu=dt+\gl^{-1}\sinh^2\eta\,d\varphi\mp\frac{\tilde\rho^2}{\Delta}\,dr\,,\\
    &l^\pm_\mu dx^\mu=dt-\gl(r^2+\gl^{-2})\,d\varphi\pm i\,\frac{\gl\tilde\rho^2}{\sinh^3\eta}\,d\eta\,,\\
    &\phif=\frac{r}{r^2+\gl^{-2}\cosh^2\eta}\,,\qquad C_{\ga(4)}\xi^{\ga(4)}=M\gl^4\frac{(\xi_1^2-\xi_2^2)^2}{(r-i\gl^{-1}\cosh\eta)^3}\,.
\end{align}
\end{subequations}
As for $b=-1$, the Kerr--Schild metric $\bar g+2M\phif kk$ is brought to \eqref{IIb:AdS:g+} with $\Delta\to\Delta-2Mr$ by a change of $t$ and $\varphi$ by functions of $r$. The resulting space-time is the hyperbolic counterpart of the super-entropic black hole, obtained from it by $\theta\to i\eta$. Its horizon $\Delta(r_+)=2Mr_+$ is non-compact and is parameterized by $\eta\in(0,\infty)$ and $\varphi$. As in the super-entropic case, $\partial_\varphi$ is null in the background, and the end $\eta\to0$ lies at infinite proper distance and plays the role of the puncture. 

\paragraph{General $b$ and the limit $b\to -0$.} As we have stressed already, the particular choice of the parameter $b< 0$ in \eqref{KG:IIa} is not important. However, having $b$ arbitrary is useful in approaching orbit $II^0$, which has no well-defined double copy via the Penrose transform. Nevertheless, the limit $II_{b\to 0}$ and $II_{a\to 0}$ can be addressed via a regularization resulting in two different space-times, despite the limiting orbits are the same $II^0$ (see Table 1a), implying the resulting regularized double copy is not fully characterized by the orbit $II^0$. We therefore consider these cases separately starting with type $II_b$.     

For type $II_{b<0}$ in the chart \eqref{IIb:emb}, the Killing vector of $K$ in \eqref{KG:IIa} is, for any $b$,
\begin{equation}\label{IIb:vb}
    v=\gl K=b\,\partial_t+\frac{\gl}{2}(1+b)\,\partial_\varphi\,.
\end{equation}
In this chart, $v$ is static only for $b=-1$. For $b=+1$, $v=\partial_t+\gl\partial_\varphi$, and the static chart is \eqref{IIb:emb+}. For any $b<0$,
\begin{equation}
    K_b=(-b)\,e^{sH}K_{-1}e^{-sH}\,,\qquad e^s=-\frac1b\,,
\end{equation}
with the boost $H$ of Appendix~{B}, which satisfies $[H,J_{01}+J_{23}]=0$ and $[H,N]=N$. Hence all double copies with $b<0$ are super-entropic black holes, seen in boosted charts.

Now, as  $b\to0$, the Killing vector \eqref{IIb:vb} tends to the null Killing vector $\tfrac\gl2\partial_\varphi=\gl N$. We take $b<0$, so that the whole family belongs to $II^-$. With
\begin{equation}
    \mathrm z:=r+i\gl^{-1}\cos\theta\,,\qquad |\mathrm z|^2=\rho^2\,,
\end{equation}
the components in the vierbein \eqref{IIb:SE:e} remain diagonal for any $b$:
\begin{subequations}\label{IIb:kappa}
\begin{align}
    &v_{\ga\dgb}=\frac{1}{\gl\rho}\begin{pmatrix}\sin^2\theta\,(1+\gl^2r^2)-b\,(1+\gl^2r^2\cos^2\theta) & 0\\ 0 & -b\,\gl^2\rho^2\end{pmatrix}\,,\\
    &\gk_{\ga\gb}=\gl^2\begin{pmatrix}\dfrac{(1+b)\chi^2-b\,\mathrm z^2}{\mathrm z} & 0\\ 0 & b\,\mathrm z\end{pmatrix}\,.
\end{align}
\end{subequations}
These reduce to \eqref{IIb:SE:vk} at $b=-1$. Consequently,
\begin{equation}
    \rf^2=b^2\mathrm z^2-b(1+b)\chi^2\,,\qquad C_2=2b^2\,,\qquad C_4=8b^4\,,
\end{equation}
and the Weyl tensor \eqref{Weyl:DC} reads
\begin{equation}\label{IIb:Weyl}
    C_{\ga(4)}\xi^{\ga(4)}=M\gl^4\,\frac{\Big(\big((1+b)\chi^2-b\,\mathrm z^2\big)\mathrm z^{-1}\xi_1^2+b\,\mathrm z\,\xi_2^2\Big)^2}{\big(b^2\mathrm z^2-b(1+b)\chi^2\big)^{\frac52}}\,.
\end{equation}
}
{ As $b\to-0$ one has $\rf^2=-b\chi^2+O(b^2)$, so $\rf\simeq\sqrt{-b}\,\chi$ is real, and $\gk_{22}=O(b)$. The two principal spinors of $\gk_{\ga\gb}$ merge, and the Weyl tensor becomes type N:
\begin{equation}\label{IIb:Weyl:lim}
    C_{\ga(4)}\xi^{\ga(4)}\Big|_{b\to-0}=\frac{M}{(-b)^{\frac52}}\,\frac{\gl^4\,\xi_1^4}{\mathrm z^2\chi}\,.
\end{equation}
}
{This is the leading term of \eqref{IIb:Weyl} at $b\to -0$. After the renormalization $M\to (-b)^{\frac{5}{2}}M$ it remains finite and nonzero.}
{
The zeroth copy and the Kerr--Schild vectors behave as
\begin{equation}\label{lim2b-zero-copy}
    \phif=\frac{1}{\sqrt{-b}\,\chi}\big(1+O(b)\big)\,,
\end{equation}
\begin{equation}\label{IIb:k:lim}
    k^\pm_\mu dx^\mu=-\frac{1}{2b}\,dt\mp\frac{1}{\sqrt{-b}}\,\frac{d\chi}{\gl^2\chi^2}+O(1)\,,\qquad l^\pm_\mu dx^\mu=-\frac{\gl^2\chi^2}{2b^2}\,dt+O\big(|b|^{-\frac32}\big)\,.
\end{equation}
The order-$(-b)^{-1/2}$ term in $k^\pm$ is fixed by the null condition, since $\bar g^{tt}=0$ and $|d\chi|^2=\gl^2\chi^2$. The leading term of the real Kerr--Schild vectors,
\begin{equation}\label{lim2b-one-copy}
    \hat{k}^{\pm}_{\mu}dx^{\mu} = -\frac{1}{2b}\, dt\,.
\end{equation}
Up to an overall function, the same holds for the complex vectors. Consequently, in this limit only one independent Kerr--Schild vector survives. With the renormalization $M=2(-b)^{\frac52}\hat M$, the Kerr--Schild term becomes
\begin{equation}
    2M\phif\,k^\pm k^\pm=\frac{\hat M}{\chi}\,dt^2+O\big(\sqrt{-b}\big)\,,
\end{equation}
since all other components are suppressed by powers of $\sqrt{-b}$. Renaming $\hat M\to M$, the limiting real Kerr--Schild solution is
\begin{equation}\label{II0:metric}
    ds^2 = d\overline{s}^2 + \frac{M}{\chi}\, dt^2\,,\qquad \chi=\sin\theta\sqrt{r^2+\gl^{-2}}\,,
\end{equation}
where the AdS metric is \eqref{IIb:AdS:g}. Combined with the conjugation below \eqref{IIb:vb}, \eqref{II0:metric} is an infinite-boost limit of the super-entropic black hole, with its mass parameter scaled to zero as $(-b)^{5/2}$.

For $b\to+0$ (orbit $II^+$), $\rf\simeq i\sqrt b\,\chi$ is imaginary, and the roles of the real and dual copies are exchanged. The real Kerr--Schild metric has no finite curved limit in this chart. Instead, the dual double copy $\tilde\phif\,ll$ of \eqref{KS_DC:def} converges to \eqref{II0:metric} with $M\to-M$.
}

\subsection{Type $II_{b=0}$: The Siklos spacetime} 

To identify the obtained solution in \eqref{II0:metric} with that available in the literature, we set $\gl = 1$ and perform the following coordinate transformation:
{\begin{equation}
    u=t\,,\qquad v=\varphi-\tfrac12 t\,,\qquad x=\frac{r\cos\theta}{\sin\theta\sqrt{r^2+1}}\,,\qquad r'=\frac{1}{\sin\theta\sqrt{r^2+1}}\,,
\end{equation}
from the coordinates $(t,r,\theta,\varphi)$ of \eqref{IIb:emb}; below, the prime on the new radial coordinate is dropped.}
It brings Eq. \eqref{II0:metric} to the standard form of a particular member of the Siklos metrics \cite{Siklos}, which describes the AdS plane wave as the Einstein $\Lambda$-vacuum. Its Petrov type is N.
\begin{equation}\label{II0:H}
    ds^2 = \frac{dr^2 + dx^2 + 2du dv + H du^2}{r^2}\,,\quad H = -r^2 -x^2 + M r^3\,,
\end{equation}
where the function $H$ satisfies the condition (see also \cite{Calvaruso:2022zad})
\begin{equation}\label{H:def}
    2H'_r - r(H''_{rr} + H''_{xx}) = 0\,.
\end{equation}
Note that while different functions $H$  that satisfy equation \eqref{H:def} describe inequivalent Siklos spacetimes\footnote{This is true up to certain gauge equivalences, as discussed in \cite{Siklos}.}, in our case, we have a specific solution given by \eqref{II0:H}. When $M=0$, Eq. \eqref{II0:H} with  $H_0 = -x^2 - r^2$ gives a gauge equivalent form of AdS space. 

As we discussed earlier, the case with $b=0$, corresponding to the following generating parameter: 
\begin{equation}\label{IIb0:K}
    K=-J_{01}+J_{02}-J_{13}+J_{23}\,,
\end{equation}
yields a centralizer with six generators. One might naively expect that the isometry of the metric \eqref{II0:H} would include six Killing vectors. However, this is not the case. The derivation of the global symmetries, as outlined in Sec. \ref{sec:Penrose}, is hindered by the degeneracy of the Penrose map. Consequently, we need to directly analyze the Killing equation for the metric \eqref{II0:H}.     

The analysis of the Killing equations $\mathcal{L}_{\xi} g = 0$ is conducted in Appendix {F}. It yields four (out of six from the center) Killing vectors. Specifically, in the convenient basis, we have the following symmetries of \eqref{II0:H}:
\begin{align}
    K &= J_{01} - J_{02} + J_{13} - J_{23} = \partial_v\,, \quad J = J_{01} + J_{23} = \partial_u\,,\\
    P_1 &= J_{34} - J_{04}  = \cos{u} \partial_x + x \sin{u}\partial_v\,, \quad 
    P_2 = J_{14} - J_{24} = -\sin{u} \partial_x + x \cos{u}\partial_v\,,
\end{align}
and their commutators
\begin{equation}
    [K, P_i] =[K,J]= 0\,,\quad [P_1, P_2] = K\,,\quad [J, P_1] = P_2\,,\quad [J, P_2] = -P_1\,.
\end{equation}
The corresponding algebra is $\mathfrak{h}_3 \ltimes \mathfrak{so}(2)$,\footnote{In a different context, the same algebra emerges as an isometry of the Nappi-Witten solution \cite{NappiWitten1993}; see \cite{Duval:1994qz}.} where  $\mathfrak{h}_3 = \langle K, P_1, P_2 \rangle$ is the Heisenberg algebra and  $ \mathfrak{so}(2) = \langle J \rangle$ is the algebra of rotation generated by a semi-simple elliptic (compact) element. This symmetry algebra can also be identified with the Lie algebra $\mathfrak{g}^{0}_{4.9}$ in the classification of \cite{ayad2026pseudoriemannianalgebraicriccisolitons} with the following identification: $e_1 = K$, $e_2 = P_1$, $e_3 = P_2$, and $e_4 = J$.

Let us comment more on the algebra of the minimally nilpotent orbit that arises as a center of \eqref{IIb0:K}. This is the six-dimensional symmetry algebra $\mathfrak{h}_3 \ltimes \mathfrak{so}(1, 2)$,  which does not correspond to a symmetry of type $II_{b=0}$, as we previously discussed. On the other hand, there appear to be no known four-dimensional Lorentzian vacuum Einstein solutions with a negative cosmological constant that {admit $\mathfrak{h}_3 \ltimes \mathfrak{so}(1, 2)$ as their isometry algebra.\footnote{Up to local isometry, the anti-Nariai spacetime $AdS_2\times H_2$ is the only $\Lambda<0$ vacuum solution of Petrov type D with six Killing vectors \cite{Stephani:2003tm, Griffiths:2009dfa}. We thank the anonymous Referee for pointing this out to us.}}

However, there is a Petrov type N solution from the Siklos family that exhibits the global symmetry algebra  $\mathfrak{h}_3 \ltimes \mathfrak{so}(1, 2)$. This solution describes a pure-radiation spacetime \cite{Calvaruso:2022zad} characterized by a non-vanishing stress-energy tensor. Interestingly, as demonstrated in Sec. \ref{max-symmetry-orbit}, this solution can be generated through a suitable ansatz using the parameter \eqref{IIb0:K}.

Moreover, the ansatz we employ allows for a certain degree of freedom, which leads us to another solution from the Siklos family that has five isometries, known as the Kaigorodov spacetime \cite{Kaigorodov}. This spacetime is a homogeneous Einstein vacuum solution of Petrov type N and features the profile $H = \mu r^3$ instead of the $H$ in \eqref{II0:H}. It is important to highlight once again that this case does not fall within the class derived in our current construction and will therefore be addressed separately in Sec. \ref{max-symmetry-orbit}.

\subsection{Type $\rom{2}_a$}\label{sec:IIa}

Unlike sub-types of $I$, type $II_a$ cannot be directly reached via Wick rotation of type $II_b$. So, let us consider type $II_a$ independently.
{
We choose the representative
\begin{equation}\label{IIa:K}
    K=-J_{01}+(1+a)J_{02}-(1-a)J_{13}+J_{23}=a\,S+N\,,\qquad B:=J_{02}+J_{13}\,,\qquad [S,N]=0\,,
\end{equation}
where $N$ is the same minimal nilpotent element as in \eqref{IIb:N}, and $S$ is a semisimple element generating boosts in the $(0,2)$ and $(1,3)$ planes. For $a\neq0$, \eqref{IIa:K} has eigenvalues $\pm a,\pm a,0$ with a nontrivial Jordan block. It is therefore conjugate to the representative of Table~1, which has the same Casimir invariants $C_2=-2a^2$ and $C_4=8a^4$ (cf.\ Appendix~{B}). The centralizer of $K$ is spanned by $S$ and $N$.

Since orbits $II_{a\neq 0}$ are equivalent up to rescaling, we pick $a=1$ for convenience.  As for type $II_b$, we straighten $K$ together with $N$. For $\gl r<1$, a convenient embedding is
\begin{subequations}\label{IIa:emb}
\begin{align}
    &X^0\pm X^2=\frac{e^{\mp\gl t}}{2\sqrt2}\Big[\chi+2R\pm2\chi\,(2\varphi-\gl t)\Big]\,,\qquad
    X^3\pm X^1=\frac{e^{\mp\gl t}}{2\sqrt2}\Big[\chi-2R\mp2\chi\,(2\varphi-\gl t)\Big]\,,\\
    &X^4=r\sinh\eta\,,\qquad \chi=\cosh\eta\sqrt{\gl^{-2}-r^2}\,,\qquad R=\frac{\gl^{-2}+r^2\sinh^2\eta}{\chi}\,.
\end{align}
\end{subequations}
It satisfies the hyperboloid condition, and
\begin{equation}
    \partial_t=-\gl K\,,\qquad \partial_\varphi=2N\,.
\end{equation}
For $\gl r>1$, the embedding is obtained by $X^0\leftrightarrow X^1$, $X^2\leftrightarrow X^3$, $2\varphi-\gl t\to-(2\varphi-\gl t)$ and $\chi=\cosh\eta\sqrt{r^2-\gl^{-2}}$. The inner coordinates follow from
\begin{subequations}
\begin{align}
    &e^{2\gl t}=\frac{X^0-X^2+X^3-X^1}{X^0+X^2+X^3+X^1}\,,\quad \chi^2=2\big(X^0+X^2+X^3+X^1\big)\big(X^0-X^2+X^3-X^1\big)\,,\\
    &r\sinh\eta=X^4\,.
  \end{align}  
\end{subequations}
The AdS metric reads
\begin{equation}\label{IIa:AdS}
    d\overline{s}^2=-\frac{\Delta_r}{\rho^2}\big(dt-\gl^{-1}\cosh^2\eta\,d\varphi\big)^2+\frac{\rho^2dr^2}{\Delta_r}+\frac{\rho^2d\eta^2}{\cosh^2\eta}+\frac{\cosh^4\eta}{\rho^2}\big(\gl^{-1}dt+(r^2-\gl^{-2})d\varphi\big)^2\,,
\end{equation}
with
\begin{equation}
    \Delta_r=\frac{(1-\gl^2r^2)^2}{\gl^2}\,,\qquad \rho^2=r^2+\gl^{-2}\sinh^2\eta\,.
\end{equation}
Formally, it follows from \eqref{IIb:AdS:g} by $(t,r,\theta,\varphi)\to(it,ir,\tfrac\pi2+i\eta,i\varphi)$. In the Carter--Pleba\'nski form, with $y=\gl^{-1}\sinh\eta\in\mathbb R$, one has $\Delta_y=(1+\gl^2y^2)^2/\gl^2>0$, while $\Delta_r$ has the double root $\gl r=1$. Compared to type $II_b$, the double root has moved from the angular function to the radial one. As before, $\partial_\varphi=2N$ is a null Killing vector.

With the vierbein
\begin{subequations}\label{IIa:e}
\begin{align}
    &\e^0=\frac{\sqrt{\Delta_r}}{\rho}\big(dt-\gl^{-1}\cosh^2\eta\,d\varphi\big)\,,\qquad \e^1=\frac{\rho\,dr}{\sqrt{\Delta_r}}\,,\\ &\e^2=\frac{\rho\,d\eta}{\cosh\eta}\,,\qquad \e^3=\frac{\cosh^2\eta}{\rho}\big(\gl^{-1}dt+(r^2-\gl^{-2})d\varphi\big)\,,
\end{align}    
\end{subequations}
the Killing vector $v=\gl K=-\partial_t$ has the components
\begin{subequations}
\begin{align}
    &v_{\ga\dgb}=\frac{1}{\gl\rho}\begin{pmatrix}|1-\gl^2r^2|+\cosh^2\eta & 0\\ 0 & |1-\gl^2r^2|-\cosh^2\eta\end{pmatrix}\,,\\
    &\gk_{\ga\gb}=\gl^2\big(r-i\gl^{-1}\sinh\eta\big)\begin{pmatrix}1&0\\0&-1\end{pmatrix}\,,
\end{align}    
\end{subequations}
and $C_2=-2$, $C_4=8$. The Kerr--Schild vectors \eqref{KS:spin}, normalized by \eqref{KS:norm}, the zeroth copy and the Weyl tensor are
\begin{subequations}\label{IIa:KS}
\begin{align}
    &k^\pm_\mu dx^\mu=dt-\gl^{-1}\cosh^2\eta\,d\varphi\mp\frac{\rho^2}{\Delta_r}\,dr\,,\\
    &l^\pm_\mu dx^\mu=dt+\gl(r^2-\gl^{-2})\,d\varphi\pm i\,\frac{\gl\rho^2}{\cosh^3\eta}\,d\eta\,,\\
    &\phif=\frac{r}{r^2+\gl^{-2}\sinh^2\eta}\,,\qquad C_{\ga(4)}\xi^{\ga(4)}=M\gl^4\frac{(\xi_1^2-\xi_2^2)^2}{(r-i\gl^{-1}\sinh\eta)^3}\,.
\end{align}
\end{subequations}
The Kerr--Schild metric $\bar g+2M\phif kk$ is brought to \eqref{IIa:AdS} with $\Delta_r\to\Delta_r-2Mr$ by a change of $t$ and $\varphi$ by functions of $r$. This is the Carter--Pleba\'nski metric with
\begin{equation}
    Q(r)=\frac{(1-\gl^2r^2)^2}{\gl^2}-2Mr\,,\qquad P(y)=\frac{(1+\gl^2y^2)^2}{\gl^2}\,,
\end{equation}
i.e., the member $\alpha=\gl^{-2}$, $\epsilon=-2$ of the family studied in \cite{Klemm:2014rda}, in which $P$ has no real roots. Such solutions were identified there as rotating generalizations of AdS black branes \cite{Klemm:1998kd}. The super-entropic black hole of type $II_b$ is the member $\alpha=\gl^{-2}$, $\epsilon=+2$ of the same family. In the present case, the horizon $Q(r_+)=0$ is non-compact, with $\eta\in\mathbb R$. 

\paragraph{General $a$ and the limit $a\to -0$.} Similarly to type $II_b$, the limit $II_{a\to 0}$ is ill-defined in the Penrose transform. To address its regularization, we keep $a<0$ arbitrarily small.  In the chart \eqref{IIa:emb}, the Killing vector of \eqref{IIa:K} is
\begin{equation}\label{IIa:va}
    v=\gl K=-a\,\partial_t+\frac{\gl}{2}(1-a)\,\partial_\varphi\,.
\end{equation}
For any $a\neq0$, $K$ is conjugate to $|a|$ times the $a=1$ representative by a boost that commutes with $S$ and rescales $N$, combined, for $a<0$, with an element that flips the sign of $a$ (Appendix~{B}). All double copies with $a\neq0$ are therefore the solution above, in different charts. As $a\to0$, \eqref{IIa:va} tends to the null Killing vector $\tfrac\gl2\partial_\varphi=\gl N$. For $\gl r<1$ and with
\begin{equation}
    \mathrm z:=r-i\gl^{-1}\sinh\eta\,,\qquad |\mathrm z|^2=\rho^2\,,
\end{equation}
the components in the vierbein \eqref{IIa:e} are, exactly in $a$,
\begin{subequations}\label{IIa:Kcomponents}
\begin{align}
    &v_{\ga\dgb}=\frac{1}{\gl\rho}\begin{pmatrix}\cosh^2\eta\,(1-\gl^2r^2)+a\,(1+\gl^2r^2\sinh^2\eta) & 0\\ 0 & -a\,\gl^2\rho^2\end{pmatrix}\,,\\
    &\gk_{\ga\gb}=\gl^2\begin{pmatrix}\dfrac{a\,\mathrm z^2-(1-a)\chi^2}{\mathrm z} & 0\\ 0 & -a\,\mathrm z\end{pmatrix}\,,
\end{align}
\end{subequations}
so that
\begin{equation}
    \rf^2=a\big(a\,\mathrm z^2-(1-a)\chi^2\big)\,,\qquad C_2=-2a^2\,,\qquad C_4=8a^4\,.
\end{equation}
The Weyl tensor \eqref{Weyl:DC} reads
\begin{equation}\label{IIa:Weyl:gen}
    C_{\ga(4)}\xi^{\ga(4)}=M\gl^4\,\frac{\Big(\big(a\,\mathrm z^2-(1-a)\chi^2\big)\mathrm z^{-1}\xi_1^2-a\,\mathrm z\,\xi_2^2\Big)^2}{\big(a(a\,\mathrm z^2-(1-a)\chi^2)\big)^{\frac52}}\,.
\end{equation}
For $a\to-0$, $\rf^2=-a\chi^2+O(a^2)$, so $\rf\simeq\sqrt{-a}\,\chi$ is real, and the Weyl tensor becomes type N:
\begin{equation}\label{IIa:Weyl:lim}
    C_{\ga(4)}\xi^{\ga(4)}\Big|_{a\to-0}=\frac{M}{(-a)^{\frac52}}\,\frac{\gl^4\,\xi_1^4}{\mathrm z^2\chi}\,.
\end{equation}
The divergent factor is absorbed by the renormalization of $M$ discussed in the next section.
}
\subsection{Type $II_{a=0}$: The Siklos spacetime}
{
In the limit $a\to-0$ of type $II_a$, the zeroth copy and the Kerr--Schild vectors in the chart \eqref{IIa:emb} behave as
\begin{equation}\label{lim2a-zero-copy}
    \phif=\frac{1}{\sqrt{-a}\,\chi}\big(1+O(a)\big)\,,
\end{equation}
\begin{equation}\label{lim2a-one-copy}
    k^\pm_\mu dx^\mu=\frac{1}{2a}\,dt\mp\frac{1}{\sqrt{-a}}\,\frac{d\chi}{\gl^2\chi^2}+O(1)\,,\qquad
    l^\pm_\mu dx^\mu=-\frac{\gl^2\chi^2}{2a^2}\,dt+O\big(|a|^{-\frac32}\big)\,.
\end{equation}
As for type $II_b$, the leading terms are proportional to the dual of the null Killing vector $N$, so only one independent Kerr--Schild vector survives. With $M=2(-a)^{\frac52}\hat M$, which keeps \eqref{IIa:Weyl:lim} finite, the Kerr--Schild metric converges to
\begin{equation}\label{IIa0:metric}
    ds^2=d\overline{s}^2+\frac{\hat M}{\chi}\,dt^2\,,\qquad \chi=\cosh\eta\sqrt{\gl^{-2}-r^2}\,,
\end{equation}
where the AdS metric is \eqref{IIa:AdS} and the corrections are of order $\sqrt{-a}$. This is a Petrov type N space-time. The Siklos coordinates are reached in one step:
\begin{equation}
    u=e^{2\gl t}\,,\qquad v=-\Big(2\varphi-\gl t+\frac{R}{\chi}\Big)\,,\qquad x=\frac{2e^{\gl t}r\sinh\eta}{\chi}\,,\qquad r'=\frac{2e^{\gl t}}{\gl\chi}\,.
\end{equation}
In these coordinates, after dropping the prime and renaming $\gl\hat M/8\to M$, the metric takes the Siklos form:
\begin{equation}\label{IIa:Siklos}
    ds^2 = \frac{1}{\gl^2 r^2}(dr^2 + dx^2 + 2 du dv + H du^2)\,, \qquad H = M \frac{r^3}{u^{\frac{5}{2}}}\,.
\end{equation}
For $\hat M=0$, the same map brings \eqref{IIa:AdS} to AdS in Poincar\'e coordinates, with $H=0$.
}
This is the Einstein-Siklos space-time that admits four isometries. In the notation of \cite{Calvaruso:2022zad}, a convenient basis of the global symmetry vector fields is 
\begin{align}   
    &K = -K_1 = -\partial_v\,, \quad P_2 = K_4 = \partial_x\,,\\
    &P_1 = K_5 = x \partial_v - u \partial_x\,, \quad 
    J  = -\half K_{6, \alpha = -2} = - 2u \partial_u - x \partial_x - r \partial_r\,.
\end{align}
These generators span the indecomposable four-dimensional Lie algebra, which can be identified with the Lie algebra $\mathfrak{g}^{-1}_{4.8}$ in the classification of \cite{ayad2026pseudoriemannianalgebraicriccisolitons} with the identification $e_1 = K$, $e_2 = P_1$, $e_3 = P_2$, $e_4 = J$. Equivalently, this algebra may be described as the semidirect sum $\mathfrak{h}_3 \rtimes \mathfrak{so}(1,1)$, where \(\mathfrak{h}_3=\langle K, P_1, P_2\rangle\) is the three-dimensional Heisenberg algebra, while the one-dimensional factor is generated by \(J\). Its action on \(\mathfrak{h}_3\) is given by the adjoint derivation $D:=\operatorname{ad}_{J}\big|_{\mathfrak h_3} = diag(0, 1, -1),$
as follows:
\begin{equation}
    [K, P_i] =[K, J]=0\,,\quad
    [P_1, P_2] = K\,, \quad 
    [J, P_1] = -P_1\,,\quad
    [J, P_2] = P_2\,.
\end{equation}
The same symmetry algebra also appears in the algebraic Ricci soliton setup for the left-invariant pseudo-Riemannian metrics on Lie groups considered in \cite{ayad2026pseudoriemannianalgebraicriccisolitons}.

\subsection{Special case with the isometry group $G_6$ and $G_5$}\label{max-symmetry-orbit}

In this section, we will examine more the puzzling cases  $II_{a=0}$ and $II_{b=0}$, representing the same orbit $II^0$ (see Table 1a). The isometry algebra -- defined as the centralizer of the corresponding global symmetry parameter -- is generated by six linearly independent Killing vector fields. This is in contrast to the actual isometries of the spacetimes obtained in the limiting procedure, which are only four and arise from the respective orbits $II_{b\to 0}$ and $II_{a\to 0}$. As previously explained, this discrepancy occurs because of the degenerate nature of the Penrose transform, which causes two of the isometries to be anomalous, while the two limiting double copies inequivalent despite $II_{a=0}=II_{b=0}$.

The type $II^0$ generators of the stabilizer algebra  can be chosen in such a way that we obtain the standard commutation relations for the Lie subalgebra \(\mathfrak{sl}(2, \mathbb{R}) \cong \so(2, 1)\):
\begin{equation}
[e_1, e_2] = -2e_3, \quad [e_2, e_3] = -2e_1, \quad [e_3, e_1] = 2e_2.
\end{equation}
with the additional elements $\langle P_1, P_2, K \rangle$:
\begin{equation}
[P_1, P_2] = -K, \quad [K, P_{1,2}] =[K,e_{i}]= 0\,,
\end{equation}
where 
\begin{equation}\label{K:G5-G6}
    K=J_{02}-J_{01}\pm (J_{13}-J_{23})\,.
\end{equation}
The remaining commutation relations form the action of \(\mathfrak{sl}(2, \mathbb{R})\) on \(P_1\) and \(P_2\):
\begin{equation}
\begin{aligned}
[e_1, P_1] &= \pm P_2\,, & [e_1, P_2] &= \pm P_1\,, \\
[e_2, P_1] &= -P_2\,, & [e_2, P_2] &= P_1\,, \\
[e_3, P_1] &= \pm P_1\,, & [e_3, P_2] &= \mp P_2\,,
\end{aligned}
\end{equation}
where the sign $\pm$ corresponds to whether they are derived from type $\rom{2}_{a = 0}$ or $\rom{2}_{b = 0}$, respectively.
From the commutation relations, it follows that the generators \(P_1\) and \(P_2\) form a two-dimensional representation of the algebra \(\mathfrak{sl}(2, \mathbb{R})\), where the generator \(e_1\) acts as a rotation between \(P_1\) and \(P_2\), and the element \(e_3\) acts as a diagonal operator.
The vector space $\mathfrak{v} = \mathrm{span}\{P_1, P_2\}$ on which \(\mathfrak{sl}(2, \mathbb{R})\) acts in the fundamental representation.
The element $K$ arises via the relation
\begin{equation}
   [P_1, P_2] = -K 
\end{equation}
as a one-dimensional central extension of the algebra. 


According to the commutation relations, the basis elements $\langle P_1, P_2, K \rangle$ correspond exactly to type $\rom{2}$ in the Bianchi classification of three-dimensional Lie algebras, which is indecomposable and is called the Lie algebra $\mathfrak{h}_3$ of the Heisenberg group, while $\mathfrak{sl}(2, \mathbb{R})$ has type $\rom{8}$ in the same classification. The complete algebra is the semidirect sum of these two algebras:
\begin{equation}\label{IIdeg:alg}
\mathfrak{g} = \mathfrak{sl}(2, \mathbb{R}) \ltimes \mathfrak{h}_3\,.
\end{equation}
This algebra describes the distinct, so called, minimal nilpotent orbits in a semisimple real algebra with the dimension dim $\mathcal{O}_{min}$ = dim $\so(2, 3) - $dim $\mathfrak{g}$ = 4.

Now, let us derive a proper coordinate chart for these two special cases as we did before. Explicitly writing the Killing field
\begin{equation}\label{K:IIdeg}
    K = (X^1 + X^2)(\partial_0 \pm \partial_3) + (X^0 \mp X^3)(\partial_2 - \partial_1)\,,
\end{equation}
introduce the new coordinates
\begin{equation}
    p = X^1 + X^2\,, \quad q = X^1 - X^2\,, \quad u = X^0 \pm X^3\,, \quad v = X^0 \mp X^3\,.
\end{equation}
In both cases, this results in
\begin{align}
    &K = p\partial_u - v \partial_q\,,\\
     &-uv -pq + X_4^2 = -1\,.
\end{align}
The canonical coordinates require the field $K$ to be $K = \frac{\partial}{\partial \psi}$, while the three other coordinates, denoted as $\phi_i$, $i=1,2,3$, do not depend on $\psi$
\begin{equation}
    K \psi =1\,, \quad K \phi_i = 0\,. 
\end{equation}
Solving the system, we can take the following solutions as coordinates:
\begin{equation}
    \psi = \frac{u}{p},\quad \chi = \frac{v}{p},\quad \phi = \frac{X^4}{p},\quad p\,.
\end{equation}
They give us the AdS metric in the form
\begin{equation}
    d\bar s^2 = -dudv - dpdq + dX_4^2 = p^2(d\phi^2 -d\chi d\psi) + \frac{dp^2}{p^2}\,.
\end{equation}
The metric can be further transformed within the canonical setting. Substituting $p = \frac{1}{z}$ and inverting the coordinates, we write out a new set of inner coordinates
\begin{equation}
    u = \frac{\psi}{z}\,,\quad v = \frac{\chi}{z}\,,\quad p = \frac{1}{z}\,,\quad 
    q = z + \frac{\phi^2 - \psi \chi}{z}\,.
\end{equation}
The generating field $K = \partial_\psi$ is still canonical, whereas the coordinates describing this case are almost Poincar\'{e} $(\psi, \chi, \phi, z)$:
\begin{equation}
    d\bar s^2 = \frac{1}{z^2}(dz^2 + d\phi^2 - d\chi d\psi)\,.
\end{equation}
The standard Poincar\'{e} form of the metric follows from defining
\begin{equation}
   \psi = t + x\,,\quad \chi = t - x\,,\quad y = \phi\,. 
\end{equation}
Consequently, in the coordinates $(t, x, y, z)$, we have:
\begin{equation}
    d\s^2 = \frac{1}{z^2}(-dt^2 + dz^2 + dx^2 + dy^2)\,,\qquad  v^{\mu} = (1, 1, 0, 0)\,,
\end{equation}
where we recall $v^{\mu}$ is the Killing vector associated with the symmetry parameter \eqref{K:IIdeg}. The convenient vierbein is diagonal 
\begin{equation}
\e^a_{\mu} = \frac{1}{z}\delta^a_{\mu}\,,\qquad  \e^{\mu}_a = z\delta^{\mu}_a\,.    
\end{equation}
The corresponding global symmetry components are given by:
\begin{equation}\label{IIdeg:K}
v_{\ga \dot{\gb}} =   \frac{1}{z} \begin{pmatrix}
      1 & 1\\
      1 & 1
  \end{pmatrix}\,,\quad
     \varkappa_{\ga \gb} = \frac{1}{z}\begin{pmatrix}
      1 & -1\\
      -1 & 1
    \end{pmatrix}\,.
\end{equation}
Both matrices are degenerate
\begin{equation}\label{IIdeg:v}
    \det v_{\al\dgb}=\det\gk_{\al\gb}=0\,.
\end{equation}
The Casimir invariants also vanish 
\begin{equation}
    C_2=C_4=0\,.
\end{equation}
The degeneracy in \eqref{IIdeg:v} makes our procedure for generating solutions via the Penrose transform ill-defined. Specifically, Eqs. \eqref{C:Dtype} and \eqref{KS:spin} require regularization. This regularization was performed in Secs. \ref{sec:IIb} and \ref{sec:IIa} with the resulting Siklos spacetimes that  have isometries as subalgebras of \eqref{IIdeg:alg}.    

Interestingly, having the degenerate global symmetry parameter \eqref{IIdeg:K}, one can construct spacetime with the exact global symmetry \eqref{IIdeg:alg} in the Kerr-Schild form.  Although the Kerr-Schild vectors \eqref{KS:spin} are not well-defined in view of the denominator $v^- v^+ = 0$, we note that the Killing vector $v^{a}$ itself is null \eqref{IIdeg:K} and geodesic
\begin{equation}
    v^\mu v_\mu=v^{\mu}\nabla_{\mu}v_{\nu}=0\,.
\end{equation}
This motivates us to identify the Kerr-Schild vector with the Killing vector $k^{\mu}=z \cdot v^\mu$. In the coordinates $(t, x, y, z)$, this gives us
\begin{equation}\label{KS=v}
k_{\mu} = \frac{1}{z}(-1, 1, 0, 0)\,.    
\end{equation}
Notice that the regularized Kerr-Schild vectors obtained in type $II_{a=0}$ and $II_{b=0}$ in \eqref{lim2b-one-copy} and \eqref{lim2a-one-copy} differ from \eqref{KS=v}. Another comment is that in this special case, the geodesic condition 
\begin{equation}
   f(z)k^\mu \nabla_\mu (f(z) k_\nu) = 0 
\end{equation}
holds also for any function $f(z)$. We thus consider the candidate for the Kerr-Schild metric
\begin{equation}
g_{\mu \nu} = \overline{g}_{\mu \nu} + M \phi k_\mu k_\nu\,,\qquad \phi = z^{\ga}\,,\quad \ga \in \mathbb{R}    
\end{equation}
with some power-law profile $\phi$ and an arbitrary constant $M$. The resulting metric satisfies Einstein's vacuum equations only for the following parameter values: $\ga = 0$ and $\al = 3$. In these cases, the metric takes the form
\begin{equation}
    ds^2 = \frac{1}{z^2}(-dt^2 + dx^2 + dy^2 + dz^2) + Mz^{\ga} \frac{(dx - dt)^2}{z^2}\,.
\end{equation}
Choosing the light-cone coordinates
\begin{equation}
u = \frac{x - t}{\sqrt{2}}\,,\quad v = \frac{x + t}{\sqrt{2}}\,,\quad r = z\,,\quad x = y\,, 
\end{equation}
we arrive at the standard form of the Siklos metric:
\begin{equation}
    ds^2 = \frac{1}{r^2}(dr^2 + dx^2 + 2dudv + Hdu^2)\,.
\end{equation}
where $H  =  M r^\ga$ is the conventional profile function appearing in the Siklos metric. There are analogs of a single copy and a zero copy represented by the function $\phi$. Their equations of motion take a characteristic form of Siklos-type geometries: 
\begin{align}
    &{\nabla}_{\mu} F^{\mu \nu} - 2 \phi^{\nu} = 0, \\
    &{\nabla}^2 \phi = 0\,,
\end{align}
where
\begin{equation}
   F_{\mu\nu} = \nabla_{[\mu} \phi_{\nu]}\,. 
\end{equation}
Notice that the current conservation of a single copy is violated by the Ricci scalar term. Recall that $\al=0$ or $\al=3$. Consider these two cases separately.
\begin{itemize}
    \item {\bf Case} $\ga = 0$ 
    
    In this case, we have 
    \begin{equation}
        H = M\,,    
    \end{equation}
    while the Weyl tensor vanishes
    \begin{equation}
        C_{\alpha(4)} = 0\,.
    \end{equation}
     The zero copy is a constant $\phi = \pm 1$. So, the spacetime is AdS.

    \item {\bf Case} $\ga = 3$: 
    
    The Siklos function $H$ and the Weyl tensor are nontrivial 
    \begin{equation}\label{Kaigorodov:prof}
        H = Mr^3\,,\qquad C_{\alpha(4)}\xi^{\alpha(4)} = 24M\frac{(\xi_1 - \xi_2)^4}{M + \frac{1}{z^3}}\,,
    \end{equation}
    where $\xi^{\al}=(\xi_1, \xi_2)$ is an arbitrary test spinor. The corresponding spacetime is described by the so-called Kaigorodov metric \cite{Kaigorodov} (a special case of the Siklos metric) with five Killing vectors; see also \cite{Calvaruso:2022zad}, \cite{Podolsky:1997ik}. As we mentioned, this metric is not included in our classification. The wave fronts of this solution are planes perpendicular to the direction $v$. It is argued to be the only nontrivial vacuum space conformal to pp-waves; see  \cite{Carrillo-Gonzalez:2017iyj}. 

    As a side remark in connection with solutions obtained by regularizing types $II_{a=0}$ and $II_{b=0}$, it is of interest to observe that the Kaigorodov profile \eqref{Kaigorodov:prof} is invariant under a translation $\partial_u$, which turns out to form a parabolic element in its stabilizer-part $\sl(2,\mathbb{R})$, while the similar isometry is hyperbolic in the limit $II_{a=0}$,  ($J_{03}+J_{12}$), and it is elliptic for $II_{b=0}$,  $(J_{01}+J_{23})$. 
\end{itemize}

Interestingly, the case $\al=-2$, while not an Einstein vacuum, corresponds to a spacetime with six isometries, which are exactly those that emerge as the center of \eqref{K:G5-G6}.  In this case, the profile function is $H = M r^{-2}$ and the metric reads
\begin{equation}
    ds^2 = \frac{1}{r^2}(dr^2 + dx^2 + 2dudv + Mr^{-2} du^2).
\end{equation}
This ansatz solves Einstein's equations with the stress tensor $T_{\mu \nu} \sim k_\mu k_\nu$, which corresponds to the radiation field propagating along the principal light-like direction $k_\mu$. Such a nonvacuum solution is known as Defrise spacetime \cite{Defrise1969}.

\section{TYPE $III$}\label{sec:III}

Type $III$ consists of two subtypes: type $III_a$, which contains two orbits $III_{a\neq 0}$ and $III_{a=0}$, and type $III_b$, which contains two orbits $III_{b\neq 0}$ and $III_{b=0}$. The respective generating elements from $\so(2,3)$ are given in Table 1. For  $a\neq 0$ and $b\neq 0$ , there are four Kerr-Schild vectors \eqref{KS:spin} that enable us to construct a family of GR solutions. This can be done using either the complex double Kerr-Schild representation for the metric \eqref{Double KS}, or the real representation using equation \eqref{metric:gen}. For both types, the centralizer that defines the global symmetries of the resulting double copies consists of two commuting isometries. However, this condition holds only when  $a\neq 0$ and $b\neq 0$.

The family admits two special subcases, corresponding to the points $a = 0$ and $b = 0$, denoted by $III^+_0=\rom{3}_{a = 0}$ and $III^-_0=\rom{3}_{b = 0}$ in Table 1a, that feature the global symmetry enhancement. A direct analysis of the corresponding centralizers shows that, in both cases, the symmetry algebra has dimension four. For type $III_{a=0}$, this  algebra is a direct sum of the pseudo Euclidean motions in two dimensions and a translation, $\iso(1,1)\oplus\u(1)$, while for type $III_{b=0}$, the algebra represents isometries of two-dimensional Euclidean plane and time translation  $\iso(2)\oplus\u(1)$. Type $III_{b=0}$ generates a black hole with a planar horizon (black brane, see, {\it e.g.,} \cite{Birmingham1999, Cai_1996}, which is known to require a negative cosmological constant $\Lambda<0$. For this sub-case the real Kerr-Schild vectors exist, while the complex ones do not. In contrast, Type $III_{a=0}$ reproduces the AdS soliton \cite{Horowitz_1998} that admits only a pair of complex Kerr-Schild vectors.  

Solutions from type $III_{a\neq 0}$ will be shown to originate from a zero-rotation contraction of Klemm, Moretti and Vanzo topological black hole (KMV) \cite{Klemm:1997ea}. Solutions from $III_{b\neq 0}$ in turn arise as an infinite rotation limit of KMV.

\subsection{Type $\rom{3}_b$: infinitely rotating KMV}
{
Consider the case with $b\neq 0$. Its generating parameter reads
\begin{equation}\label{K:IIIb}
     K = J_{02} - b J_{34} + J_{10}=N-b\,J_{34}\,,\qquad N:=J_{02}-J_{01}\,,
\end{equation}
where $N$ is nilpotent of Jordan type, $N^3=0\neq N^2$. Apart from $K$ itself, the centralizer contains a single generator
\begin{equation}
G =  J_{34}\,,\qquad [K,G]=[N,G]=0\,.
\end{equation}
In terms of $u=X^1-X^2$, $v=X^1+X^2$ and $p=X^0$, with $-uv-p^2+(X^3)^2+(X^4)^2=-\gl^{-2}$, the vector fields are
\begin{equation}\label{K,G:IIIb}
    K_b = (p\partial_u - v \partial_p) + b(X^4 \partial_3 - X^3 \partial_4)\,,\qquad G = X^4 \partial_3 - X^3 \partial_4\,.
\end{equation}
The eigenvalues of $K$ are $0,0,0,\pm ib$, and the Casimir invariants are $C_2=b^2$ and $C_4=b^4$. By Appendix~{B}, $K_b=|b|\,e^{sH}K_{\pm1}e^{-sH}$ with $H=-J_{12}$ and $e^s=1/|b|$, and the sign of $b$ is removed by a reflection. Hence all $III_b$ with $b\neq0$ give the same double copy up to the mass normalization, and we set $b=1$.

To introduce the appropriate inner coordinates, we straighten $K$ with $b=1$ together with the compact generator $G=J_{34}$:
\begin{equation}\label{IIIb:straight}
    -\gl K\,t=1\,,\quad K\,r=K\,\theta=K\,\varphi=0\,,\qquad G\,\varphi=-1\,,\quad G\,t=G\,r=G\,\theta=0\,.
\end{equation}
The joint invariants of $K$ and $G$ are $X^1+X^2$ and $(X^3)^2+(X^4)^2$. A solution of \eqref{IIIb:straight} is
\begin{subequations}\label{IIIb:emb}
\begin{align}
    &X^0=\gl\,r\,t\cosh\theta\,,\qquad X^1+X^2=-r\cosh\theta\,,\\
    &X^2-X^1=\frac{\gl^{-2}\cosh^2\theta+r^2\sinh^2\theta-\gl^2r^2t^2\cosh^2\theta} {r\cosh\theta}\,,\\
    &X^4+iX^3=\gl^{-1}\sqrt{1+\gl^2r^2}\,\sinh\theta\;e^{i(\gl t-\varphi)}\,,
\end{align}
\end{subequations}
with $r>0$ and $\theta>0$. It satisfies the hyperboloid condition identically, and
\begin{equation}
    \partial_t=-\gl K_1\,,\qquad \partial_\varphi=-J_{34}\,,\qquad N_3=-\gl^{-1}\partial_t-\partial_\varphi\,.
\end{equation}
The AdS metric reads
\begin{equation}\label{IIIb:AdS}
    d\overline{s}^2=-\frac{\Delta_r}{\rho^2}\big(dt+\gl^{-1}\sinh^2\theta\,d\varphi\big)^2+\frac{\rho^2dr^2}{\Delta_r}+\frac{\rho^2d\theta^2}{\cosh^2\theta}+\frac{\cosh^2\theta\sinh^2\theta}{\rho^2}\big(\gl^{-1}dt-(r^2+\gl^{-2})d\varphi\big)^2\,,
\end{equation}
with
\begin{equation}
    \Delta_r=r^2(1+\gl^2r^2)\,,\qquad \rho^2=r^2+\gl^{-2}\cosh^2\theta\,.
\end{equation}
At $\theta=0$ the compact generator $J_{34}$ has fixed points, $X^3=X^4=0$. Near $\theta=0$, the $(\theta,\varphi)$ part of \eqref{IIIb:AdS} at fixed $t$ is $(r^2+\gl^{-2})(d\theta^2+\theta^2d\varphi^2)$, so $\theta=0$ is a regular axis with $\varphi\sim\varphi+2\pi$. As shown below, \eqref{IIIb:emb} are the Boyer--Lindquist coordinates of the rotating topological black hole of Appendix~{D} in the limit of infinite rotation.

With the vierbein
\begin{subequations}
\begin{align}
    &\e^0=\frac{\sqrt{\Delta_r}}{\rho}\big(dt+\gl^{-1}\sinh^2\theta\,d\varphi\big)\,,\qquad \e^1=\frac{\rho\,dr}{\sqrt{\Delta_r}}\,,\\
    &\e^2=\frac{\rho\,d\theta}{\cosh\theta}\,,\qquad \e^3=\frac{\cosh\theta\sinh\theta}{\rho}\big(\gl^{-1}dt-(r^2+\gl^{-2})d\varphi\big)\,,
\end{align}
\end{subequations}
the Killing vector $v=\gl K_1=-\partial_t$, i.e., $v^\mu=(-1,0,0,0)$, has the components
\begin{subequations}\label{K:BBb}
\begin{align}
    &v_{\ga\dgb}=-\frac{1}{\rho}\begin{pmatrix}\sqrt{\Delta_r}+\gl^{-1}\cosh\theta\sinh\theta&0\\0&\sqrt{\Delta_r}-\gl^{-1}\cosh\theta\sinh\theta\end{pmatrix}\,,\\
    &\gk_{\ga\gb}=\gl^2\big(r-i\gl^{-1}\cosh\theta\big)\begin{pmatrix}1&0\\0&-1\end{pmatrix}\,,
    \end{align}
\end{subequations}
so that $\rf=r-i\gl^{-1}\cosh\theta$ and the Casimir invariants are $C_2=1$ and $C_4=1$. The Kerr--Schild vectors \eqref{KS:spin}, normalized by \eqref{KS:norm}, the zeroth copy \eqref{zero} and the Weyl tensor \eqref{Weyl:DC} are
\begin{subequations}\label{IIIb:KS}
\begin{align}
    &k^\pm_\mu dx^\mu=dt+\gl^{-1}\sinh^2\theta\,d\varphi\mp\frac{\rho^2}{\Delta_r}\,dr\,,\\
    &l^\pm_\mu dx^\mu=dt-\gl(r^2+\gl^{-2})\,d\varphi\pm i\,\frac{\gl\rho^2}{\cosh^2\theta\sinh\theta}\,d\theta\,,\\
    &\phif=\frac{r}{r^2+\gl^{-2}\cosh^2\theta}\,,\qquad C_{\ga(4)}\xi^{\ga(4)}=M\gl^4\frac{(\xi_1^2-\xi_2^2)^2}{(r-i\gl^{-1}\cosh\theta)^3}\,.\label{Weyl:BBb}
\end{align}
\end{subequations}
The standard change of coordinates
\begin{equation}
dt\to dt\pm\gl^{-2}(1+\gl^2r^2)\zeta\,dr\,,\qquad  d\varphi\to d\varphi\pm\gl^{-1}\zeta\,dr\,,\qquad \zeta={2}Mr/\big(\Delta_r(\Delta_r-{2}Mr)\big)    
\end{equation}
removes the $dr$ cross terms in the metric. It brings the Kerr--Schild metric {$g=\bar g+2M\phif kk$} to \eqref{IIIb:AdS} with
\begin{equation}\label{IIIb:full}
    \Delta_r\ \to\ r^2(1+\gl^2r^2)-{2}Mr\,.
\end{equation}

The metric obtained this way can be obtained as a certain contraction supplemented by sending the rotation paramter to infinity of the rotating topological black hole of Klemm, Moretti and Vanzo \cite{Klemm:1997ea}. In other words, this solution can be obtained by taking the limit of type $I_d$ (Appendix~{D}). Indeed, consider the $I_d$ metric in Boyer--Lindquist form,
\begin{multline}\label{IIIb:KMV}
    ds^2=-\frac{\Delta_r}{\rho^2\Xi^2}\big(dt+a\sinh^2\theta\,d\varphi\big)^2+\frac{\rho^2dr^2}{\Delta_r}+\frac{\rho^2d\theta^2}{\Delta_\theta}+\frac{\Delta_\theta\sinh^2\theta}{\rho^2\Xi^2}\big(a\,dt-(r^2+a^2)d\varphi\big)^2\,,\\
    \Delta_r=(r^2+a^2)(\gl^2r^2-1)-2\eta r\,,\qquad \Delta_\theta=1+\gl^2a^2\cosh^2\theta\,,\qquad \Xi=1+\gl^2a^2\,,\qquad \rho^2=r^2+a^2\cosh^2\theta\,,
\end{multline}
where $\eta$ is a massive parameter. The metric \eqref{IIIb:AdS}, \eqref{IIIb:full} resembles \eqref{IIIb:KMV}, but it does not follow from it by fixing $a$ at any particular value. It follows from a limit $a\to\infty$ combined with a rescaling of the coordinates and of the mass. Let $s=\sqrt{\gl^2a^2-1}$ and change the coordinates and the mass parameter of \eqref{IIIb:KMV} as
\begin{equation}\label{IIIb:KMVmap}
    r\to s\,r\,,\quad \cosh\theta\to\frac{s}{\gl a}\cosh\theta\,,\quad \varphi\to\frac{\gl a\,\Xi}{s^3}\,\varphi\,,\quad t\to\frac{\Xi}{s}\Big(t-\frac{\varphi}{\gl}\Big)+\frac{\gl a^2\Xi}{s^3}\,\varphi\,,\quad {\eta=s^3M}\,.
\end{equation}
This corresponds to the rescaling $K\to K/s$ of the $I_d$ generator. For any $a>\gl^{-1}$, Eq. \eqref{IIIb:KMV} becomes exactly
\begin{multline}\label{IIIb:KMVexact}
    ds^2=-\frac{\Delta_r+\tilde\gamma}{\rho^2}\big(dt+\gl^{-1}\sinh^2\theta\,d\varphi\big)^2+\frac{\rho^2dr^2}{\Delta_r+\tilde\gamma}+\frac{\rho^2\sinh^2\theta\,d\theta^2}{F}+\frac{F}{\rho^2}\big(\gl^{-1}dt-(r^2+\gl^{-2})d\varphi\big)^2\,,\\
    F=\cosh^2\theta\sinh^2\theta+\gl^2\tilde\gamma\,,\qquad \tilde\gamma=-\frac{a^2}{(\gl^2a^2-1)^2}\,,
\end{multline}
with 
\begin{equation}
\Delta_r=r^2(1+\gl^2r^2)-{2}Mr\,,\qquad \rho^2=r^2+\gl^{-2}\cosh^2\theta\,.    
\end{equation}
As $a\to\infty$, $\tilde\gamma\to0$ and \eqref{IIIb:KMVexact} tends to the type $III_b$ double copy \eqref{IIIb:AdS}, \eqref{IIIb:full}. The KMV axis $F=0$ moves to $\theta=0$, which becomes the regular axis of $J_{34}$. Hence, the type $III_b$ double copy is the contraction of KMV black hole at infinite rotation, $\gl a\to\infty$, with the mass parameter growing as {$\eta\simeq(\gl a)^3M$}, so that the limit remains well-defined. Therefore, the limit is a contraction of the orbit $I_d$ onto $III_b$.
}

\subsection{Type $III_{b=0}\,$: black brane}
In the limiting case $b\rightarrow 0$, the orbit \eqref{K:IIIb} generated by 
\begin{equation}\label{K:BB}
   K = J_{02} - J_{01} 
\end{equation}
becomes a pure nilpotent subregular orbit with $ dim (\mathcal{O}) = 6$; see \cite{McGovern1993NilpotentOrbits}. Its centralizer gets bigger compared to the case when $b\neq 0$. Specifically, apart from \eqref{K:BB}, there are three more generators 
\begin{equation}
    G_1 = J_{34}\,,\quad G_2 = J_{14} - J_{24}\,,\quad G_3 = J_{31} - J_{32}
\end{equation}
instead of a single $J_{34}$ for $b\neq 0$.
They define the algebra of motions of the Euclidean plane. Combined with $K$, the full algebra is \(\u(1) \oplus \iso(2)\), which is type $\rom{7}_0$ in the Bianchi classification \cite{Bergshoeff:2003ri}:
\begin{equation}
    [K, G_i] = 0\,,\quad [G_1, G_2] = G_3\,,\quad [G_3, G_1] = G_2\,,\quad [G_2, G_3] = 0\,.
\end{equation}
{
To derive the double copy, we straighten $K$ together with the Euclidean motions of the horizon plane. We look for inner coordinates $(\psi,z,\chi,\phi)$, with $x=\chi\cos\phi$ and $y=\chi\sin\phi$, such that
\begin{equation}
    K\psi=1\,,\qquad G_1\phi=1\,,\qquad G_2\,x=-1\,,\qquad G_3\,y=1\,,
\end{equation}
while $z$ is invariant under the whole centralizer. A solution is the planar embedding
\begin{subequations}\label{BB:emb}
\begin{align}
    &X^0=-\frac{\psi}{\gl z}\,,\qquad X^1+X^2=-\frac{1}{\gl z}\,,\qquad X^2-X^1=\frac{z}{\gl}+\frac{\chi^2-\psi^2}{\gl z}\,,\\
    &X^3=\frac{\chi\sin\phi}{\gl z}\,,\qquad X^4=\frac{\chi\cos\phi}{\gl z}\,,
\end{align}
\end{subequations}
which satisfies the hyperboloid condition identically. In these coordinates,
\begin{equation}
    K=\partial_\psi\,,\qquad G_1=\partial_\phi\,,\qquad G_2=-\partial_x\,,\qquad G_3=\partial_y\,,
\end{equation}
and the AdS metric takes the Poincar\'e form
\begin{equation}\label{AdS:IIIb}
d\overline{s}^2 = \frac{1}{\gl^2 z^2}\left(-d\psi^2 + dz^2 + d\chi^2 + \chi^2 d\phi^2\right)\,.
\end{equation}
With the vierbein $\e^a_{\ \mu}=(\gl z)^{-1}\mathrm{diag}(1,1,1,\chi)$, the Killing vector $v=\gl K$, i.e., $v^\mu=\gl(1,0,0,0)$, has the components
\begin{equation}
    v_{\ga\dgb}=\frac{1}{z}\begin{pmatrix}1&0\\0&1\end{pmatrix}\,,\qquad
    \gk_{\ga\gb}=\frac{\gl}{z}\begin{pmatrix}1&0\\0&-1\end{pmatrix}\,,
\end{equation}
so that $C_2=C_4=0$ and $\rf=(\gl z)^{-1}$.  
}

In this case, the only real pair of Kerr-Schild vectors survives when $b=0$. Specifically, 
\begin{equation}
    \gl k^{\pm}_\mu dx^{\mu} = -d \psi \pm dz\,.
\end{equation}
The zero copy also simplifies
\begin{equation}
    \phif = \gl z\,,
\end{equation}
as well as the Weyl tensor 
\begin{equation}
    C_{\alpha(4)}\xi^{\alpha(4)} = M z^3{\left(\xi_1^2 - \xi_2^2\right)^2}\,,
\end{equation}
where in the latter the factor $\gl^7$ is absorbed into $M$. The metric is obtained using the Kerr-Schild form \eqref{Kerr-Schild}. In so doing, one can redefine the coordinates
    \begin{align}\label{AIII:transform}
       \chi = \sqrt{x^2 + y^2}\,,\quad \tan \phi = \frac{y}{x}\,,\quad  d\psi = \gl dt + \frac{{2}\gl M z^3 }{1-{2}\gl M z^3}dz\,. 
    \end{align}
Introducing also the radial distance $r=(\gl z)^{-1}$, we arrive at the familiar metric
\begin{equation}\label{black brane}
    ds^2 =  -f(r)dt^2 + f(r)^{-1} dr^2 + r^2(dx^2 + dy^2)\,, \quad f(r) = \gl^2 r^2 - \frac{{2}M}{r}
\end{equation}
of a black hole with a planar horizon in AdS. Written in this standard form, it represents the so called AIII-metric with a cosmological constant \cite{Podolsk__2017}, where the parameters in \eqref{A-metrics} (see Appendix {E}) are set to
\begin{equation}
    \epsilon_0 = 1\,,\quad  \epsilon_2 = 0\,,\quad p = r\,,\quad M \rightarrow {-M}\,. 
\end{equation}
The Weyl curvature now reads
\begin{equation}
C_{\alpha(4)}\xi^{\alpha(4)} = M \frac{(\xi_1^2 - \xi_2^2)^2}{r^3}\,.
\end{equation}
Interestingly, in a black brane geometry, the scalar function $\phif$ enjoys 
\begin{equation}
    \boldsymbol{\nabla}^2\phif + 2\gl^2 \phif + {2}M \phif^4 = 0
\end{equation}
that can be verified by a direct computation, where the operator $\boldsymbol{\nabla}^2$ is built from the full metric \eqref{black brane}, rather than the background.

\subsection{Type $\rom{3}_a$: KMV at zero rotation limit}
{
For this type, the generating parameter reads
\begin{equation}\label{K:IIIa}
     K = J_{30} - a J_{14} + J_{23}=N-a\,J_{14}\,,\qquad N:=J_{23}-J_{03}\,,
\end{equation}
where $N$ is nilpotent of Jordan type $N^3=0\neq N^2$ and $J_{14}$ is a boost. For $a\neq0$, the centralizer is spanned by $K$ and
\begin{equation}
    G =  J_{14}\,, \qquad [K_a,G] = [N,G]=0\,.
\end{equation}
In terms of $u=X^0-X^2$, $v=X^0+X^2$ and $p=X^3$, with $-uv-(X^1)^2+p^2+(X^4)^2=-\gl^{-2}$,
\begin{equation}\label{K,G:IIIa}
    K_a = (p\partial_u + v \partial_p) + a(X^4 \partial_1 + X^1 \partial_4)\,,\qquad G = X^4 \partial_1 + X^1 \partial_4\,.
\end{equation}
The eigenvalues of $K_a$ are $0,0,0,\pm a$, and $C_2=-a^2$, $C_4=a^4$. By Appendix~{B}, $|a|$ is removed by the boost $H_a=-J_{02}$ and the sign of $a$ by a reflection, so we set $a=1$.

\paragraph{Wick rotation.} Types $III_a$ and $III_b$ are two real forms of the same complex orbit. The complex linear map
\begin{equation}\label{IIIa:Phi}
    \big(X^0,X^1,X^2,X^3,X^4\big)\ \to\ \big(-iX^3,\;X^0,\;X^2,\;iX^1,\;X^4\big)
\end{equation}
preserves $\eta_{\ua\ub}X^{\ua}X^{\ub}$, and it maps $iK_{III_b}\mapsto K_{III_{a}}|_{a=b}$, and the eigenvalues $\pm ib$ of \eqref{K:IIIb} become the real eigenvalues $\mp b$. Applied to the embedding \eqref{IIIb:emb}, the map \eqref{IIIa:Phi} amounts to the change of inner coordinates
\begin{equation}\label{IIIa:Wick}
    t\to i\,t\,,\qquad \varphi\to-i\,\varphi\,,\qquad r\to\gl^{-1}\sinh\theta\,,\qquad \cosh\theta\to\gl r\,.
\end{equation}
Besides the Wick rotation of the Killing coordinates, it exchanges the radial coordinate with the hyperbolic angle. For $\gl r>1$ and $\theta>0$, the result is the real embedding
\begin{subequations}\label{IIIa:emb}
\begin{align}
    &X^3=-\gl\,r\,t\sinh\theta\,,\qquad X^0+X^2=-r\sinh\theta\,,\\
    &X^2-X^0=\frac{r^2\cosh^2\theta-\gl^{-2}\sinh^2\theta+\gl^2r^2t^2\sinh^2\theta}{r\sinh\theta}\,,\\
    &X^4\pm X^1=\gl^{-1}\sqrt{\gl^2r^2-1}\,\cosh\theta\;e^{\pm(\gl t+\varphi)}\,.
\end{align}
\end{subequations}
For $\gl r<1$, one replaces $X^4\pm X^1$ by $X^1\pm X^4$ and $\sqrt{\gl^2r^2-1}$ by $\sqrt{1-\gl^2r^2}$. The chart straightens $K$ together with the boost $G$:
\begin{equation}
    \partial_t=-\gl K_1\,,\qquad \partial_\varphi=J_{14}\,,\qquad N_3'=-\gl^{-1}\partial_t+\partial_\varphi\,.
\end{equation}
The AdS metric is the Wick rotation of \eqref{IIIb:AdS}:
\begin{equation}\label{IIIa:AdS}
    d\overline{s}^2=-\frac{\Delta_r}{\rho^2}\big(dt+\gl^{-1}\cosh^2\theta\,d\varphi\big)^2+\frac{\rho^2dr^2}{\Delta_r}+\frac{\rho^2d\theta^2}{\sinh^2\theta}+\frac{\sinh^2\theta\cosh^2\theta}{\rho^2}\big(\gl^{-1}dt-(r^2-\gl^{-2})d\varphi\big)^2\,,
\end{equation}
with
\begin{equation}
    \Delta_r=r^2(\gl^2r^2-1)\,,\qquad \rho^2=r^2+\gl^{-2}\sinh^2\theta\,.
\end{equation}
Compared with \eqref{IIIb:AdS}, the factors $1+\gl^2r^2$ and $\cosh\theta$ are replaced by $\gl^2r^2-1$ and $\sinh\theta$. Now $\Delta_r$ has a simple root at $\gl r=1$, the Killing horizon of $K$ in AdS, while the angular function has a double root at $\theta=0$, so there is no axis. Since $G$ is a boost, $\varphi$ is non-compact; it can be compactified only in a quotient of AdS, as for topological black holes.

With the vierbein
\begin{subequations}
\begin{align}
    &\e^0=\frac{\sqrt{\Delta_r}}{\rho}\big(dt+\gl^{-1}\cosh^2\theta\,d\varphi\big)\,,\qquad \e^1=\frac{\rho\,dr}{\sqrt{\Delta_r}}\,,\\
    &\e^2=\frac{\rho\,d\theta}{\sinh\theta}\,,\qquad \e^3=\frac{\sinh\theta\cosh\theta}{\rho}\big(\gl^{-1}dt-(r^2-\gl^{-2})d\varphi\big)\,,
\end{align}
\end{subequations}
the Killing vector $v=\gl K_1=-\partial_t$ has the components
\begin{subequations}
\begin{align}
    &v_{\ga\dgb}=-\frac{1}{\rho}\begin{pmatrix}\sqrt{\Delta_r}+\gl^{-1}\sinh\theta\cosh\theta&0\\0&\sqrt{\Delta_r}-\gl^{-1}\sinh\theta\cosh\theta\end{pmatrix}\,,\\
    &\gk_{\ga\gb}=\gl^2\big(r-i\gl^{-1}\sinh\theta\big)\begin{pmatrix}1&0\\0&-1\end{pmatrix}\,,
    \end{align}
\end{subequations}
so that 
\begin{equation}
   \rf=r-i\gl^{-1}\sinh\theta\,,\qquad C_2=-1\,,\qquad C_4=1\,.
\end{equation}
The Kerr--Schild vectors \eqref{KS:spin}, normalized by \eqref{KS:norm}, the zeroth copy \eqref{zero} and its dual \eqref{dual:zero} are
\begin{subequations}\label{IIIa:KS}
\begin{align}
    &k^\pm_\mu dx^\mu=dt+\gl^{-1}\cosh^2\theta\,d\varphi\mp\frac{\rho^2}{\Delta_r}\,dr\,,\\
    &l^\pm_\mu dx^\mu=dt-\gl(r^2-\gl^{-2})\,d\varphi\pm i\,\frac{\gl\rho^2}{\sinh^2\theta\cosh\theta}\,d\theta\,,\\
    &\phif=\frac{r}{r^2+\gl^{-2}\sinh^2\theta}\,,\qquad \tilde\phif=\frac{\gl^{-1}\sinh\theta}{r^2+\gl^{-2}\sinh^2\theta}\,.
\end{align}
\end{subequations}
The change of coordinates
\begin{equation}
dt\to dt\pm\gl^{-2}(\gl^2r^2-1)\zeta\,dr\,,\qquad d\varphi\to d\varphi\pm\gl^{-1}\zeta\,dr\,,\qquad \zeta={2}Mr/\big(\Delta_r(\Delta_r-{2}Mr)\big)\,,    
\end{equation}
brings the Kerr--Schild metric $\bar g+{2}M\phif kk$ to \eqref{IIIa:AdS} with
\begin{equation}\label{IIIa:mass}
    \Delta_r\ \to\ r^2(\gl^2r^2-1)-{2}Mr\,,
\end{equation}
and the Weyl tensor \eqref{Weyl:DC} is
\begin{equation}
C_{\ga(4)}\xi^{\ga(4)}=\frac{M\gl^4(\xi_1^2-\xi_2^2)^2}{(r-i\gl^{-1}\sinh\theta)^3}\,.    
\end{equation}
Similarly, imaginary $\theta$-dependent shifts of $t$ and $\varphi$ bring the dual copy $\bar g+{2}N\tilde\phif\,ll$ to \eqref{IIIa:AdS} with
\begin{equation}\label{IIIa:nut}
    \sinh^2\theta\cosh^2\theta\ \to\ \sinh^2\theta\cosh^2\theta+{2}\gl N\sinh\theta
\end{equation}
in the last two terms. The map \eqref{IIIa:Wick} exchanges the two copies. It takes the mass copy \eqref{IIIb:full} of type $III_{b=1}$ exactly to the dual copy \eqref{IIIa:nut} with $N=-M$, and the dual copy of $III_{b=1}$ to the mass copy \eqref{IIIa:mass}. Thus the mass of one type becomes the NUT-like parameter of the other.

\paragraph{Interpretation.} The obtained solution with \eqref{IIIa:mass} can be interpreted as the contraction of KMV black hole \eqref{IIIb:KMV} at zero rotation. Indeed, let us change in \eqref{IIIb:KMV} the coordinates and the mass parameter as
\begin{equation}\label{IIIa:KMVmap}
    \cosh\theta\to\frac{\sinh\theta}{\gl a}\,,\qquad \varphi\to\gl a\,\Xi\,\varphi\,,\qquad t\to\Xi\Big(t+\frac{\varphi}{\gl}\Big)+\gl a^2\,\Xi\,\varphi\,,\qquad {\eta=M}\,,
\end{equation}
with $r$ unchanged. For any $a$, \eqref{IIIb:KMV} becomes exactly \eqref{IIIa:AdS} with
\begin{equation}\label{IIIa:KMVexact}
    \Delta_r\to(r^2+a^2)(\gl^2r^2-1)-{2}Mr\,,\qquad \sinh^2\theta\cosh^2\theta\to\cosh^2\theta\,\big(\sinh^2\theta-\gl^2a^2\big)\,,
\end{equation}
the latter in the last two terms, as in \eqref{IIIa:nut}. The KMV axis lies at $\sinh\theta=\gl a$. As $a\to0$, \eqref{IIIa:KMVexact} tends to \eqref{IIIa:AdS}, \eqref{IIIa:mass}, and the KMV axis moves to $\theta=0$. Unlike the infinite-rotation limit of Sec.~7.1, the mass and the radial coordinate are not rescaled. The limit is taken at fixed $a\cosh\theta_{\rm KMV}$, i.e., with the KMV hyperbolic angle $\theta_{\rm KMV}\to\infty$. It therefore zooms into the asymptotic end of the KMV horizon. The naive limit $a\to0$ at fixed $\theta_{\rm KMV}$ gives instead the static hyperbolic black hole of Sec.~\ref{static-hyp-bh}. Both limiting generators have the eigenvalues $0,0,0,\pm1$, but the static one is semisimple, while \eqref{IIIa:KMVmap} keeps the nilpotent part $N$.

At the level of orbits, the $I_d$ generator of the KMV black hole has eigenvalues $0,\pm\alpha,\pm i\beta$ with $\beta^2-\alpha^2=\gl^2a^2-1$ and $\alpha^2\beta^2=\gl^2a^2$. As $a\to0$, $\beta\to0$ and $\alpha\to1$. The rotation pair $\pm i\beta$ merges with the zero eigenvalue into the block of $N$, while the boost pair $\pm\alpha$ survives as $J_{14}$. Accordingly, the Casimir invariants of Appendix~{D}, $C_2=\gl^2a^2-1$ and $C_4=1-6\gl^2a^2+\gl^4a^4$, tend to the values $-1$ and $1$ of $III_{a=1}$. Types $III_a$ and $III_b$ are thus the two contractions of the KMV orbit $I_d$: zero rotation keeps the boost and gives $III_a$, while infinite rotation keeps the rotation and gives $III_b$.
}

\subsection{Type $III_{a=0}\,$: AdS soliton}
When $a=0$ the generating parameter \eqref{K:IIIa} reduces to 
\begin{equation}
    K = J_{30} - J_{32}\,,
\end{equation}
forming the so-called a pure nilpotent subregular orbit \cite{McGovern1993NilpotentOrbits}.
Its centralizer is bigger compared to the case when $a\neq 0$. Specifically, the generators 
\begin{equation}
    G_1 = J_{14}\,,\quad G_2 = J_{04} - J_{24}\,,\quad G_3 = J_{10} - J_{12}
\end{equation}
define an algebra of motions of 2-dimensional Minkowski space. Along with $K$, the full algebra is \(\u(1) \oplus \iso(1,1)\), which is type $\rom{6}_0$ in the Bianchi classification; see {\it e.g.,}  \cite{Bergshoeff:2003ri}
\begin{equation}
    [K, G_i] = 0\,,\quad [G_1, G_2] = G_3\,,\quad [G_3, G_1] = -G_2\,,\quad [G_2, G_3] = 0\,.
\end{equation}
{
The double copy follows from type $III_{b=0}$ by the Wick rotation of Sec.~7.3. Applied to the planar embedding \eqref{BB:emb}, it amounts to
\begin{equation}\label{IIIa0:Wick}
    \psi\to i\,\psi\,,\qquad \phi\to-i\,\phi\,,
\end{equation}
with $z$ and $\chi$ unchanged, in agreement with \eqref{IIIa:Wick}. The Euclidean plane of the black brane becomes a two-dimensional Minkowski plane, with Rindler coordinates $(\chi,\phi)$. The result is the real embedding
\begin{subequations}\label{IIIa0:emb}
\begin{align}
    &X^3=\frac{\psi}{\gl z}\,,\qquad X^0+X^2=-\frac{1}{\gl z}\,,\qquad X^2-X^0=\frac{z}{\gl}+\frac{\chi^2+\psi^2}{\gl z}\,,\\
    &X^1=-\frac{\chi\sinh\phi}{\gl z}\,,\qquad X^4=\frac{\chi\cosh\phi}{\gl z}\,,
\end{align}
\end{subequations}
in which
\begin{equation}
    K=\partial_\psi\,,\qquad G_1=-\partial_\phi\,,\qquad G_2=-\partial_x\,,\qquad G_3=-\partial_t\,,
\end{equation}
where $x=\chi\cosh\phi$ and $t=\chi\sinh\phi$ are the Minkowski coordinates of the plane. The AdS metric \eqref{AdS:IIIb} becomes:
\begin{equation}
    d\overline{s}^2=\frac{1}{\gl^2z^2}\left(-\chi^2d\phi^2+dz^2+d\chi^2+d\psi^2\right)=\frac{1}{\gl^2z^2}\left(-dt^2+dx^2+dz^2+d\psi^2\right)\,.
\end{equation}
With the vierbein $\e^a_{\ \mu}=(\gl z)^{-1}\mathrm{diag}(\chi,1,1,1)$ in the order $(\phi,z,\chi,\psi)$, the Killing vector $v=\gl K$, i.e., $v^\mu=\gl(0,0,0,1)$, has the components
\begin{equation}
    v_{\ga\dgb}=\frac{1}{z}\begin{pmatrix}1&0\\0&-1\end{pmatrix}\,,\qquad
    \gk_{\ga\gb}=\frac{\gl}{z}\begin{pmatrix}1&0\\0&1\end{pmatrix}\,,
\end{equation}
so that $C_2=C_4=0$ and $\rf=-i(\gl z)^{-1}$ is purely imaginary. Hence the zeroth copy \eqref{zero} vanishes, and only the dual one \eqref{dual:zero} survives. 
}

In this limiting case, only the complex pair of Kerr-Schild vectors remains, while the real ones blow up 
\begin{equation}
    \gl l^{\pm}_\mu dx^{\mu} = -d\psi \pm i dz\,.
\end{equation}
The nontrivial zero copy emerges from \eqref{dual:zero} and is given by 
\begin{equation}
    \Tilde{\phif} = \gl z\,.
\end{equation}
The Weyl tensor is
\begin{equation}
    C_{\alpha(4)}\xi^{\alpha(4)} = M z^3{\left(\xi_1^2 + \xi_2^2\right)^2}\,,
\end{equation}
which can be formally reproduced from the complex Kerr-Schild metric
\begin{equation}\label{IIIa=0:g}
    g_{\mu \nu} = \overline{g}_{\mu \nu} + {2}M \Tilde{\phif} l_{\mu} l_{\nu}
\end{equation}
with a real parameter $M$. However, the metric can be brought to a real form by the proper Wick rotation of the black brane metric \eqref{black brane}. The final result is  
\begin{equation}
        ds^2 =  f(r)dy^2 + f(r)^{-1} dr^2 + r^2(dx^2 - dt^2)\,, \qquad f(r) = \gl^2 r^2 + \frac{{2}M}{r}\,,
\end{equation}

The resulting spacetime is of Petrov type $D$ and has the topology $\mathbb{R}\times S^1$. It describes the AdS soliton; see 
\cite{Horowitz_1998}. The soliton is neither of constant curvature nor supersymmetric, but it is known to have minimal energy under small metric perturbations \cite{Galloway_2003}. The analytic continuation of two coordinates $t, y \to it, iy$ brings us back to the black brane \eqref{black brane}. 

This construction also furnishes an example of a $B\rom{3}$-metric characterized by the parameters; see Eq. \eqref{B-metrics}
\begin{equation}
    \epsilon_0 = -1\,,\quad  \epsilon_2 = 0\,,\quad p = r\,.
\end{equation}
It is also curious to note that the zero copy on such a background satisfies
\begin{equation}
    \boldsymbol{\nabla}^2\Tilde{\phif} + 2\gl^2 \Tilde{\phif} - {2}M \Tilde{\phif}^4 = 0\,,
\end{equation}
where we recall that $\boldsymbol{\nabla}^2$ is the d'Alembert operator with respect to the full metric \eqref{IIIa=0:g}.  

\section{TYPE $\rom{5}$}\label{sec:V}

This type is generated by a maximally degenerate orbit characterized by a Jordan form of size five, known as the principal nilpotent orbit, which has a dimension of dim $\mathcal{O} = 8$ \cite{McGovern1993NilpotentOrbits}. Despite its high degree of degeneracy, this orbit leads to a Petrov type D solution that takes on the Kerr-Schild form. Moreover, it allows for four Kerr-Schild vectors, consisting of two real vectors and two complex conjugates. As a result, the metric can accommodate two massive parameters: a mass-like parameter and a NUT-like parameter, if necessary, as indicated in \eqref{metric:gen}.

{
We show below that the resulting double copy is the most degenerate member of the Carter-Pleba\'nski family. All roots of its structure functions coincide, so it carries no kinematical parameter. The natural coordinates are therefore the Carter-Pleba\'nski ones, and we introduce them directly.

The orbit representative is given by the following zero-parameter form:
\begin{equation}\label{K:V}
   K = J_{10} + J_{30} - J_{12} - J_{14} + J_{23} + J_{34}\,,\qquad K^5=0\neq K^4\,.
\end{equation}
Its centralizer is spanned by the odd powers of $K$, i.e., by $K$ and
\begin{equation}\label{G:V}
    G =   J_{01} - J_{03} + J_{12} + J_{23}=-\half K^3\,,\qquad [G, K] = 0\,.
\end{equation}
Since $(K^3)^2=0$ and $K^3$ has rank two, $G$ is the minimal nilpotent element of Jordan type $[2,2,1]$, i.e., of type $II^0$, and both Casimir invariants vanish for $K$ and $G$.

In terms of the embedding coordinates, the Killing vector fields read
\begin{subequations}\label{V:KG}
\begin{align}
    &K=(X^1-X^3)(\partial_0-\partial_2)-(X^0+X^2)(\partial_1+\partial_3)-X^4(\partial_1-\partial_3)-(X^1+X^3)\partial_4\,,\\
    &G=-(X^1+X^3)(\partial_0-\partial_2)+(X^0+X^2)(\partial_1-\partial_3)\,.
\end{align}
\end{subequations}
Both annihilate $X^0+X^2$. Their norms and scalar product are
\begin{equation}\label{V:norms}
    |K|^2=(X^1+X^3)^2-4(X^0+X^2)X^4\,,\qquad \eta(K,G)=2(X^0+X^2)^2\,,\qquad |G|^2=0\,,
\end{equation}
so the vector field $G$ is null everywhere. We look for inner coordinates $(\tau,\sigma,r,y)$ that straighten $K$ and $G$:
\begin{equation}\label{V:straight}
    -\gl K\tau=1\,,\quad K\sigma=Kr=Ky=0\,,\qquad G\sigma=\frac\gl2\,,\quad G\tau=Gr=Gy=0\,,
\end{equation}
so that $\partial_\tau=-\gl K$ and $\partial_\sigma=2\gl^{-1}G=-\gl^{-1}K^3$. The joint invariants of $K$ and $G$ are $X^0+X^2$ and $|K|^2$, and we take $r$ and $y$ to be functions of them only. A solution of \eqref{V:straight} is
\begin{subequations}\label{V:inner}
\begin{align}
    &ry=\frac{2}{\gl}\,(X^0+X^2)\,,\qquad r^2-y^2=4(X^0+X^2)X^4-(X^1+X^3)^2\,,\\
    &\gl\tau=\frac{X^1+X^3}{2(X^0+X^2)}\,,\qquad \sigma=-\frac{1}{2ry}\Big(X^3-X^1+\frac{r^2-y^2}{ry}\,\tau\Big)-\frac{\gl^4\tau^3}{6}\,.
\end{align}
\end{subequations}
The normalization of $r$ and $y$ in the first line is the unique one for which the metric has no $dr\,dy$ term. The inverse transformation gives the embedding coordinates, which are polynomial in $\tau$ and $\sigma$, as expected for a nilpotent generator:
\begin{subequations}\label{V:emb}
\begin{align}
    &X^0+X^2=\frac{\gl ry}{2}\,,\qquad X^1+X^3=\gl^2\tau\,ry\,,\qquad X^4=\frac{r^2-y^2}{2\gl ry}+\frac{\gl^3\tau^2ry}{2}\,,\\
    &X^3-X^1=-\frac{r^2-y^2}{ry}\,\tau-2\sigma ry-\frac{\gl^4\tau^3ry}{3}\,,\\
    &X^2-X^0=-\frac{(r^2+y^2)^2}{2\gl^3r^3y^3}+\frac{\gl(r^2-y^2)}{ry}\,\tau^2+4\gl ry\,\tau\sigma+\frac{\gl^5ry}{6}\,\tau^4\,.
\end{align}
\end{subequations}
The hyperboloid condition holds identically. The chart covers $X^0+X^2\neq0$ and we take $r,y>0$.

The pullback of the AdS metric takes the familiar Carter-Pleba\'nski form
\begin{equation}\label{V:AdS}
    d\s^2=-\frac{\Delta_r}{\Sigma}(d\tau+y^2d\sigma)^2+\frac{\Delta_y}{\Sigma}(d\tau-r^2d\sigma)^2+\frac{\Sigma\,dr^2}{\Delta_r}+\frac{\Sigma\,dy^2}{\Delta_y}\,,
\end{equation}
with
\begin{equation}
    \Delta_r=\gl^2r^4\,,\qquad \Delta_y=\gl^2y^4\,,\qquad \Sigma=r^2+y^2\,.
\end{equation}
Compared with the general Carter-Pleba\'nski functions $\Delta_r=\gl^2r^4+\epsilon r^2+\gamma$ and $\Delta_y=\gl^2y^4-\epsilon y^2+\gamma$ of Sec.~\ref{CP}, this is the case $\epsilon=\gamma=0$. In this case $g_{\sigma\sigma}=-\epsilon r^2y^2+\gamma(r^2-y^2)$ vanishes identically, in agreement with $|G|^2=0$ in \eqref{V:norms}.

To construct the respective double copy, we use the vierbein
\begin{equation}
    \e^0=\frac{\gl r^2}{\sqrt\Sigma}(d\tau+y^2d\sigma)\,,\quad \e^1=\frac{\sqrt\Sigma}{\gl r^2}\,dr\,,\quad \e^2=\frac{\sqrt\Sigma}{\gl y^2}\,dy\,,\quad \e^3=\frac{\gl y^2}{\sqrt\Sigma}(d\tau-r^2d\sigma)\,,
\end{equation}
the Killing vector $v=\gl K=-\partial_\tau$, i.e., $v^\mu=(-1,0,0,0)$, has the Lorentz components
\begin{equation}
    v_{\ga\dgb}=-\frac{\gl}{\sqrt\Sigma}\begin{pmatrix}r^2+y^2&0\\0&r^2-y^2\end{pmatrix}\,,\qquad
    \gk_{\ga\gb}=\gl^2(r-iy)\begin{pmatrix}1&0\\0&-1\end{pmatrix}\,,
\end{equation}
so that
\begin{equation}
    \rf=r-iy\,,\qquad C_2=0\,,\qquad C_4=0\,.
\end{equation}
Unlike the other nilpotent orbits $II^0$, $III_{a=0}$ and $III_{b=0}$, the principal nilpotent orbit has the generic form $\gk_{\ga\gb}\propto\mathrm{diag}(1,-1)$ with $\rf\neq0$. Hence all four Kerr-Schild vectors \eqref{KS:spin} are well defined. Normalized by \eqref{KS:norm}, they are
\begin{equation}\label{V:KS}
    k^\pm_\mu dx^\mu=d\tau+y^2d\sigma\mp\frac{\Sigma}{\gl^2r^4}\,dr\,,\qquad
    l^\pm_\mu dx^\mu=d\tau-r^2d\sigma\pm i\,\frac{\Sigma}{\gl^2y^4}\,dy\,.
\end{equation}
The zeroth copy \eqref{zero}, its dual \eqref{dual:zero} and the Weyl tensor \eqref{Weyl:DC} are
\begin{equation}\label{V:Weyl}
    \phif=\frac{r}{r^2+y^2}\,,\qquad \tilde\phif=\frac{y}{r^2+y^2}\,,\qquad C_{\ga(4)}\xi^{\ga(4)}=M\gl^4\frac{(\xi_1^2-\xi_2^2)^2}{(r-iy)^3}\,.
\end{equation}
These are the universal Carter-Pleba\'nski expressions of Sec.~\ref{CP} at $\epsilon=\gamma=0$. The Weyl tensor is of Petrov type D.

The Kerr-Schild metric \eqref{Kerr-Schild}, $g=\bar g+{2}M\phif\,k^\pm k^\pm$, is brought to the standard Carter-Pleba\'nski form by the change of coordinates
\begin{equation}
    d\tau\to d\tau\pm r^2\zeta\,dr\,,\qquad d\sigma\to d\sigma\pm\zeta\,dr\,,\qquad \zeta=\frac{{2}Mr}{\Delta_r(\Delta_r-{2}Mr)}\,.
\end{equation}
It preserves $d\tau-r^2d\sigma$ and removes the $dr$ cross terms, so that \eqref{V:AdS} is recovered with $\Delta_r\to\Delta_r-{2}Mr$. Similarly, the dual copy $\bar g+{2}N\tilde\phif\,l^\pm l^\pm$ is brought to \eqref{V:AdS} with $\Delta_y\to\Delta_y+{2}Ny$ by imaginary $y$-dependent shifts of $\tau$ and $\sigma$. Together, the type $V$ double copy is
\begin{equation}\label{V:metric}
    ds^2=-\frac{Q(r)}{r^2+y^2}(d\tau+y^2d\sigma)^2+\frac{P(y)}{r^2+y^2}(d\tau-r^2d\sigma)^2+\frac{(r^2+y^2)\,dr^2}{Q(r)}+\frac{(r^2+y^2)\,dy^2}{P(y)}\,,
\end{equation}
\begin{equation}\label{V:QP}
    Q(r)=\gl^2r^4-{2}Mr\,,\qquad P(y)=\gl^2y^4+{2}Ny\,.
\end{equation}
This is the non-accelerating Pleba\'nski-Demia\'nski metric \cite{Plebanski:1975xfb,Plebanski:1976gy,Griffiths:2005qp} with $\Lambda=-3\gl^2$, $\epsilon=\gamma=0$, {mass $m=M$ and NUT parameter $n=N$}. We call it the degenerate Carter-Pleba\'nski metric. For $M=N=0$, both structure functions have a quadruple root at the origin. The relation of this metric to the other orbits is discussed in Sec.~\ref{sec:CPall}.
}

{
\section{Carter-Pleba\'nski master metric}\label{sec:CPall}
Most double copies of this paper with two commuting isometries are real sections of a single local metric, the Carter-Pleba\'nski metric with $\Lambda=-3\gl^2$:
\begin{equation}\label{CPall:metric}
    ds^2=-\frac{Q(r)}{r^2+y^2}(d\tau+y^2d\sigma)^2+\frac{P(y)}{r^2+y^2}(d\tau-r^2d\sigma)^2+\frac{(r^2+y^2)\,dr^2}{Q(r)}+\frac{(r^2+y^2)\,dy^2}{P(y)}\,,
\end{equation}
\begin{equation}\label{CPall:QP}
    Q(r)=\Delta_r(r)-{2}Mr\,,\qquad P(y)=\Delta_y(y)+{2}Ny\,,\qquad \Delta_r=\gl^2r^4+\epsilon r^2+\gamma\,,\qquad \Delta_y=\gl^2y^4-\epsilon y^2+\gamma\,.
\end{equation}
For $M=N=0$ it is AdS. In every case, $v=\gl K=-\partial_\tau$, the Kerr-Schild vectors are
\begin{equation}
    k^\pm_\mu dx^\mu=d\tau+y^2d\sigma\mp\frac{\Sigma}{\Delta_r}dr\,,\qquad l^\pm_\mu dx^\mu=d\tau-r^2d\sigma\pm i\frac{\Sigma}{\Delta_y}dy\,,
\end{equation}
and $\rf=r-iy$, $\phif=r/\Sigma$, $\tilde\phif=y/\Sigma$, with $\Sigma=r^2+y^2$. The mass $M$ comes from the real copy and the NUT-like parameter $N$ from the dual one, as in Secs.~\ref{CP} and \ref{sec:V}. What distinguishes the orbits is only the background structure function. The following observation fixes it.

\paragraph{Structure functions from the generator.} In all Carter-Pleba\'nski charts of this paper, the structure function is fixed by the generating parameter through its characteristic polynomial (see also \cite{Didenko:2009tc}):
\begin{equation}\label{CPall:Delta}
   \Delta_r(r)=\frac{\det(\gl r-K)}{\gl^3r}=\gl^2r^4+C_2\,r^2+\frac{C_4-C_2^2}{4\gl^2}\,,\qquad \Delta_y(y)=\Delta_r(iy)\,.
\end{equation}
Here the Casimir invariants are normalized as in Appendix~{D} and Secs.~6--8. Thus $\epsilon=C_2$ and $\gamma=(C_4-C_2^2)/(4\gl^2)$. If the nonzero eigenvalues of $K$ are $\pm\mu_1$ and $\pm\mu_2$, then
\begin{equation}\label{CPall:roots}
    \Delta_r(r)=\frac{(\gl^2r^2-\mu_1^2)(\gl^2r^2-\mu_2^2)}{\gl^2}\,,\qquad \Delta_y(y)=\frac{(\gl^2y^2+\mu_1^2)(\gl^2y^2+\mu_2^2)}{\gl^2}\,,
\end{equation}
so that
\begin{equation}
\epsilon=-(\mu_1^2+\mu_2^2)\,,\qquad \text{and}\qquad \gamma=\mu_1^2\mu_2^2/\gl^2\,.  
\end{equation}
These are real for any real $K$: either $\mu_1^2$ and $\mu_2^2$ are both real, or they are complex conjugate, as for the quartet $\mu_{1,2}=\alpha\pm i\beta$ of type $I_a$. For example, the KMV black hole of type $I_d$ has $\mu_1=1$ and $\mu_2=i\gl a$, which gives $\Delta_r=(\gl^2r^2-1)(r^2+a^2)$, in agreement with Appendix~{D}. This yields a dictionary between the orbit data and the geometry:
\begin{itemize}
    \item a real pair of eigenvalues $\pm\mu$ (a boost in $K$) gives the real roots $r=\pm\mu/\gl$ of $\Delta_r$, i.e., Killing horizons of AdS when the roots are simple;
    \item an imaginary pair $\pm i\beta$ (a rotation in $K$) gives the real roots $y=\pm\beta/\gl$ of $\Delta_y$, i.e., axes of the corresponding rotation when the roots are simple;
    \item a complex quartet gives no real roots at all;
    \item a nontrivial Jordan block gives coinciding roots. In particular, a zero eigenvalue of multiplicity three or five gives a double or quadruple root at $r=0$ and $y=0$, which lies at infinite proper distance. For type $V$, moreover, $g_{\sigma\sigma}=-\epsilon r^2y^2+\gamma(r^2-y^2)$ vanishes identically, and $\partial_\sigma$ is a null Killing vector.
\end{itemize}
The rescaling $(\tau,\sigma,r,y,M,N)\to(s^{-1}\tau,s^{-3}\sigma,sr,sy,s^3M,s^3N)$ corresponds to $K\to K/s$ and acts as $(\epsilon,\gamma)\to(s^{-2}\epsilon,s^{-4}\gamma)$. Up to this rescaling, the two-parameter families of type $I$ depend on one ratio, types $II$ and $III$ are discrete points, and type $V$ is the fixed point $\epsilon=\gamma=0$.

Summarizing, the following Table~\ref{tab:CPall} lists the background structure functions of all orbits with a two-dimensional centralizer, together with the real section and the resulting double copy. For comparison, it also includes the symmetry-enhanced nilpotent orbits $III_{b=0}$ and $III_{a=0}$.
\begin{table}[h!]
\centering
\resizebox{\linewidth}{!}{
\begin{tabular}{@{}lllll@{}}
\toprule
\textbf{Orbit} & \textbf{Eigenvalues} & $\boldsymbol{\gl^2\Delta_r(r)}$ & $\boldsymbol{\gl^2\Delta_y(y)}$ & \textbf{Double copy} \\
\midrule
$I_c$ & $\pm ib_1,\pm ib_2,0$ & $(\gl^2r^2+b_1^2)(\gl^2r^2+b_2^2)$ & $(\gl^2y^2-b_1^2)(\gl^2y^2-b_2^2)$ & Kerr-AdS, Carter-Pleba\'nski (Secs.~\ref{sec: Kerr}, \ref{CP}) \\
$I_d$ & $\pm a,\pm ib,0$ & $(\gl^2r^2-a^2)(\gl^2r^2+b^2)$ & $(\gl^2y^2+a^2)(\gl^2y^2-b^2)$ & rotating topological black hole (KMV) \\
$I_b$ & $\pm a_1,\pm a_2,0$ & $(\gl^2r^2-a_1^2)(\gl^2r^2-a_2^2)$ & $(\gl^2y^2+a_1^2)(\gl^2y^2+a_2^2)$ & Carter-Pleba\'nski, $t\leftrightarrow i\phi$ \\
$I_a$ & $\pm\alpha\pm i\beta,0$ & $\big|\gl^2r^2-(\alpha+i\beta)^2\big|^2$ & $\big|\gl^2y^2+(\alpha+i\beta)^2\big|^2$ & generalized Carter-Pleba\'nski \\
\midrule
$II_b$ & $\pm ib,\pm ib,0$ & $(\gl^2r^2+b^2)^2$ & $(\gl^2y^2-b^2)^2$ & super-entropic black hole for $b<0$ (Sec.~\ref{sec:IIb})\\
{} & {} & {} & {} & hyperbolic super-entropic for $b>0$\\
$II_a$ & $\pm a,\pm a,0$ & $(\gl^2r^2-a^2)^2$ & $(\gl^2y^2+a^2)^2$ & rotating black brane of \cite{Klemm:2014rda} (Sec.~\ref{sec:IIa}) \\
\midrule
$III_b$ & $\pm ib,0,0,0$ & $\gl^2r^2(\gl^2r^2+b^2)$ & $\gl^2y^2(\gl^2y^2-b^2)$ & KMV at infinite rotation (Sec.~7.1) \\
$III_a$ & $\pm a,0,0,0$ & $\gl^2r^2(\gl^2r^2-a^2)$ & $\gl^2y^2(\gl^2y^2+a^2)$ & KMV at zero rotation (Sec.~7.3) \\
\midrule
$V$ & $0,0,0,0,0$ $[5]$ & $\gl^4r^4$ & $\gl^4y^4$ & degenerate Carter-Pleba\'nski (Sec.~\ref{sec:V}) \\
\midrule
$III_{b=0}$ & $0,0,0,0,0$ $[3,1,1]$ & $\gl^4r^4$ & $\gl^4y^4$ & black brane$^{*}$ (Sec.~7.2) \\
$III_{a=0}$ & $0,0,0,0,0$ $[3,1,1]$ & $\gl^4r^4$ & $\gl^4y^4$ & AdS soliton$^{*}$ (Sec.~7.4) \\
\bottomrule
\end{tabular}
}
\caption{Background structure functions \eqref{CPall:Delta} of the Carter-Pleba\'nski metric \eqref{CPall:metric} for the orbits with two commuting isometries. The double copy is \eqref{CPall:metric} with $Q=\Delta_r-{2}Mr$ and $P=\Delta_y+{2}Ny$ on the indicated real section.\\ {\footnotesize $^{*}$ Symmetry-enhanced nilpotent orbits with a four-dimensional centralizer. They share the structure functions of type $V$, but the chart \eqref{CPall:metric} does not exist for them. Their double copies are the limits $b\to0$ and $a\to0$ of types $III_b$ and $III_a$ taken in the charts of Secs.~7.1 and 7.3, respectively.}}
\label{tab:CPall}
\end{table}
The Kerr-type charts of Secs.~6 and 7 are related to \eqref{CPall:metric} by simple maps. For example, 
\begin{equation}
y=\gl^{-1}\cosh\theta\,,\qquad \tau=t-\varphi/\gl\,,\qquad \sigma=\gl\varphi\qquad\text{for}\qquad III_b\,,  
\end{equation}
and
\begin{equation}
y=\gl^{-1}\sinh\theta\,,\qquad \tau=t+\varphi/\gl\,,\qquad \sigma=\gl\varphi\qquad\text{for}\qquad III_a  
\end{equation}
Type $V$ is written directly in the form \eqref{CPall:metric}.

\paragraph{Semisimple versus non-semisimple degenerations.} The structure functions depend only on the eigenvalues, not on the Jordan structure. Hence each degenerate row of Table~\ref{tab:CPall} has a semisimple counterpart with the same $\Delta_r$ and $\Delta_y$:
\begin{itemize}
    \item $I_c$ with $b_1=\pm b_2$ shares the functions of $II_b$;
    \item $I_b$ with $a_1=\pm a_2$ shares those of $II_a$;
    \item $I_c$ with $b_1=0$ or $b_2=0$ shares those of $III_b$;
    \item $I_b$ with $a_1=0$ or $a_2=0$ shares those of $III_a$;
    \item the nilpotent orbits $II^0$, and $III^\pm_0$ (see Table 1a) share those of $V$.
\end{itemize}
The two cases behave differently. If $K$ is regular, i.e., its centralizer is two-dimensional, the chart \eqref{CPall:metric} exists and the double copy is a genuine Carter-Pleba\'nski metric. For the degenerate rows this happens precisely when $K$ is not semisimple: the super-entropic black hole, the rotating black brane, the KMV limits, and the degenerate metric of type $V$. If $K$ is not regular, either semisimple with degenerate eigenvalues or one of the nilpotent orbits $II^0$, $III^\pm_0$, the centralizer is enhanced, the chart \eqref{CPall:metric} degenerates, and the double copy is reached only through a scaling limit. Examples are Schwarzschild from Kerr with $y=a\cos\theta$, the hyperbolic black hole, the hyperbolic Taub-NUT-AdS spacetime at critical rotation, and the black brane and AdS soliton. For $II^0$, the double copies are not even of type D, but the Siklos waves of type N. This explains why the same critical limit $\gl a\to\pm1$ of Kerr-AdS gives either the super-entropic black hole or the hyperbolic Taub-NUT-AdS spacetime (Sec.~5.7, Table~\ref{tab:limits}). It also explains why $b\to0$ in type $III_b$ gives either the degenerate metric of type $V$, when taken in Carter-Pleba\'nski coordinates, or the black brane, when taken in the KMV-adapted chart.

\paragraph{Contractions.} Since \eqref{CPall:metric} depends continuously on $(\epsilon,\gamma)$, the limits between orbits are limits within one family of metrics, at fixed or suitably rescaled coordinates. Type $V$, at $\epsilon=\gamma=0$, is the common endpoint: it is the limit $\epsilon,\gamma\to0$ of every row of Table~\ref{tab:CPall} at fixed $(\tau,\sigma,r,y,M,N)$. Types $III_a$ and $III_b$ are the two contractions $\gamma\to0$ of the KMV family $I_d$, at zero and infinite rotation, respectively. At the level of orbits, these statements reflect the closure ordering of adjoint orbits: the principal nilpotent orbit lies in the closure of all regular orbits.
}

\section{Summary and discussions}\label{sec:conc} 
In this paper, we tackle the classification of classical double copies of the form \eqref{Weyl:DC}, as well as their higher-spin counterpart multicopies \eqref{HS:Weyl DC}. To achieve this, we utilize findings from \cite{Didenko:2009tc} regarding double copies and from \cite{Didenko:2008va, Didenko:2009td} concerning multicopies, where it was demonstrated that these copies are derived from the background global symmetry parameter, which, in our context, is AdS spacetime.

The vacuum solutions of Einstein's equations for double copies and the linearized solutions of Fronsdal's spin $s$ equations, arise from a deformation of the AdS isometry condition. The deformation parameters\footnote{In \cite{Didenko:2009tc}, the deformation also included an electrodynamic field, corresponding to Einstein-Maxwell double copies with spin one charges.} are represented by spin $s$ ``charges'' $M_s$ and $\bar M_s$. Following \cite{Didenko:2009td}, the deformation is realized through the Penrose transform, which employs an AdS isometry element $K\in \so(2,3)\sim\sp(4)$. Below, we briefly summarize some key properties of the solutions obtained in this manner:
\begin{itemize}
    \item The number of distinct solutions corresponds to the number of adjoint $\so(2,3)$ orbits. According to \cite{Holst:1997tm}, there are four main types of orbits (see Table 1) for AdS in four dimensions, each characterized by specific eigenvalues. This makes the proposed classification reminiscent of the one presented in \cite{Banados:1992gq} for three dimensions, where solutions of topological 3D gravity, including the BTZ black hole, arise from the AdS isometry. These 3D solutions are naturally associated with and classified according to the adjoint orbits of $\so(2,2)$. However, a key difference is that the four-dimensional double copies and multicopies are not topological.

    \item Focusing primarily on double copies, both the curvature and the metric are expressed in terms of the AdS symmetry parameter $K$. Specifically, the Weyl tensor is represented through its Lorentz component, which describes the AdS covariant derivative of the background Killing vector that generates the spacetime. In general, the metric takes on a Kerr-Schild form (which may be double if  $M\neq \bar M$), with the Kerr-Schild vectors also constructed from $K$. By construction, the metric depends on two constants,  $I_1$ and $I_2$, which are Casimir invariants of the $\so(2,3)$ algebra. Specifically, whenever $K$ is regular the double copy is the Carter-Pleba\'nski metric \eqref{CPall:metric} with structure functions
    \[
    \Delta_r=\det(\gl r-K)/(\gl^3r)=\gl^2r^4+I_1r^2+(I_1^2+I_2)/(2\gl^2)\,,\qquad  \Delta_y(y)=\Delta_r(iy)\,.
    \]

    \item The algebraic type of the Weyl double copies is Petrov type D, with a few exceptions that represent type N. In the case of type D, the generators of global symmetries $G_i$, commute with the generating vector field $K\in\so(2,3)$, satisfying the relation $[G_i, K]=0$. The size of the isometry group depends on the specific orbit $K$: it can have either two, four, or six generators. The four-generator case arises when $K$ is not regular, i.e., when it is semisimple with coinciding eigenvalues, or nilpotent but not principal. The Casimir invariants alone do not determine this: for instance, types $II_b$ and $I_c$ with $b_1=\pm b_2$ share the same invariants, but only the latter has enhanced symmetry. The origin of global symmetries is also analogous to those in 3D gravity. 

    \item Every solution derived from the $\so(2,3)$ generator features two commuting isometries. This property enables us to define canonical coordinates, where two of these coordinates correspond to the vector fields of the commuting isometries. As a result, the metric coefficients do not depend on these two coordinates. However, this does not fully determine the choice of coordinates, as there is still freedom to linearly mix the commuting vector fields and to reparameterize the remaining two coordinates.
\end{itemize}
Our method can also be viewed as a step toward establishing a systematic approach to the zoo of vacuum Einstein solutions. In addition, it offers a straightforward way to understand their isometries. For example, we find that the AdS-Kerr black hole, which is known to possess two Killing vectors, exhibits three different critical limits $\lambda |a|\to 1$, one of which results in four isometries representing {the hyperbolic Taub--NUT--AdS solution} \cite{Hennigar:2015wxa} with a fine-tuned, in terms of the cosmological constant, NUT-parameter (see Sec. \ref{Sec:3lims}). Other limits include super-entropic black hole \cite{Hennigar:2014cfa} and a Siklos wave \cite{Siklos}.

The solutions that have been classified so far fall into the following types based on the adjoint $\so(2,3)$ orbits:

\begin{enumerate}
    \item[I] Type $I$ corresponds to semisimple $K$ and includes four sub-types $I_a$, $I_b$, $I_c$, $I_d$, distinguished by their eigenvalues. Type $I_c$, generated by $K=J_{10}+\gl aJ_{23}$, produces the Kerr-AdS black hole. For $a=0$ and $\gl a=\pm1$ the symmetry enhances to four isometries, giving the Schwarzschild and the hyperbolic Taub-NUT-AdS solutions, respectively. Types $I_b$ and $I_d$ are Wick rotations of the Carter-Pleba\'nski metric \cite{Carter:1968rr, Plebanski:1975xfb}. Type $I_d$ is the rotating topological black hole of Klemm, Moretti and Vanzo \cite{Klemm:1997ea}. The less familiar type $I_a$, which we call the generalized Carter-Pleba\'nski solution, is a real section of the complexified Kerr-AdS metric with the complex rotation parameter on the circle $|a_K|=\gl^{-1}$. It fills the region $\epsilon^2<4\gl^2\gga$ of the Carter-Pleba\'nski parameters.

    \item[II] Type $II$ corresponds to two coinciding pairs of eigenvalues with a nilpotent part. The orbit $II^-$ ($b<0$) gives the super-entropic black hole \cite{Hennigar:2014cfa} and $II^+$ ($b>0$) its hyperbolic counterpart, while $II$ ($a\neq0$) gives the rotating black brane of \cite{Klemm:2014rda}. The degenerate orbit $II^0$ yields a divergent Penrose transform. Its regularized limits $a\to0$ and $b\to0$ produce two inequivalent Siklos AdS waves of Petrov type N \cite{Siklos}.
    
    \item[III] Type $III$ also has two sub-types: $III_a$ and $III_b$. Type $III_{b=0}$ describes the AdS black brane with a planar horizon and $III_{a=0}$ the AdS soliton, related by a double Wick rotation. For nonzero parameters, $III_b$ and $III_a$ are the two contractions of the Klemm-Moretti-Vanzo rotating topological black hole: at infinite and at zero rotation, respectively, with $\gga\to0$ in Carter-Pleba\'nski variables. The two subtypes are related by a Wick rotation exchanging the radial and angular Carter-Pleba\'nski coordinates, which maps the mass of one into the NUT-like parameter of the other.

    \item[V] Type $V$ is maximally degenerate and has its both Casimir invariants vanishing. Despite this, the associated generator allows for a well-defined double copy that takes on the Kerr-Schild form and possesses two isometries. The corresponding space-time is the degenerate Carter-Pleba\'nski metric with $\epsilon=\gga=0$, with all roots of the structure functions $\Delta_r$ and $\Delta_y$ coincide. It is the common limit of all Carter-Pleba\'nski families.  
\end{enumerate}

Let us comment further on the global symmetries of the solutions we have examined. Our formalism derives these symmetries from the center of a given generator of $\so(2,3)$. However, this holds true only when the Penrose transform is well-defined. For types $II_{a=0}$ and $II_{b=0}$, this is not the case -- the Penrose transform becomes divergent.
Fortunately, we can manage this divergence by considering $a=b=\gep$ and renormalizing the deformation parameter$M\sim (1/\gep)^{\Delta}$, where an appropriate power $\Delta$ ensures that the limit $\gep\to 0$ is well-defined. 

The center of both types $II_{a,b=0}$ contains six generators; however, only four of these contribute to the true isometries of double copies, as we have specifically verified. The remaining two generators are considered ``anomalous''. In this regard, it is also noteworthy that by utilizing the ansatz in terms of the global symmetry parameters from types $II_{a,b=0}$, one can construct spacetimes with either five (Kaigorodov spacetime \cite{Kaigorodov}) or six (Defrise spacetime \cite{Defrise1969}) isometries, although the latter solution is non-vacuum.

There are several aspects that call for further investigation. For instance, the classification proposed in this paper does not encompass all double copies of the form \eqref{Weyl:DC}. In four dimensions, the complex structure involved in the isomorphism $\so(2,3)\sim\sp(4)$ enables the generation of Weyl double copies from the non-Hermitian AdS global symmetry parameter $K_{AB}$. This is illustrated in Sec. \ref{miss} for the Pleba\'nski-Demia\'nski metric (see also \cite{Didenko:2009tc, Luna:2018dpt}). We argue that this parameter should not be arbitrary; it is expected to commute with its conjugate, as indicated in \eqref{com:conj}. If this condition holds true generally, it would allow for an extension of the presented classification.

Section \ref{sec:CPall} shows that all orbits with two commuting isometries fit into the Carter-Pleba\'nski (non-accelerating Pleba\'nski-Demia\'nski) family, with structure functions given by the characteristic polynomial of $K$. It remains to understand the global structure of the less familiar members, in particular type $I_a$ and the degenerate type $V$, and the role of acceleration, i.e., the full Pleba\'nski-Demia\'nski metric \cite{Plebanski:1976gy}, in relation to our classification. 

In terms of vacuum Petrov type D solutions, Kinnersley \cite{Kinnersley:1969zza} has classified them for the case of $\Lambda=0$. { However, the case when $\Lambda\neq 0$ is more complex}\footnote{Some solutions in general relativity, such as black branes and Siklos spacetimes, do not permit the limit $\Lambda\to 0$.}. { For $\Lambda\neq0$, the general type D vacuum solution was obtained by exhaustive integration in \cite{Debever}. The well-known Pleba\'nski--Demia\'nski metric \cite{Plebanski:1976gy} covers only the case of expanding repeated principal null congruences, while the non-expanding type D solutions of Kundt's class are obtained from it only by separate degenerate limits \cite{Griffiths:2005qp} or found independently \cite{Hacyan}. What we have in mind here is not the integration of the field equations, but a classification organized by the global AdS symmetry, in which inequivalent solutions are labeled by adjoint $\so(2,3)$ orbits and their isometries by the corresponding centralizers. It would be interesting to relate this classification to the many-parameter metric of \cite{Debever}.} In this regard, it is also of interest to reconsider our analysis for $\Lambda=0$, where the respective algebra is $\iso(1,3)$ instead of $\so(2,3)$. Notice that the former is not semi-simple. Additionally, we expect a natural higher-dimensional generalization of the proposed classification with the global symmetry parameter $\so(2,d-1)$ for $\Lambda<0$ or $\so(1,d)$ for $\Lambda>0$. Some results in this regard are available for $d=5$ in \cite{Didenko:2011ir}.

Another point worth mentioning is the equivalence of certain GR solutions in non-Lorentzian spacetimes. For instance, in \cite{Crawley:2021auj}, it was observed that the rotation parameter of the Kerr-Taub-NUT black hole can be eliminated in the self-dual case. From our perspective, we anticipate various accidental isomorphisms of the $\so(2,3)$ orbits related to complex similarity transformations, which may have interpretations in non-Lorentzian spaces. Exploring this further could yield interesting insights.

Finally, our classification directly applies to the classical higher-spin multicopies \cite{Didenko:2008va, Didenko:2022qxq}. However, unlike in the case of GR, the considered multicopies are not linearly exact {(see, however, footnote~\ref{foot:HSlocal})}. Despite this, next-to-leading order corrections still appear in the form of the multicopy, as demonstrated for type $III_{b=0}$ in \cite{Didenko:2021vdb}. Consequently, the formalism we have developed is well-suited for analyzing order-by-order corrections. In particular, it would be interesting to find an analog of the AdS plane wave solutions in higher-spin theory, similar to the Siklos metric in GR corresponding to types $II_{a=0}$ and $II_{b=0}$.

\section*{Acknowledgments}
V.D. would like to thank Per Sundell for the warm hospitality at the Universidad Arturo Prat in Chile during the final stage of this work. V.D. also expresses his gratitude to Vladimir Lukash and Elena Mikheeva for providing us with the rare reference \cite{Siklos}. We are grateful for the financial support from the Foundation for the Advancement of Theoretical Physics and Mathematics ``BASIS''.
   
\section{Appendix}

\newcounter{appendix}
\setcounter{appendix}{0}
\refstepcounter{appendix} 

\renewcommand{\theequation}{\Alph{appendix}.\arabic{equation}}
\setcounter{equation}{0}
 \renewcommand{\thesection}{\Alph{appendix}.}

\addcontentsline{toc}{section}{\,\,\,\,\,A. Spinorial formalism}

\section*{A. Spinorial formalism}
\label{sec:spinorial-dic}

In this section, we elaborate on the four-dimensional spinor dictionary that establishes the isomorphism between the two algebras \(\mathfrak{so}(2,3)\) and \(\mathfrak{sp}(4, \mathbb{R})\). Conveniently, one uses the van der Waerden symbols to bridge tensor and spinor representations.

Any tensor can be represented in terms of spinors using the two-component van der Waerden symbols
\begin{equation}
    (\sigma_a)_{\alpha \dot{\beta}} = (I, \sigma^i),\quad (\overline{\sigma}_{a})^{\dot{\alpha} \beta} = (-I, \sigma^i)
\end{equation}
that satisfy the following relations:
\begin{equation}
(\sigma_a)_{\alpha \dot{\gamma}}(\overline{\sigma}_{b})^{\dot{\gamma} \beta} = \eta_{ab}{\delta_{\alpha}}^{\beta} + {{(\sigma_{ab})}_{\alpha}}^{\beta}\,,\quad {{(\sigma_{ab})}_{\alpha}}^{\beta}=\half {(\sigma_a \overline{\sigma}_{b} - \sigma_b \overline{\sigma}_{a})_{\alpha}}^{\beta}\,,
\end{equation}
where \(I\) is the identity matrix. The metric in \(\mathfrak{so}(1,3)\) is chosen mostly plus \( \eta = (- + + +) \), and \(\sigma^i\) is the standard basis of Pauli matrices
\begin{equation}
   {(\sigma_{a})}_{\alpha \dot{\beta}} = \left\{
\begin{pmatrix} 1 & 0\\ 0 & 1 \end{pmatrix},  
\begin{pmatrix} 0 & 1\\ 1 & 0 \end{pmatrix},  
\begin{pmatrix} 0 & -i\\ i & 0 \end{pmatrix},  
\begin{pmatrix} 1 & 0\\ 0 & -1 \end{pmatrix}  
\right\}\,, \quad a = (0,1,2,3)\,.
\end{equation}
We raise and lower spinor indices according to the conventions $A_{\gb} = \epsilon_{\ga \gb} A^\ga,\ A^{\gb} = \epsilon^{\gb \ga} A_\ga$.

For a space-time vector \( v^a = (v^0, v^1, v^2, v^3) \), the corresponding spinor representation is a Hermitian matrix
\begin{equation}
     v_{\alpha \dot{\beta}} := v^{a} (\sigma_{a})_{\alpha \dot{\beta}} = \begin{pmatrix}
          v^0 + v^3 & v^1 - iv^2\\
          v^1 + iv^2 & v^0 - v^3
      \end{pmatrix}\,,
\end{equation}
whose determinant is the squared negative length of the vector
\begin{equation}\label{A:det}
    \det v_{\alpha \dot{\beta}}:=\frac{1}{2}v_{\al\dgb}v^{\al\dgb} = (v^0)^2 - (v^1)^2 - (v^2)^2 - (v^3)^2 = -v_a v^a\,.
\end{equation}
The inverse transform is given by 
\begin{equation}
    v^a = \frac{1}{2} (\overline{\sigma}^{a})^{\dot{\beta} \alpha} v_{\alpha \dot{\beta}}\,.
\end{equation}
Consequently, the space-time Killing vector \( v^\mu \) and the self-dual part of antisymmetric rank-two tensor $\gk_{ab}$ -- the fields that we extensively use in our analysis -- are expressed as
\begin{equation}
    v_{\alpha \dot{\beta}} = (\sigma_{a})_{\alpha \dot{\beta}} \e^a_{\mu} v^{\mu}\,,\qquad     \varkappa_{\ga \gb} = -\frac{1}{4} \varkappa^{ab}(\sigma_{ab})_{\alpha \beta}\,,
\end{equation}
where $\e_{\mu}^{a}$ is the vierbein field.
The duality between the two algebras is realized via five antisymmetric \(4 \times 4\) matrices \((\Gamma_{\underline{I}})_{AB}\), generating the Clifford algebra
\begin{equation}
    (\Gamma_{\underline{I}})_{A}{}^{C} (\Gamma_{\underline{J}})_{C}{}^{B} + (\Gamma_{\underline{J}})_{A}{}^{C} (\Gamma_{\underline{I}})_{C}{}^{B} = 2\eta_{\underline{IJ}} \delta^B_A\,,\quad \eta_{\underline{IJ}} = (--+++)\,,
\end{equation}
where the gamma matrices are
\begin{equation}
    (\Gamma^{\underline{I}})_{A}{}^{B} =
i \begin{pmatrix}
    0 & (\sigma^a)_{\alpha}{}^{\dot{\beta}} \\
    (\bar{\sigma}^a)_{\dot{\alpha}}{}^{\beta} & 0
\end{pmatrix}\,,\quad
(\Gamma^{\underline{IJ}})_{AB} = (\Gamma^{[\underline{I}} \Gamma^{\underline{J}]})_{AB} =
\begin{pmatrix}
    (\sigma^{ab})_{\alpha\beta} & 0 \\
    0 & (\bar{\sigma}^{ab})_{\dot{\alpha} \dot{\beta}}
\end{pmatrix}\,.
\end{equation}

The global symmetry parameter  \( \Omega_{\underline{IJ}}\) of AdS space, which takes values in \(\mathfrak{so}(2,3)\)  has the following realization as an \(\mathfrak{sp}(4, \mathbb{R})\) element:
\begin{equation}
    K_{AB} := -\frac{1}{4} \Omega_{\underline{IJ}} (\Gamma^{\underline{IJ}})_{AB} =
    \begin{pmatrix}
        \gl^{-1} \varkappa_{\ga \gb} &  \upsilon_{\ga \dot{\gb}} \\
         \upsilon_{\gb \dot{\ga}} & \gl^{-1} \overline{\varkappa}_{\dot{\ga} \dot{\gb}}
    \end{pmatrix}\,,
\end{equation}
where the \(\mathfrak{sp}(4, \mathbb{R})\) metric is
\begin{equation}
    \epsilon_{AB} = -\epsilon_{BA} =
    \begin{pmatrix} \epsilon_{\ga \gb} & 0\\ 0 & \epsilon_{\dot{\alpha} \dot{\beta}} \end{pmatrix}\,.
\end{equation}

{
\renewcommand{\theequation}{\Alph{appendix}.\arabic{equation}}
\addtocounter{appendix}{1} \setcounter{equation}{0}
\addtocounter{section}{1}
\addcontentsline{toc}{section}{\,\,\,\,\,B. Discrete orbits of types $II$ and $III$}

\section*{B. Discrete orbits of types $II$ and $III$}\label{app:discrete}
In this Appendix, we show that, up to the normalization of the generating parameter $K$, the representatives of types $II_{a,b}$ and $III_{a,b}$ in Table 1 with arbitrary $a$ and $b$ are conjugate to the discrete representatives of Table~\ref{tab:discrete}. Strictly speaking, different values of $|a|$ or $|b|$ belong to different adjoint orbits, since the Casimir invariants \eqref{Casimirs} depend on them. However, under $K\to cK$ one has $I_1\to c^2I_1$ and $I_2\to c^4I_2$, and the double copy changes only by a rescaling of the mass parameter. It is therefore natural to classify generating parameters up to conjugation and normalization.

We use two kinds of group elements. First, the reflections $R_s=\mathrm{diag}(s_0,\dots,s_4)$, $s_{\underline A}=\pm1$, of the embedding coordinates belong to $O(2,3)$ and act on the generators as
\begin{equation}\label{G:refl}
R_s\,J_{\underline{AB}}\,R_s^{-1}=s_{\underline A}s_{\underline B}\,J_{\underline{AB}}\,.
\end{equation}
Second, if a boost generator $H$ satisfies $[H,N]=N$ for some element $N$, then $e^{tH}Ne^{-tH}=e^{t}N$. In each case below, $K$ is decomposed as $K=cS+N$ with $c\neq0$ and $[S,N]=0$, where $S$ is a fixed semisimple generator and $N$ is nilpotent, and $H$ is chosen so that $[H,S]=0$ and $[H,N]=N$. Then $e^{tH}(\sigma S+N)e^{-tH}=\sigma S+e^{t}N$, and therefore
\begin{equation}\label{G:resc}
cS+N=|c|\,e^{tH}\big(\mathrm{sign}(c)\,S+N\big)e^{-tH}\,,\qquad t=-\ln|c|\,.
\end{equation} All relations below were also verified with explicit $5\times5$ matrices of the vector representation of $\so(2,3)$ satisfying \eqref{ads:algebra}.

\paragraph{Type $II_b$.} Here $K_b=bS+N$ with
\begin{equation}
S=J_{01}+J_{23}\,,\qquad N=-J_{01}+J_{23}+J_{02}-J_{13}\,,\qquad [S,N]=0\,,\quad N^2=0
\end{equation}
in the vector representation. The boost $H=-\frac12(J_{03}+J_{12})$ satisfies $[H,S]=0$ and $[H,N]=N$, so that by \eqref{G:resc}
\begin{equation}
K_b=|b|\,e^{tH}K_{\mathrm{sign}(b)}e^{-tH}\,,\qquad t=-\ln|b|\,.
\end{equation}
Hence every $II_b$ with $b>0$ is a rescaled conjugate of $K_{b=1}$, and every $II_b$ with $b<0$ is a rescaled conjugate of $K_{b=-1}$.

The cases $b>0$ and $b<0$ are not conjugate, even allowing $K\to cK$ with any $c\neq0$.
To distinguish the two signs, consider the real symmetric form $A_K=\eta K_sK_n$, built from the semisimple and nilpotent parts of the Jordan decomposition $K=K_s+K_n$. Its uniqueness gives $A_{cgKg^{-1}}=c^2g^{-T}A_Kg^{-1}$ for every $g\in O(2,3)$ and $c\neq0$. For $K_b$ one has $K_s=bS$, $K_n=N$, and
\begin{equation}
    x^T A_{K_b} x = -b \left[ (x^0 + x^3)^2 + (x^1 + x^2)^2\right].
\end{equation}
It has rank two and is negative (positive) semidefinite for $b>0$ ($b<0$), so its inertia distinguishes II$^+$ and II$^-$ under conjugation combined with any nonzero real rescaling. Equivalently, one can restrict to the generalized eigenspace $E_\mu(K_b)$, $\mu=i|b|$, where $K_s=\mu$, and consider the Hermitian form
\begin{equation}
\mathcal H(x)=i\,\eta\big(\bar x,(K_b-\mu)x\big) = \frac{\bar{x}^T A_{K_b} x}{|b|}\,,\qquad x\in E_\mu\,.
\end{equation}
For $K \to -K$, complex conjugation identifies the two eigenspaces and preserves the inertia of their Hermitian forms.
Geometrically, they generate the two complementary real regions of the same Carter-Pleba\'nski metric (on $\gl^{2}Y^2 > 1$ for $b > 0$ and $0\le \gl^{2}Y^2<1$ for $b<0$).

\paragraph{Type $II_a$.} Here $K_a=aB+N_a$ with
\begin{equation}
B=J_{03}+J_{12}\,,\qquad N_a=J_{01}-J_{02}-J_{13}+J_{23}\,,\qquad [B,N_a]=0\,,\quad N_a^2=0\,.
\end{equation}
The boost $H_a=\frac12(J_{03}-J_{12})$ satisfies $[H_a,B]=0$ and $[H_a,N_a]=N_a$, so that $K_a=|a|\,e^{tH_a}K_{\mathrm{sign}(a)}e^{-tH_a}$ with $t=-\ln|a|$. Unlike type $II_b$, the sign of $a$ can also be removed. The element
\begin{equation}\label{B8}
g=e^{\frac\pi2J_{01}}\,e^{-\frac\pi2J_{23}}\in SO_0(2,3)\,,\qquad g\,K_{a=1}\,g^{-1}=K_{a=-1}\,,
\end{equation}
a rotation by $\pi/2$ in the timelike plane $(X^0,X^1)$ combined with a rotation by $-\pi/2$ in the spacelike plane $(X^2,X^3)$, exchanges the two signs. The reason is that the eigenvalues $\pm a$ are real, and the generalized eigenspaces $E_{\pm a}$ are real totally null planes paired by $\eta$; this pairing carries no sign. Hence all $II_a$ with $a\neq0$ form the single class $II$.

\paragraph{Type $II^0$.} The degenerate members of both families are nilpotent with Jordan type $[2,2,1]$. The reflection $X^0\to-X^0$, {\it i.e.,} $R_s$ with $s=(-1,1,1,1,1)$ in \eqref{G:refl}, maps $II_{a=0}=N_a$ to $II_{b=0}=N$, and positive multiples of $N$ are conjugate by $e^{tH}$. Hence $II_{a=0}=II_{b=0}=II^0$. Note that the double copies obtained in the limits $a\to0$ and $b\to0$ are nevertheless different Siklos spacetimes, since the regularized limit depends on the direction of approach; see Sec.~\ref{sec:II}.

\paragraph{Type $III$.} For $III_b$, $K_b=-bJ_{34}+N_3$ with $N_3=J_{02}-J_{01}$, which commutes with $J_{34}$ and is nilpotent with Jordan type $[3,1,1]$. The boost $H=-J_{12}$ satisfies $[H,J_{34}]=0$ and $[H,N_3]=N_3$, so $|b|$ is removed as in \eqref{G:resc}. The sign of $b$ is removed by the reflection $R_s$ with $s=(-1,-1,-1,1,-1)$, which belongs to $SO_0(2,3)$, flips $J_{34}$ and leaves $N_3$ invariant. Similarly, for $III_a$, $K_a=-aJ_{14}+N_3'$ with $N_3'=J_{23}-J_{03}$; the boost $H_a=-J_{02}$ removes $|a|$, and $R_s$ with $s=(-1,-1,-1,-1,1)\in SO_0(2,3)$ removes the sign of $a$. In both cases the nonzero eigenvalues are simple, so no sign invariant analogous to $\mathcal H$ exists. Hence all $III_a$ with $a\neq0$ form the single class $III^+$, and all $III_b$ with $b\neq0$ the single class $III^-$. Finally, $III^+_0=III_{a=0}$ and $III^-_0=III_{b=0}$ are not conjugate: for $v\notin\ker K^2$ the sign of $\eta(v,K^2v)$ is an invariant of conjugation, of $K\to-K$ and of rescaling, and it equals $-1$ for $III^+_0$ and $+1$ for $III^-_0$.

\paragraph{Negative rescaling.} For $c<0$, the map $K\to cK=-|c|K$ is realized within $O(2,3)$. For $II_b$ and $III_b$, the reflection $R_s$ with $s=(1,-1,-1,1,-1)$ gives $R_sK_bR_s^{-1}=-K_b$. For $II_a$, the same conjugation to $-K_a$ is realized by $gR_s$, with $g$ as in \eqref{B8}. In these three cases, conjugation reverses time orientation: it reverses the nilpotent part of $K$, which generates a nowhere vanishing causal Killing field on $AdS$. For $III_a$, the spatial rotation by $\pi$ in the $(X^3, X^4)$-plane gives $R_s K_a R_s^{-1} = -K_a$ with $s = (1, 1, 1, -1, -1) \in SO_0(2, 3)$, and preserves time orientation. Thus allowing negative rescaling makes no further identifications.
}

\renewcommand{\theequation}{\Alph{appendix}.\arabic{equation}}
\addtocounter{appendix}{1} \setcounter{equation}{0}
\addtocounter{section}{1}
\addcontentsline{toc}{section}{\,\,\,\,\,C. More on the Killing projectors}

\section*{C. More on the Killing projectors}

The key feature of the Kerr-Schild multicopy is the manifest construction \cite{Didenko:2009tc} of the Kerr-Schild vectors using the AdS global symmetry parameter $K_{AB}$. This parameter includes its Lorentz components, being an AdS Killing vector $v_{\al\dgb}$ and its covariant derivative, which is represented by the Killing-Yano field  $\gk_{\al\gb}$  (along with its Hermitian conjugate $\bar\gk_{\dal\dgb}$). 

The proposed Kerr-Schild construction relies on utilizing the only available vector in  $K_{AB}$, which is the Killing vector $v_{\al\dgb}$. The global symmetry parameter allows us to extract lightlike directions from this Killing vector. This extraction is achieved through the use of rank-one projectors, defined in equation \eqref{proj:def}, which are constructed from the Killing-Yano field. Let us revisit their definition here
\begin{equation}
    {\pi^{\pm}}_{\ga \gb} = \frac{1}{2}\left(\epsilon_{\ga \gb} \pm \frac{\varkappa_{\ga \gb}}{\sqrt{-\varkappa^2}}\right)\,,\qquad {\overline{\pi}^{\pm}}_{\dot{\ga} \dot{\gb}} = \frac{1}{2}\left(\epsilon_{\dot{\ga} \dot{\gb}} \mp \frac{\overline{\varkappa}_{\dot{\ga}\dot{\gb}}}{\sqrt{-\overline{\varkappa}^2}}\right)\,,\qquad     \overline{{\pi^{\pm}}}_{\ga \gb} =     {\overline{\pi}^{\pm}}_{\dot{\ga} \dot{\gb}}\,,
\end{equation}
satisfying
\begin{equation}
    {\pi^{\pm}}_{\ga}^{\ \ \gga}\pi^{\pm}_{\gga \gb} = \pi^{\pm}_{\ga \gb}\,,\qquad     {\pi^{\pm}}_{\ga}^{\ \ \gga}\pi^{\mp}_{\gga \gb} = 0\,.
\end{equation}
These projectors decompose the two-dimensional (anti)holomorphic spinor space into a direct sum of two one-dimensional subspaces. Indeed, for an arbitrary spinor $\eta_\ga$, we have
\begin{equation}
    \eta^{\pm}_{\ga} = {\pi^{\pm}}_{\ga}^{\ \gb}\eta_{\gb}\,,\qquad \eta^+ + \eta^- = \eta\,,\qquad {\pi^{\mp}}_{\ga}^{\ \gb} \eta^{\pm}_{\gb} = 0\,.
\end{equation}
This enables one to introduce a basis of four lightlike vectors
\begin{equation}
    e_{I, \ga \dot{\ga}}:= (k^+_{\ga \dot{\ga}}, k^-_{\ga \dot{\ga}}, l^{+}_{\ga \dot{\ga}}, l^{-}_{\ga \dot{\ga}})
\end{equation}
consisting of two real $k^\pm$ and two complex conjugate\footnote{Recall that there are no four independent real lightlike vectors in the Lorentz signature.} vectors $l^{-+}_{\ga \dot{\ga}} = (l^{+-}_{\ga \dot{\ga}})^{\dag}$, as defined in \eqref{KS:spin}. The notation used is as follows:
\begin{subequations}\label{B:notation}
\begin{align}
    &v^2 = \frac{v^{\ga \dot{\gb}}v_{\ga \dot{\gb}}}{2} = v^- v^+ + v^{-+} v^{+-} = \det{v_{\alpha \dot{\beta}}}\,,\qquad 
    v^- v^+ = \frac{1}{2} v^-_{\ga \dot{\ga}} v^{+ \ga \dot{\ga}}\,,\\
    &\varkappa^2 = \frac{1}{2} \varkappa_{\ga \gb} \varkappa^{\ga \gb} = \det{\varkappa}\,,\qquad \bar\varkappa^2 = \frac{1}{2} \bar\varkappa_{\dal \dgb} \bar\varkappa^{\dal \dgb} = \det{\bar\varkappa}
\end{align}    
\end{subequations}
Notice the normalization chosen in the definition \eqref{KS:spin} is not relevant for the vectors to be lightlike. However, it becomes important when it comes to the geodesity condition \eqref{KS:geo}. 

A consequence of the constructed projectors is that these vectors satisfy the null condition and the geodesic equation. The null condition is obvious: it is sufficient to write the vectors in some basis of spinors $(\xi, \eta)$:
\begin{equation}
    k^+_{\ga \dot{\ga}} \sim \xi_\ga \xi_{\dot{\ga}}\,,\qquad k^-_{\ga \dot{\ga}} \sim \eta_\ga \eta_{\dot{\ga}}\,,\qquad l^{+}_{\ga \dot{\ga}} \sim \xi_{\ga} \overline{\eta}_{\dot{\ga}}\,,\qquad l^{-}_{\ga \dot{\ga}}\sim \overline{\xi}_{\dot{\ga}} \eta_{\ga}\,.
\end{equation}
To verify the geodesity condition, it is necessary to use the system of equations \eqref{ads:param-spin} to identify covariant derivatives of the Kerr-Schild vectors as found in \cite{Didenko:2009tc}. All multicopy consequences including \eqref{1:DC}, \eqref{1:DC-dual}, \eqref{2:DC} and \eqref{Ks:multicopy} follow from these relations.

\renewcommand{\theequation}{\Alph{appendix}.\arabic{equation}}
\addtocounter{appendix}{1} \setcounter{equation}{0}
\addtocounter{section}{1}
\addcontentsline{toc}{section}{\,\,\,\,\,D. $I_d:$ Rotating topological black hole}

\section*{D. $I_d:$ Rotating topological black hole}\label{sec: topological-kerr}
In this Appendix, we revisit type $I_d$ to highlight its relation to the Kerr solution via analytic continuation and to show its relevance to the so called topological black holes in AdS. For the case $I_d$ with nonzero constants $a$ and $b$ from Table 1, we can adopt the embedding coordinates from the Kerr solution through analytic continuation:
\begin{equation}
    t \rightarrow it\,,\quad r \rightarrow ir\,,\quad \theta \rightarrow i \theta\,,\quad \phi \rightarrow \phi\,,\quad a \rightarrow ia
\end{equation}
Equivalently, this means an analytic continuation of the embedding coordinates that preserves the signature of $\eta_{\ua\ub}$: 
$X^0 \leftrightarrow i X^2$, $X^1 \leftrightarrow i X^4$ and replacing $a \rightarrow ia$. In this way, we have the Killing field of $I_c$ and its centralizer get transformed into 
\begin{equation}\label{K:Kerr-top}
    K = J_{24} + \gl a J_{03}\,,\qquad G = J_{24} - \gl a J_{03}\,, \qquad [K, G] = 0\,.
\end{equation}
Notice, if the coordinates and the parameter $a$ are within the range  $|\gl r| > 1$, $r<a$, $\theta > 0$, and  $|\gl a| > 1$, then the norm of the Killing field $K$ is negative $KK < 0$.

The two generators \eqref{K:Kerr-top} form an Abelian subalgebra of $\so(2, 3)$:
\[
u(1) \oplus u(1) \subset \so(2,3)\,.
\]
To exhibit the Abelian structure explicitly, we construct such coordinates $(t, r, \theta, \varphi)$ in which the Abelian centralizer of $K$ is canonical 
\begin{align}
    J_{03} = -\gl^{-1} \partial_t\,,\qquad J_{24} = - \partial_{\varphi}\,.
\end{align}  
This problem is solved using the  Boyer-Lindquist hyperbolic coordinates 
\begin{subequations}\label{coord:BL2}  
\begin{align}
    &X^0 = \gl^{-1} \sqrt{\frac{(\gl^2 r^2 - 1)(1 + \gl^2 a^2\cosh{\theta}^2)}{1 + \gl^2 a^2}}\sinh{\gl t}\,,\\
    &X^3 = \gl^{-1} \sqrt{\frac{(\gl^2 r^2 - 1)(1 + \gl^2 a^2\cosh{\theta}^2)}{1 + \gl^2 a^2}}\cosh{\gl t}\,, \\
    &X^2 = \sqrt{\frac{a^2+r^2}{1 + \gl^2 a^2}}\sinh{\theta}\sin{\varphi}\,, \\
    &X^4 = \sqrt{\frac{a^2+r^2}{1 + \gl^2 a^2}}\sinh{\theta}\cos{\varphi}\,,\\
    &X^1 = r \cosh{\theta}\,,
\end{align} 
\end{subequations}
in which the AdS line element takes the following form:
\begin{equation}\label{AdS:BL-hyp}
    d\overline{s}_{AdS}^2 = -\frac{(-1+\gl^2 r^2)\Delta_{\theta}\, dt^2}{1+\gl^2 a^2} + \frac{\rho^2\, dr^2}{(-1+\gl^2 r^2)(r^2+a^2)} + \frac{\rho^2\, d\theta^2}{\Delta_{\theta}} + \frac{(r^2 + a^2)\sinh^2{\theta}\, d\varphi^2}{1+\gl^2 a^2}\,,
\end{equation}
where 
\begin{equation}
    \rho^2 := r^2 + a^2\cosh^2{\theta}\,,\qquad
    \Delta_{\theta} := 1+\gl^2 a^2\cosh^2{\theta}\,.
\end{equation}
We now choose the following vierbein $\e^m = \e^m_{\mu} dx^{\mu}$:
\begin{align}
    &\e^0 = \sqrt{\frac{-1+\gl^2 r^2}{1+\gl^2 a^2}}\sqrt{\Delta_{\theta}} dt\, , \qquad     
    \e^1 = \frac{\rho}{\sqrt{(r^2+a^2)(-1+\gl^2 r^2)}} dr\,, \\
    &\e^2 = \frac{\rho}{\sqrt{\Delta_{\theta}}}d\theta\, ,
    \qquad    \e^3 = \sqrt{\frac{r^2 + a^2}{1+\gl^2 a^2}}\sinh{\theta} d\varphi\, .
\end{align}
 The Killing vector in the chosen spacetime coordinates $v^{\mu}= (-1, 0, 0, -\gl^2 a)$ generates the global symmetry parameter $K_{AB}$ with the following Lorentz components:
\begin{align}
    v_{\ga \dot{\gb}}  &=  
    \frac{1}{\sqrt{1+\gl^2 a^2}}
    \begin{pmatrix}
        \sqrt{-1+\gl^2 r^2}\sqrt{\Delta_{\theta}} + \gl^2 a\sqrt{r^2+a^2}\sinh{\theta}&  0 \\
        0 & \sqrt{-1+\gl^2 r^2}\sqrt{\Delta_{\theta}} - \gl^2 a\sqrt{r^2+a^2}\sinh{\theta} \\ 
    \end{pmatrix}\label{v:top-Kerr}\\  
    \varkappa_{\ga \gb} &= 
    \frac{\gl^2 (r-ia\cosh{\theta})}{\rho \sqrt{1+\gl^2 a^2}}
    \begin{pmatrix}    \sqrt{\Delta_{\theta}}\sqrt{a^2+r^2} + a\sinh{\theta}\sqrt{-1+\gl^2r^2} &  0 \\
        0 &  -\sqrt{\Delta_{\theta}} \sqrt{a^2+r^2} + a\sinh{\theta}\sqrt{-1+\gl^2 r^2} 
\end{pmatrix}\,.\label{gk:top-Kerr}
\end{align}
From here, we find that
\begin{equation}
     \sqrt{-\varkappa^2}:=\sqrt{-\det\gk_{\al\gb}} = \gl^2 (r - ia\cosh{\theta})\,.
\end{equation}
This allows us to derive the zeroth copy using \eqref{zero}
\begin{equation}
    \phif  =  \frac{r}{r^2 + a^2\cosh^2{\theta}}
\end{equation}
and find the Casimir invariants  
\begin{equation}
    C_2 = -1 + \gl^2 a^2\,, \qquad 
    C_4 = 1 - 6 \gl^2 a^2 + \gl^4 a^4\,.
\end{equation}
It is of interest to note that the zeroth copy satisfies a nonlinear version of the scalar equation when written in the full metric, rather than in the background AdS 
\begin{equation}
    \boldsymbol{\nabla}^2\phif + 2\gl^2\phif + 2M \nabla_{\gamma}(\phif^{\beta \gamma} \boldsymbol\nabla_{\beta} \phi) = 0\,,
\end{equation}
where $\phif^{\beta \gamma} = \phif k^{\beta}k^{ \gamma}$.
Using \eqref{v:Kerr} and \eqref{gk:Kerr}, we extract the Kerr-Schild vectors from \eqref{KS:spin}:
\begin{subequations}
    \begin{align}
       &k^{\pm}_{\mu}dx^{\mu} = \frac{1+\gl^2 a^2\cosh^2{\theta}}{1+\gl^2a^2}dt \pm \frac{r^2+a^2\cosh^2{\theta}}{(-1+\gl^2 r^2)(a^2 + r^2)}dr - \frac{a\sinh^2{\theta}}{1+\gl^2 a^2}d\varphi\,,\\ 
       &l^{\pm}_{\mu}dx^{\mu} = \frac{-1+\gl^2 r^2}{1+\gl^2a^2}dt \pm \frac{i}{a \sinh{\theta}} \frac{r^2+a^2\cosh^2{\theta}}{(1+\gl^2 a^2\cosh^2{\theta})}d\theta - \frac{a^2+r^2}{a(1+\gl^2 a^2)}d\varphi\,. 
    \end{align}
\end{subequations}

The self-dual part of the Weyl tensor is 
\begin{equation}\label{Weyl:BL-hyp}
    C_{\alpha(4)}(\xi^{\alpha})^4 = M \frac{\left( a\sinh{\theta}\sqrt{-1+\gl^2 r^2}(\xi_1^2 + \xi_2^2) + \sqrt{(a^2 + r^2)\Delta_{\theta}}(\xi_1^2 - \xi_2^2)\right )^2}{\rho^2 (r-ia\cosh{\theta})^3 (1+\gl^2 a^2)}\,.
\end{equation}

As discussed in \cite{Vanzo:1997gw, Klemm:1997ea}, such a solution with a nonzero rotation parameter $a\neq 0$ forms a topological family of rotating solutions where the horizon section has negative curvature. Spacelike 2-surfaces formed by $(\theta, \phi)$ are Riemann surfaces of genus $g > 1$, and after a compactification, the metric describes higher genus black holes. The metric does not have global axial symmetry associated with the vector field $\partial_{\phi}$, but it still acts as a locally exact symmetry.

According to our classification, this solution is a representation of the orbit $\rom{1}_d$. Specifically, in the limit $a = 0$ the metric becomes a static topological black hole that we already described in Sec. \ref{static-hyp-bh}. Unlike the Kerr solution in AdS, where the rotation parameter $a$ is bounded above by value of the order of the AdS radius $\gl^{-1}$ with the limiting value $\gl a \rightarrow 1$ corresponding to a critically spinning case, in the topological case, such a limit reveals no specific behavior.

Similarly to the static black hole  solution with a planar horizon, which describes a black brane, there are rotating solutions with a flat (toroidal) horizon ($g=1$), which may also be generated with our technique.

\renewcommand{\theequation}{\Alph{appendix}.\arabic{equation}}
\addtocounter{appendix}{1} \setcounter{equation}{0}
\addtocounter{section}{1}
\addcontentsline{toc}{section}{\,\,\,\,\,E. Kerr–Schild form for $A$- and $B$-metrics}

\subsection*{E. Kerr–Schild form for $A$- and $B$-metrics}\label{A:B-metrics}
Here, we consider the generalized $A$ and $B$ metrics that include a nonzero cosmological constant, as introduced in \cite{Podolsk__2017}.
We show how the Kerr-Schild form is related to the metrics of these types.

\paragraph{Kerr–Schild form of the $A$-metrics}

The explicit Kerr-Schild form for static, axially symmetric black-hole solutions in AdS\textsubscript{4} with horizons of different topology are given by:
\begin{equation}
    ds^2 = \overline{g}_{\mu \nu} + {2}M\phi k_{\mu} k_{\nu}\,,
\end{equation}
    where the background metric is defined in \eqref{AdS:AI}, \eqref{AdS:AII}
\begin{equation}
    d\overline{s}^2 = - f_{\epsilon}dt^2 + f_{\epsilon}^{-1} dr^2 + r^2 d \Omega^2 \,, 
\end{equation}
where 
\begin{equation}
    f_{\epsilon} = \epsilon_2 + \gl^2 r^2
\end{equation}
and
\begin{equation}
    d\Omega^2 = 
    \begin{cases}
    d\theta^2 + \sin^2{\theta}  d\varphi^2\quad &\text{for}\quad \epsilon_2 = 1\,,\\
    dx^2 + dy^2\quad &\text{for}\quad \epsilon_2 = 0\,,\\
    d\chi^2 + \sinh^2{\chi} d \varphi^2\quad &\text{for}\quad \epsilon_2 = -1\,,
    \end{cases}
\end{equation}
while the Kerr-Schild vector and a scalar $\phi$ read
\begin{equation}
    k_\mu dx^{\mu} = dt + f_{\epsilon}^{-1} dr\,,\qquad \phi = \frac{1}{r}\,.
\end{equation}
By introducing a new $t$ coordinate 
\begin{equation}
    dt \to dt + {2}M \phi f^{-1}_{\epsilon} (f_{\epsilon} - {2}M \phi)^{-1}dr\,,
\end{equation}
we pass from the Kerr-Schild representation to the Schwarzschild type. One can readily verify that the resulting metric takes the familiar canonical form 
\begin{equation}
    ds^2 = - (f_{\epsilon} - {2}M \phi)dt^2 + \frac{d r^2}{(f_{\epsilon} - {2}M \phi)} + r^2 d\Omega^2 \,, 
\end{equation}
To facilitate comparison with conventions in the literature, it is customary to perform the {sign change \(M \rightarrow -M\)} and the substitution $r \rightarrow p$ and $y \rightarrow \varphi$. Let us also introduce
\begin{align}
    q =
    \begin{cases}
    \cos{\theta} &\text{for}\quad \epsilon_2 = 1\,,\\
    x\quad &\text{for}\quad \epsilon_2 = 0\,,\\
    \cosh{\chi}\quad &\text{for}\quad \epsilon_2 = -1\,.
    \end{cases}
\end{align}
After the additional transformation is performed, we arrive at the so called $A$-metrics that belong to a subset of the large Pleba\'nski-Demia\'nski family \cite{Plebanski:1976gy} -- namely, the type D Robinson-Trautman class of expanding solutions with non-zero expansion. In unified form, $A$-metrics with negative $\gl^2 = -\frac{\Lambda}{3}$ are written as (see {\it e.g.,} $\cite{Podolsk__2017}$): 
\begin{equation}\label{A-metrics}
    ds^2 = -(\epsilon_2 + \gl^2 p^2 + \frac{2M}{p})dt^2 + (\epsilon_2 + \gl^2 p^2  + \frac{2M}{p})^{-1}dp^2 + \frac{p^2}{\epsilon_0 - \epsilon_2 q^2} dq^2 + p^2 (\epsilon_0 - \epsilon_2 q^2) d\varphi^2
\end{equation}
The Schwarzschild solution is obtained by setting the discrete parameters $\epsilon_0 = \epsilon_2 = 1$ in the metric above, and it is called the {\it $A\rom{1}$-metric}. The spacetime metric of a static black hole with a hyperbolic horizon takes the form reached with the discrete parameters $\epsilon_0 = \epsilon_2 = -1$, describing the {\it A$\rom{2}$-metric}. Lastly, the canonical form of {\it $A\rom{3}$-metric} with discrete parameters $\epsilon_0 = 1$ and $\epsilon_2 = 0$  describes a black hole with a planar horizon in AdS. The non-zero expansion $\boldsymbol{\theta}$ allows us to write the classical Kerr-Schild ansatz with an existing real zero copy $\phi \sim \boldsymbol\theta = \nabla_{\mu}k^{\mu}$.

\paragraph{Kerr–Schild form of the $B$-metrics}

The next class of metrics belongs to the non-expanding subfamily of the
Pleba\'nski-Dema\'nski type D solutions. In the zero-expansion limit $\boldsymbol\theta = 0$ of the repeated principal null congruences, it belongs to the Kundt type D class; its $\gamma = 0$ sub-case known as {\it B-metrics} \cite{Podolsk__2017, Podolsk__2018}:
\begin{equation}\label{B-metrics}
    ds^2 = (\epsilon_2 + \gl^2 p^2 + \frac{2M}{p})d\varphi^2 + (\epsilon_2 + \gl^2 p^2 + \frac{2M}{p})^{-1}dp^2 + \frac{p^2}{\epsilon_0 - \epsilon_2 q^2} dq^2 - p^2 (\epsilon_0 - \epsilon_2 q^2) dt^2
\end{equation}
$A$ and $B$ metrics both have a symmetry algebra of dimension four and the Weyl curvature 
\begin{equation}
    \Psi_2 = \frac{M}{p^3}\,.
\end{equation}
By analogy with the $A$-metrics, it is customary to distinguish several subtypes depending on the value of the discrete curvature parameter \(\epsilon_2\). Setting \(\epsilon_2 = 1\) yields the {\it \(BI\)-metric}, \(\epsilon_2 = -1\) the {\it \(BII\)-metric}, and finally \(\epsilon_2 = 0\) the {\it \(BIII\)-metric}. 

$B$-metrics admit the limit to $AdS$ as $M \rightarrow 0$\footnote{The case of $BII$ is subtle in that the limit $M\to 0$ does not respect the signature of the metric for a cosmological constant  $\Lambda>0$. For $\Lambda<0$, the AdS limit can be achieved within a specific range of coordinates; see \cite{Podolsk__2017}.}. We have not found a standard real Kerr-Schild representation over AdS $g = \bar g + M \phi k k$, with a real null geodesic Kerr-Schild vector $k$ for the $B$-metrics. Instead, the representation we found admits a complex-conjugate pair of Kerr-Schild vectors. 

To show the relation between canonical $B$-metrics and their Kerr-Schild form, we start with the background metric:
\begin{equation}
    d\overline{s}^2 =  f_{\epsilon}dt^2 + f_{\epsilon}^{-1} dr^2 + r^2 d \Omega^2 \,, 
\end{equation}
where 
\begin{equation}
    f_{\epsilon} = \epsilon_2 + \gl^2 r^2
\end{equation}
and
\begin{equation}
    d\Omega^2 = 
    \begin{cases}
    -d\chi^2 + \cosh^2{\chi}  d\varphi^2\quad &\text{for}\quad \epsilon_2 = 1\,,\\
    -dx^2 + dy^2\quad &\text{for}\quad \epsilon_2 = 0\,,\\
    d\chi^2 - \cosh^2{\chi} d \varphi^2\quad &\text{for}\quad \epsilon_2 = -1\,.
    \end{cases}
\end{equation}
The metric admits the following complex-conjugate pair of Kerr-Schild vectors and a scalar zeroth copy:
\begin{equation}
    l_\mu dx^{\mu} = dt + i f_{\epsilon}^{-1} dr\,,\qquad \Tilde{\phif} = \frac{1}{r}\,.
\end{equation}
By performing a change of variable $t$
\begin{equation}
    dt \to dt - i{2}M \Tilde{\phif} f^{-1}_{\epsilon} (f_{\epsilon} + {2}M \Tilde{\phif})^{-1}dr\,,
\end{equation}
the metric can be brought into the desired canonical form,
\begin{equation}
    ds^2 = (f_{\epsilon} + {2}M \phi)dt^2 + \frac{d r^2}{(f_{\epsilon} + {2}M \phi)} + r^2 d\Omega^2 \,, 
\end{equation}
The relation to the \(B\)-metrics is then established through the identifications {(with the same $M$)} $r \rightarrow p$ together with:
\begin{align}
    q =
    \begin{cases}
    \sinh{\chi} &\text{for}\quad \epsilon_2 = 1\,,\\
    x\quad &\text{for}\quad \epsilon_2 = 0\,,\\
    \sinh{\chi}\quad &\text{for}\quad \epsilon_2 = -1\,.
    \end{cases}
\end{align}
We thus conclude that the {\it $BI$-metric} corresponds to the parameter choice $\epsilon_0 = -1$, $\epsilon_2 = 1$, with $\varphi \leftrightarrow t$; the {\it $BII$-metric} corresponds to $\epsilon_0 = 1$, $\epsilon_2 = -1$ again with $\varphi \leftrightarrow t$; and the {\it $BIII$-metric} corresponds to $\epsilon_0 = -1$, $\epsilon_2 = 0$, with $\varphi = t$, and $t = y$.

\paragraph{AdS as $A$ and $B$ space} 

AdS spacetime admits the following $B$-form
\begin{equation}\label{AdS:B-form}
    ds^2 = p^2(-Qdt^2 + \frac{dq^2}{Q}) + \frac{dp^2}{P} + P\gl^{-2}d\phi^2
\end{equation}
with
\begin{equation*}
    Q = \epsilon_0 - \epsilon_2 q^2,\quad P = \epsilon_2+\gl^2p^2\,,
\end{equation*}
where $\epsilon_0\,, \epsilon_2 = \pm 1, 0$ are two independent discrete parameters. Recall, the Gaussian curvature $\epsilon_2$ of the Lorentzian 2-surfaces on which $p$ and $\phi$ are constants determines three subclasses of the $B$-metrics.

The direct counterpart of this family is the $A$-metric, which can be obtained by formal substitutions of the coordinates $p\rightarrow r$, $t\rightarrow i \phi$, and $a \phi \rightarrow i t$. 
\begin{equation}
    ds^2 = p^2(Qd\phi^2 + \frac{dq^2}{Q}) + \frac{dp^2}{P} - P dt^2\,,
\end{equation}
where
\begin{equation*}
    P = \epsilon_2+\gl^2r^2.
\end{equation*}
As it was discussed in \cite{Podolsk__2017}, the $i$ ``formally'' interchanges the temporal and spatial character of the coordinates $t$ and $\phi$. 

Topology of the space \eqref{AdS:B-form} is a warped product of two two-dimensional manifolds: two-dimensional de Sitter space $dS_2$, two-dimensional Minkowski space $M_2$, and two-dimensional anti-de Sitter space $AdS_2$ (corresponding to $\epsilon_2 = 1, 0, -1$, respectively), spanned by the coordinates $(t, q)$, together with the hyperbolic space $H^2$ with warp factor $p^2$. The other discrete parameter $\epsilon_0$ only changes the specific coordinate foliation of the two dimensional manifold spanned by $(t, q)$, but does not change the topology of space-time.

There are two manifest symmetries of the A and B metrics generated by the vector fields $\partial_t$ and $\partial_{\phi}$.


\renewcommand{\theequation}{\Alph{appendix}.\arabic{equation}}
\addtocounter{appendix}{1} \setcounter{equation}{0}
\addtocounter{section}{1}
\addcontentsline{toc}{section}{\,\,\,\,\,F. Killing fields for type $II_{b=0}$ spacetime}

\section*{F. Killing fields for type $II_{b=0}$ spacetime}\label{Kill-equation}
    Here we will shortly provide a step by step solution of the Killing equation
    \begin{equation}
        \mathcal{L}_{\xi}g_{\mu \nu} = \xi^\rho \partial_{\rho} g_{\mu \nu} + g_{\rho \nu} \partial_{\mu} \xi^{\rho} + g_{\mu \rho} \partial_{\nu} \xi^\rho = 0
    \end{equation}
corresponding to the Siklos type metric obtained in \eqref{II0:H}, which is defined using coordinates $x^{\mu} = (u, v, x, r)$. The nonzero metric components are
    \begin{equation}
        g_{rr} = g_{xx} = g_{uv} = \frac{1}{r^2}\,,\qquad g_{uu} = \frac{H}{r^2}\,. 
    \end{equation}
    A general Killing field is parametrized by certain functions $U = U(x^{\mu})$, $V = V(x^{\mu})$, $X = X(x^{\mu})$ and  $R = R(x^{\mu})$ 
    \begin{align}
        \xi = U \partial_u + V \partial_v + X \partial_x + R \partial_r\,.      
    \end{align}
    Overall, we have ten independent components of the Killing equations. So, we solve them step by step. The nontrivial equations are
    \begin{subequations}\label{Killing-eqs}
    \begin{align}
        &\mathcal{L}_{\xi}g_{vv} = 0 \quad\Rightarrow\quad  U_v = 0 \quad\Rightarrow\quad U = U(u, x, r)\,, \\
        &\mathcal{L}_{\xi}g_{vr} = 0 \quad\Rightarrow\quad  U_r + R_v = 0 \quad\Rightarrow\quad R = R_0(u, x, r) - v U_r\,,\\
        &\mathcal{L}_{\xi}g_{vx} = 0 \quad\Rightarrow\quad U_x + X_v = 0 \quad\Rightarrow\quad X = X_0(u, x, r) - v U_x\,,\\
        &\mathcal{L}_{\xi}g_{uv} = 0 \quad\Rightarrow\quad U_u + V_v - \frac{2}{r}R = 0\,, \\        
        &\mathcal{L}_{\xi}g_{xx} = 0 \quad\Rightarrow\quad R - r X_x = 0\,, \\   &\mathcal{L}_{\xi}g_{rr} = 0 \quad\Rightarrow\quad R - r R_r = 0\,, \\
        &\mathcal{L}_{\xi}g_{rx} = 0 \quad\Rightarrow\quad R_x + X_r = 0\,.
    \end{align}
    \end{subequations}  
These equations can be easily solved. From the first three equations, we have
\begin{equation}
    \partial_r U_x = 0 \quad\Rightarrow\quad  U_x = f(u, x)
\end{equation}
Integration by $x$ amounts to a factorized representation in terms of two functions
\begin{equation}\label{U:FG}
    U(u, x, r) = F(u, x) + G(u, r)\,,\qquad F_x = f\,.
\end{equation}
Substituting the functions $R$ and $X$ constrained in the second and third equations \eqref{Killing-eqs} into the last three equations $\eqref{Killing-eqs}$ gives us the following system:
\begin{equation}
    U_r = r U_{xx}\,,\qquad U_r = r U_{rr}\,.
\end{equation}
Now we use \eqref{U:FG} and substitute it into $U_r = r U_{xx}$ with the following result:
\begin{equation}
    G_r(u, r) = r  F_{xx}(u, x)\quad \Rightarrow\quad F_{xx}(u, x) = a(u)
\end{equation}
for some $a(u)$. Consequently, 
\begin{equation}
    F(u, x) = \frac{a(u)}{2} x^2 + b(u) x + c(u)
\end{equation}
and
\begin{equation}
    G_r = r a(u) \quad\Rightarrow\quad G(u, r) = \frac{a(u)}{2}r^2 + d(u)\,.
\end{equation}
The $U_r = r U_{rr}$ equation is satisfied automatically. Collecting everything together, we arrive at
\begin{equation}
    U = \frac{a(u)}{2} (x^2 + r^2) + b(u) x + e(u)\,,\qquad U_x = ax + b\,,\qquad U_r = a r\,.
\end{equation}
From the last three equations in $\eqref{Killing-eqs}$, we further have
\begin{equation}
    R - r R_r = 0 \quad\Rightarrow\quad R_0 = r A(u, x)\,,\qquad U_r = r U_{xx}\,,
\end{equation}
\begin{equation}
    R - r X_x = 0 \quad\Rightarrow\quad \partial_x X_0 = A(u, x)\,, \qquad U_r = r U_{rr}\,.
\end{equation}
As before, integration by $x$ leads to the factorized form
\begin{equation}
    X_0(u, x, r) = S(u, x) + B(u, r)\,,\qquad S_x = A\,.
\end{equation}
From the last equation in $\eqref{Killing-eqs}$ we obtain 
\begin{equation}
    R_x + X_r = 0 \quad\Rightarrow\quad B_r(u, r) + r A_x(u, x) = 0 \quad\Rightarrow\quad A_x = \alpha(u)\,.
\end{equation}
for some $\alpha(u)$, then
\begin{equation}
    A = \alpha(u) x + \beta(u) \Rightarrow S = \frac{\alpha(u)}{2} x^2 + \beta(u) x + \epsilon(u) 
\end{equation}
and
\begin{equation}
    B = - \frac{\alpha(u)}{2}r^2 + \delta(u)\,.
\end{equation}
Eventually, we find
\begin{align}
    &R = r (\alpha(u) x + \beta(u)) - v r a(u)\,,\\
    &X = \frac{\alpha(u)}{2}(x^2 - r^2) + \beta(u)x + \gga(u) - v(a(u) x + b(u) u))\,.
\end{align}
For brevity, we omit the arguments of the functions  $\alpha, \beta, \gamma, a, b, e$, assuming, {\it e.g.,} $a':=\partial_u a$.
Substituting the obtained functions into the equation $V_v = \frac{2}{r}R - U_u$ gives us
\begin{equation}
    V_v = 2(\alpha x + \beta) - 2a v - \frac{a'}{2}(x^2 + r^2) - b'x - e'\,.
\end{equation}
Integration by $v$, we find
\begin{equation}
    V = -a v^2 + v(2(\alpha x+ \beta) - \frac{a'}{2}(x^2 + r^2) - b'x - e') + V_0(u, x, r)\,.
\end{equation}

Let us now proceed to examine the component  $ux$ of the Killing equation
\begin{equation}
    \mathcal{L}_{\xi}g_{ux} = 0 \quad\Rightarrow\quad X_u + H U_x + V_x = 0\,.
\end{equation}
Using the explicit form of $X_u, U_x, V_x$, we collect the coefficient in front of $v$
\begin{equation}
    2 \alpha - 2a' x - 2b' = 0 \quad\Rightarrow\quad a'(u) = 0\,,\qquad \alpha(u) = b'(u)\,.
\end{equation}
along with $v$ independent
\begin{equation}
    (V_0)_r = - (\frac{\alpha'}{2}(x^2-r^2) + \beta' x + \gamma') - H(ax + b)\,.
\end{equation}
Similarly, with the component $ur$
\begin{equation}
    \mathcal{L}_{\xi}g_{ur} = 0 \quad\Rightarrow\quad R_u + H U_r + V_r = 0\,.
\end{equation}
Setting $U_r = a r$, $R_u = r(\alpha' x + \beta')$, and  $V_r = (V_0)_r$, one finds
\begin{equation}
    (V_0)_r = -r (\alpha' x+ \beta') - ar H\,.
\end{equation}
The consistency $(V_0)_{xr} = (V_0)_{rx}$ implies
\begin{equation}
    2 \alpha' r - (ax + b)H_r + ar H_x = 0\,.
\end{equation}
Substituting the Siklos profile function $H = -(x^2 + r^2) + \mu r^3$, we have 
\begin{equation}
    2\alpha' + 2b - 3\mu r(ax + b) = 0 \quad\Rightarrow\quad a = b = \alpha' = 0\,.
\end{equation}
From $\alpha = b'$, it follows that $\alpha = 0$. These conditions lead us to the following formulas:
\begin{equation}
    U = e\,,\qquad X = \beta x + \gamma\,,\qquad R = r \beta\,.
\end{equation}
The component $uv$ simplifies $V_v = \frac{2}{r}R - U_u$:
\begin{equation}
    V = v(2\beta - e') + V_0\,,\qquad (V_0)_x = -(\beta' x+ \gamma')\,,\qquad (V_0)_r = -r \beta'\,.
\end{equation}
Performing integration, we obtain 
\begin{equation}
    V_0 = -\frac{\beta'}{2}(x^2 + r^2) - \gamma' x + s(u)\,.
\end{equation}
Finally, the final component $(uu)$ of the Killing equation yields 
\begin{equation}
    \mathcal{L}_{\xi}g_{uu} = 0 \quad\Rightarrow\quad X H_x + R(H_r - \frac{2H}{r}) + 2HU_u + 2V_u = 0\,.
\end{equation}
Having the functions $H_x$, $H_r$, $U_u$, $V_u$, we collect the coefficient in front of $v$:
\begin{equation}
    2\beta' - e'' = 0\,,
\end{equation}
along with the $v$ independent
\begin{equation}
    \beta + 2 e' = 0\,,\qquad \beta'' = -2e'\,,\qquad \gamma'' + \gamma = 0\,,\qquad s' = 0\,.
\end{equation}
Solving these conditions, one finds
\begin{equation}
    \beta = 0,\quad e = e_0 = const,\quad s =s_0= const,\quad \gamma'' + \gamma = 0\,,
\end{equation}
where the harmonic equation gives
\begin{equation}
    \gamma = c_1 \cos{u} + c_2 \sin{u}
\end{equation}
Finally, we substitute the obtained results into the Killing vector
\begin{equation}
    \xi = e_0 \partial_u + s_0 \partial_v + \gamma(u) \partial_x - \gamma'(u) x \partial_x\,,\qquad \gamma = c_1 \cos{u} + c_2 \sin{u}\,.
\end{equation}
From the expression above, we see that there are four linearly independent Killing vectors that can be chosen in the following form:
\begin{align}
    &K = \partial_v\,, \qquad J = \partial_u\,,\\
    &P_1 = \cos{u} \partial_x + x \sin{u} \partial_v\,,\qquad P_2 = \sin{u}\partial_x -  x \cos{u} \partial_v\,.
\end{align}

    
\end{document}